\documentclass{aa} 

\usepackage{graphicx}
\usepackage{txfonts}
\usepackage{colortbl}

\newcommand{\hi}{\textsc{H\,i}}
\newcommand{\hb}{H$\beta$}
\newcommand{\ha}{H$\alpha$}
\newcommand{\hii}{\textsc{H\,ii}}
\newcommand{\hei}{He\,\textsc{i}}
\newcommand{\siii}{[S\,\textsc{iii}]}

\newcommand{\oii}{[O\,\textsc{ii}]}

\newcommand{\sii}{\textsc{[S\,ii]}}

\newcommand{\feiii}{[Fe\,\textsc{iii}]}

\newcommand{\cliii}{[Cl\,\textsc{iii}]}
\newcommand{\oi}{[\textsc{O\,i}]}

\newcommand{\niA}{\textsc{[N\,i]}}

\newcommand{\nii}{\textsc{[N\,ii]}}

\newcommand{\ariii}{[Ar\,\textsc{\,iii]}}
\newcommand{\oiii}{\textsc{[O\,iii]}}
\newcommand{\heii}{He\,\textsc{ii}}

\usepackage{xcolor}

\definecolor{olivegreen}{rgb}{0.02, 0.8, 0.24}
\newcommand\stoutAna{\bgroup\markoverwith{\textcolor{olivegreen}{\rule[0.5ex]{2pt}{0.4pt}}}\ULon}

\definecolor{crimson}{rgb}{0.90, 0.08, 0.20}
\newcommand\stoutAnaRed{\bgroup\markoverwith{\textcolor{crimson}{\rule[0.5ex]{2pt}{0.4pt}}}\ULon}

\definecolor{orange}{rgb}{1.00, 0.55, 0.00} 
\newcommand\stoutAnaOran{\bgroup\markoverwith{\textcolor{orange}{\rule[0.5ex]{2pt}{0.4pt}}}\ULon}

\definecolor{light-gray}{gray}{0.90}

\begin{document}

   \title{\object{UM\,462}, a local green pea galaxy analogue under the MUSE magnifying glass}
   \author{Ana Monreal-Ibero
          \inst{1}
          \and
          Peter M. Weilbacher
          \inst{2}
          \and
          Genoveva Micheva
          \inst{2}
          \and
          Wolfram Kollatschny 
          \inst{3}
          \and
          Michael Maseda
          \inst{4}
          }
    \institute{
    Leiden Observatory, Leiden University, PO Box 9513, 2300 RA Leiden, The Netherlands\\
    \email{monreal@strw.leidenuniv.nl}
    \and 
    Leibniz-Institute for Astrophysics Potsdam (AIP), An der Sternwarte 16, 14482 Potsdam, Germany
    \and
    Institut f\"ur Astrophysik und Geophysik, Universit\"at G\"ottingen, Friedrich-Hund Platz 1, 37077 G\"ottingen, Germany
    \and
    Department of Astronomy, University of Wisconsin-Madison, 475 N. Charter St., Madison, WI 53706 USA
             }

   \date{Received 10 January, 2023; accepted 03 May, 2023}


  \abstract
   {Stellar feedback in high-redshift galaxies plays an important, if not dominant, role in the re-ionisation epoch of the Universe.  Because of their extreme star formation (SF), the relatively closer
green pea (GP) galaxies are postulated as favorite local laboratories, and analogues to those high-redshift galaxies. However,  at their typical redshift of $z\sim0.2$, the most intimate interaction between stars and the surrounding interstellar medium cannot be disentangled. Detailed studies of blue compact dwarf (BCD) galaxies  sharing properties with GP galaxies are necessary to anchor our investigations on them. 
   }
   {We want to study in detail UM\,462, which is a BCD with emission line ratios and equivalent widths, stellar mass, and metallicity similar to those observed in GP galaxies, and thus it is ideally suited as a corner stone and reference galaxy.}
   {We use high-quality optical  integral field spectroscopy data obtained with MUSE on the ESO Very Large Telescope.}
   {The electron density ($n_e$) and temperature ($T_e$) were mapped. Median $T_e$ decreases according to the sequence \siii$\rightarrow$\nii$\rightarrow$\hei. Furthermore, $T_e$(\siii) values are $\sim$13\,000~K, and uniform within the uncertainties over an area of $\sim20^{\prime\prime}\times8^{\prime\prime}$ ($\sim$1.4 kpc$\times$0.6 kpc). 
The total oxygen abundance by means of the direct method is 12+$\log$(O/H)$\sim$8.02 and homogenous all over the galaxy within the uncertainties, which is in stark contrast with the metallicities derived from several strong line methods.
This result calls for a systematic study to identify the best strategy to determine reliable metallicities at any location within a galaxy.
  The strong line ratios used in the BPT diagrams and other ratios tracing the ionisation structure were mapped. They are compatible with plasma ionised by massive hot stars. However, there is a systematic excess in the \oi/\ha\, ratio, suggesting an additional mechanism or a complex relative configuration of gas and stars.
  The velocity field for the ionised gas  presents receding velocities in the east and approaching velocities in the west and south-west with velocity differences of $\Delta v \sim 40$~km~s$^{-1}$, but it is not compatible with simple rotation. The most striking feature is a velocity stratification in the area towards the north with redder velocities in the high ionisation lines and bluer velocities in the low ionisation lines. This is the only area with velocity dispersions clearly above the MUSE instrumental width, and it is surrounded by two $\sim$1 kpc-long structures nicknamed 'the horns'. We interpret the observational evidence in that area as a fragmented super-bubble fruit of the stellar feedback and it may constitute a preferred channel through which Lyman continuum photons from the youngest generation of stars can escape.
The galaxy luminosity is dominated by a young (i.e. $\sim$6 Myr) stellar population that contributes only 10~\% to the stellar mass, as derived from the modelling of the stellar continuum. The most recent SF seems to propagate from the outer to the inner parts of the galaxy, and then from east to west. We identified a supernova remnant  and Wolf-Rayet stars -- as traced by the red bump -- that support this picture. The direction of the propagation implies the presence of younger Wolf-Rayet stars at the maximum in \ha. These may be detected by deep observations of the blue bump (not covered here).
  }
   {The ensemble of results exemplifies the potential of 2D detailed spectroscopic studies of dwarf star-forming galaxies at high spatial resolution as a key reference for similar studies on primeval galaxies.}


   \maketitle
%


\section{Introduction}

The Galaxy Zoo project \citep{Lintott08} revealed a group of compact, extremely star-forming galaxies at low redshift ($z\sim0.2$), named 'green peas' (GPs) due to their point-like appearance and green colours in the Sloan Digital Sky Survey (SDSS) plates.
Their greenish aspect is due to extremely strong emission lines in the optical, in particular \oiii$\lambda$5007, with equivalent widths (EWs) of typically several hundreds to $\sim$1000~\AA. 
Their mass is relatively low   (M$\sim10^{8.5}-10^{10.0}$~M$_\odot$) and they have  one of the highest star formation (SF) rates in the local Universe, reaching values of up to $\sim$10~M$_\odot$~yr$^{-1}$ \citep{Cardamone09}. 
\citet{Amorin10} demonstrated that these were low-metallicity galaxies ($\bar{Z}\sim0.2~Z_\odot$). Later on, \citet{Izotov11} showed that the original 80 GPs were part of a more general population extending over a larger redshift range. They identified $\sim$800 of these such objects in the SDSS and named them luminous compact galaxies.

Because of their characteristics, GP galaxies constitute a group of galaxies similar to high-redshift galaxies.
Thus, they can act as  nearby (by comparison) laboratories where one can study the role of the SF in environments as extreme as those occurring in the early Universe. 
This SF crucially influences the subsequent evolution of the host galaxy at different levels, in a far from obvious manner. Specifically, massive stars release huge amounts of momentum, energy, ionising photons, and processed material during their short lives (via stellar winds) and early deaths (via supernova explosions), which has an impact in the neighbouring interstellar medium (ISM). Depending on the geometry of the system, this may extend to the whole host galaxy, its circumgalactic medium, or even beyond, creating outflows of processed material that may escape from the galaxy (i.e. superwinds) or fall back as recycled material ready to form new stars (i.e. fountains).  This feedback, which is complex by construction \citep{Tumlinson17,SanchezAlmeida14}, regulates further SF in the galaxy and modulates its chemical evolution. 
In addition,  low-mass star-forming galaxies are one of the favorite candidates as sources of the Lyman continuum (LyC) radiation, responsible  for the re-ionisation of the Universe after the cosmic ‘Dark Ages’  \citep[e.g.][]{Duncan15,Bouwens15b,Bouwens16}.
At low redshift, systems with a large escape fraction of ionising flux are scarce \citep[e.g.][]{Borthakur14} with  \object{Haro 11} being perhaps the most famous example of a local LyC leaker \citep{Bergvall06}.
In contrast, the GP galaxies (or luminous compact galaxies) display a particularly high LyC detection rate \citep[e.g.][]{Izotov16,Izotov21,Flury22a,Flury22b}.

High-dispersion spectroscopy shows that GP galaxies present complex emission line profiles with distinct kinematics components with different physical and chemical properties \citep[e.g.][]{Amorin12,Hogarth20}.  Integral field spectroscopy (IFS) observations have also revealed different kinematic components in these systems \citep[e.g.][]{Bosch19}. They also suggest that some of these galaxies, but not all, may have gone through a minor merger or interaction event. Thus minor mergers are a possible but not the necessary cause of the strong burst of SF \citep{Lofthouse17}.

The above results suggest that GP  galaxies possess a complex internal structure where several mechanisms shaping the further evolution of the galaxy and its milieu may live together. In addition to the heavy SF, these may include metal-poor gas inflows, processed material outflows, and channels of low H\,\textsc{i} column density through which LyC photons could leak. 
However, at the typical distances at which GP galaxies are found, the angular scale ($\sim$3~kpc arcsec$^{-1}$) is not enough to resolve their inner structure and the most intimate interaction between the sites of extreme SF and the surrounding ISM. This is nevertheless very much needed to get an adequate understanding on the impact of this interaction. Spectroscopic studies at higher spatial resolution, allowing for a proper mapping  of the physical and chemical properties of the gas and stars, as well as the kinematics, are necessary to put an anchor on our findings on GP galaxies. High-quality IFS observations of nearby systems similar to GP galaxies are thus instrumental for that.

Local GP analogues can be found among nearby blue compact dwarf (BCD) galaxies \citep[see e.g. ][ for reviews on these galaxies]{Kunth00,Annibali22,Henkel22}. These are nearby low-metallicity gas-rich galaxies, undergoing an episode of intense SF. They have low ($M_B \gtrsim -18$) total luminosity and  typically large ($\sim10^8-10^9$~M$_\odot$) amounts of neutral hydrogen \citep{Thuan81}. Once it was thought that they could be genuinely young galaxies forming their first generation of stars. However, deep imaging observations have shown that they contain an extended underlying host population \citep[see e.g.][ and references therein]{Amorin09}.
As a consequence of their massive and on-going SF, all of them present spectra with strong emission lines, similar to those of \hii\, regions, many of them with EW($\lambda$5007) comparable to those of GP galaxies. Even, if they may be fainter than original GP galaxies, their proximity allows us to see the inner workings of a GP(-like) galaxy with an unattainable level of detail in the original GP sample.

\begin{table}[th!]
     \centering
     \caption[]{Basic data for UM\,462. \label{basicdata}}
             \begin{tabular}{ccccccccc}
\hline
\hline
            \noalign{\smallskip}
Parameter & Value & Ref.\\
           \noalign{\smallskip}
           \hline
           \noalign{\smallskip}
Name               & UM~462                & (a)\\
Other designations & UGC 06850; Mrk 1307; & (a)\\
    & SDSS~115237$-$022806  & \\
RA (J2000.0)       & 11h52m37.2s             & (a)\\
Dec (J2000.0)       & $-$02d28m09.9s            & (a)\\
$z$                & 0.003527                  & (b)\\
$D$(Mpc)           & 14.4                       & (b)\\
scale (pc/$^{\prime\prime}$) &  70         &    (b)\\
$E(B-V)_{Gal}$ & 0.017 &    (c)\\
Morphology & Irr & (d) \\
c(\hb) & 0.19$\pm$0.04 & (b)\\
12+$\log$(O/H)  & 8.03$\pm^{0.04}_{0.07}$ & (b) \\
$Z/Z_\odot$        & $\sim$0.22$^{(\ast)}$ &  (b)\\
$\log$(N/O) & $-$1.85$\pm^{0.14}_{0.15}$  &  (b)\\
$M_B$              & $-16.36$                 & (e)\\ 
$g-r$              & 0.08                    & (e)\\
$M_{HI} (M_\odot)$ & $3.3\times10^8$ & (e)\\
$M_\star (M_\odot)$ & $2.1\times10^8$ & (e)\\
SFR ($M_\odot$ yr$^{-1}$) & 0.11 & (e)\\
$U-B$              & $-0.66\pm0.15$   & (f)\\
$B-V$              & $0.36\pm0.05$    & (f)\\
$V-R$              & $0.08\pm0.25$    & (f)\\
$V-I$              & $0.07\pm0.19$    & (f)\\
$V-K$              & $1.77\pm0.07$    & (f)\\
$H-K$              & $0.39\pm0.07$    & (f)\\
%
            \noalign{\smallskip}  
            \hline
         \end{tabular}

\begin{flushleft}
$^{(\ast)}$ We assumed $12+\log(\mathrm{O/H})_\odot = 8.69$ \citep{Asplund21}.\\
$^{\mathrm{(a)}}$ NASA/IPAC Extragalactic Database (NED).\\
$^{\mathrm{(b)}}$ \citet{James10}.\\
$^{\mathrm{(c)}}$ \citet{Schlafly11}.\\
$^{\mathrm{(d)}}$ \citet{Ann15}.\\
$^{\mathrm{(e)}}$ \citet{Paudel18}.\\
$^{\mathrm{(f)}}$ \citet{Micheva13}.\\
\end{flushleft}
\label{tabbasicdata}
\end{table}
        
IFS has been widely used to study nearby BCDs in the last $\sim$15 years \citep[e.g.][]{GarciaLorenzo08,Bordalo09}, allowing us, for instance, to map the physical properties and chemical content of the ionised gas in these systems \citep[e.g.][]{MonrealIbero12a,James13b,Kumari17}, to identify outflows \citep[e.g.][]{MonrealIbero10a,Micheva19}, or to evaluate the content and distribution of massive stars in a relatively exotic evolutive phase, such as the Wolf-Rayet (WR) phase \citep[e.g.][]{Westmoquette13,MonrealIbero17b}.
In view of the characteristics of the IFS-based instruments to date, works so far have mostly focussed on the innermost regions of the galaxy, more specifically the main sites of SF. Thanks to the advent of the Multi-Unit Spectroscopic Explorer (MUSE) \citep{Bacon10}, with a field of view comparable to the angular size of many BCDs, we can now study these archetypical objects with unprecedented detail, in a more  comprehensive manner, and paying attention to the phenomena at play beyond these main sites of SF, including its outermost part and the CGM. This allowed the community, for instance, to identify possible channels for Lyman continuum escape in the halo of SBS 0335-52E \citep{Herenz17} or highly ionised outflow cones in ESO338-IG04 \citep{Bik15}, sometimes reaching scales of $\sim$15~kpc, which is well beyond the size of the galaxy, as in SBS 0335-52E \citep{Herenz23}.
Likewise, spatially resolved line diagnostics together with emission morphology gave hints that more than one ionisation mechanism is exciting the outer regions in Haro\,14 \citep{Cairos22}. Here, we present one step forward in this direction,  by analysing MUSE data of one of these prototypical BCDs.

With  an EW(\oiii$\lambda$5007) 
$\sim$250~\AA\, \citep{Moustakas06}, \object{UM\,462} constitutes an example of these BCDs with tremendous EWs, similar to GP galaxies. Its basic characteristics are compiled in Table \ref{tabbasicdata}.
From the morphological point of view, this low-metallicity galaxy was classified as $iE$ by \citet{Cairos01}, meaning that it displays a complex inner structure with several star-forming regions superposed on an external regular envelope. While the galaxy looks pretty compact \citep[effective radius, $r_{e,B}\sim0.5$~kpc,][]{Amorin09},
  it actually presents a very low surface brightness structure that extends up to $\sim$8~kpc in the north-south direction and up to $\sim$4~kpc in the east-west direction, for the adopted distance here \citep{Micheva13}.
Spectroscopically, this galaxy has been classified as a WR galaxy, with the detection of the so-called blue bump in its spectrum \citep{Izotov98}. This classification has been disputed by \citet{James10}, who did not  detect WR features in the integrated spectrum of the galaxy.

\citet{Telles95} suggested that, together with \object{UM\,461} at a projected distance of $\sim$17$^\prime$ ($\sim$70~kpc at the assumed distance), they form an interacting pair of galaxies. However, this was not supported by later radio observations \citep{vanZee98}. Besides, even if some kind of interaction with \object{UM\,461} occurred, the estimated crossing time (700 Myr) was well longer than the age estimated for the starburst. Another possible cause of the massive SF could be a merger event.
\citet{Mezcua14} show that  \object{UM\,462}  contains two nuclei separated by $\lesssim$600~pc, and that the host galaxy can be fitted well by two components, suggestive of being in the coalescence stage of a (major) merger.

So far,  \object{UM\,462} has been the subject of two studies based on IFS data. \citet{Nicholls14a} used the spectra of the two main sites of SF (together with the spectra of other galaxies) to asses the validity of the strong line methods and the so-called direct method to derive oxygen and nitrogen abundances.  In a more dedicated study, \citet{James10} presented detailed 2D spectroscopic mapping of its central region ($\sim13^{\prime\prime}\times13^{\prime\prime}$). They identified at least four star-bursting regions, where they derived the electron temperature and densities as well as several ionic and total abundances, and presented a highly disturbed velocity field for the ionised gas. None of these spectroscopic studies covered the galaxy beyond the main sites of SF. 

 \begin{figure}[ht]
 \centering
\includegraphics[angle=0,  trim=80 12 60 35, width=0.35\textwidth, clip=,]{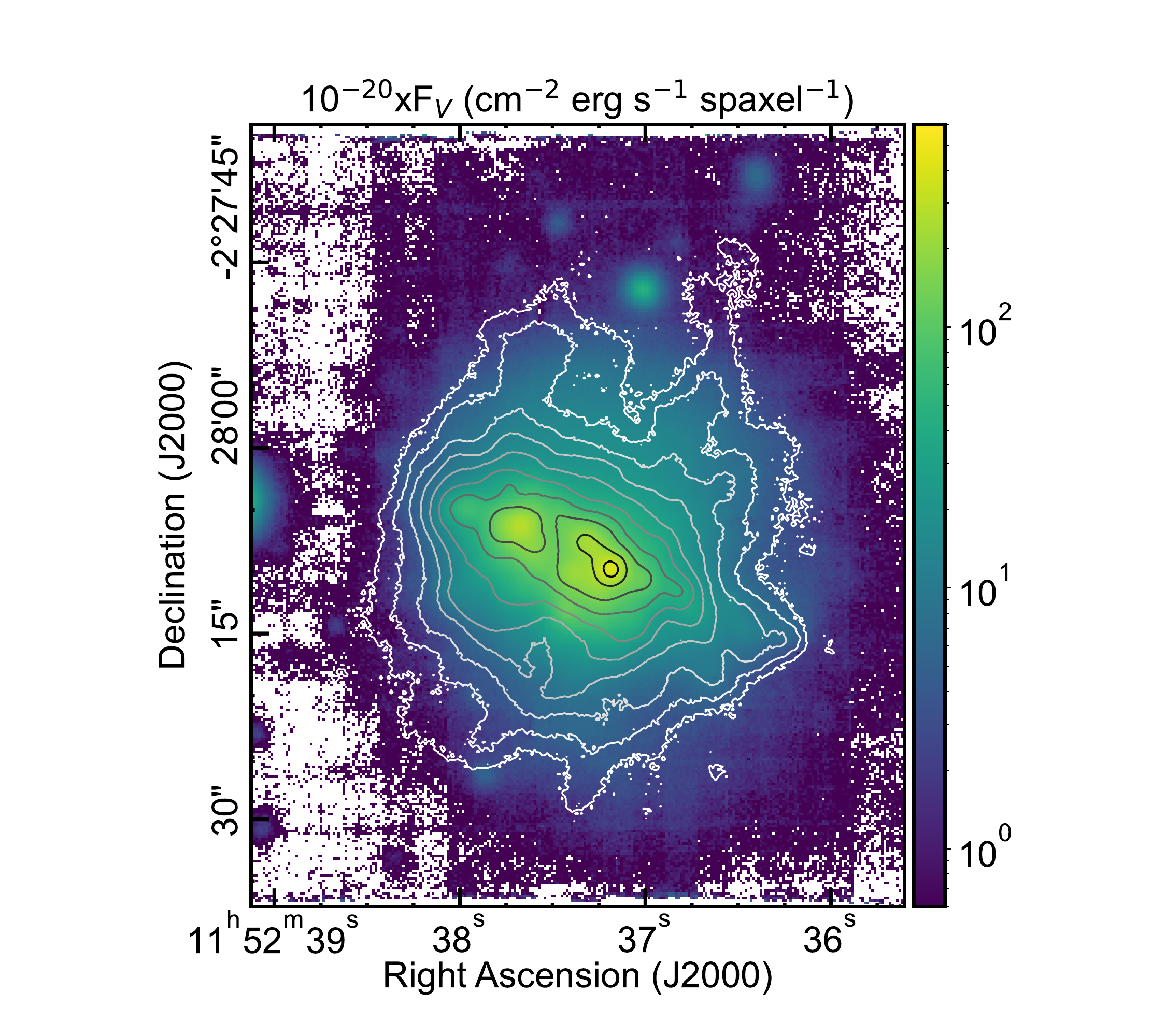}
 \caption{
Reconstructed image of \object{UM~462} made by simulating the action of a broad  filter in the $V$ band (see main text for details).
A map in \ha\, flux made by line fitting on a spaxel-by-spaxel basis is overplotted with ten evenly spaced contours (in logarithmic scale) ranging from 1.26$\times$10$^{-18}$~erg~cm$^{-2}$~s$^{-1}$~spaxel$^{-1}$ to 1.26$\times$10$^{-15}$~erg~cm$^{-2}$~s$^{-1}$~spaxel$^{-1}$.  \label{apunV}
}
\end{figure}

Here, we present the analysis of deep optical MUSE data covering the whole main body of the galaxy. The underlying question driving this research is: How would a GP(-like) galaxy look if we could resolve it at scales $\lesssim$100~pc? The high quality of this unique dataset allows us to obtain spatially resolved information of the physical (i.e. extinction, electron densities and temperatures, degree of ionisation), and chemical properties (ionic and atomic abundances) as well as kinematics of the ISM well beyond the main site of SF.  Likewise, we are able to get a basic overview of the characteristics of the underlying stellar population. All together this allows us to gain insight into the impact of the SF on the host galaxy and beyond.

The paper is organised as follows. Sect. \ref{secdata} describes the observations and technical details about the data reduction that led to the creation of the datacube. In Sect. \ref{secdataproc}, we present the general data processing necessary to extract the required information for the analysis. Sect. \ref{secresults} contains our results regarding the extinction structure in the system, the physical and chemical properties, ionisation structure, and kinematics of the ionised gas, as well as an overview of the underlying stellar population. Sect. \ref{secdiscussion} includes a discussion on the validity of different strong line methods to derive metallicities locally within a galaxy, the presence of specific stellar populations, namely, a possible supernova remnant (SNR) and  WR stars, and a discussion on the origin of the ionised gas structures observed in the outer part of the galaxy. Finally, Sect. \ref{secconclusions} sumarises our main findings.

\section{The data \label{secdata}}

UM\,462 was observed with MUSE \citep{Bacon10} as a backup target for the Guaranteed Time Observations (GTO) programme on the 17 April 2018, in four 90\degr\ rotated exposures of 500\,s each, between 02:15 and 03:01 UTC. The wide-field nominal mode of MUSE without AO support (WFM-NOAO-N) was used for the exposures and relevant calibration. An illumination flat-field exposure was taken right before the first exposure. The seeing reported by the DIMM monitor varied between 1\farcs3 and 2\farcs2. The sky conditions were clear without any clouds reported more than 20\degr\ above the horizon. The standard star LTT\,3218 was observed at the start of the night. Standard daytime calibrations were otherwise used.

   \begin{figure}[ht]
   \centering

\includegraphics[angle=0, trim=40 120 30 115, width=0.24\textwidth, clip=,]{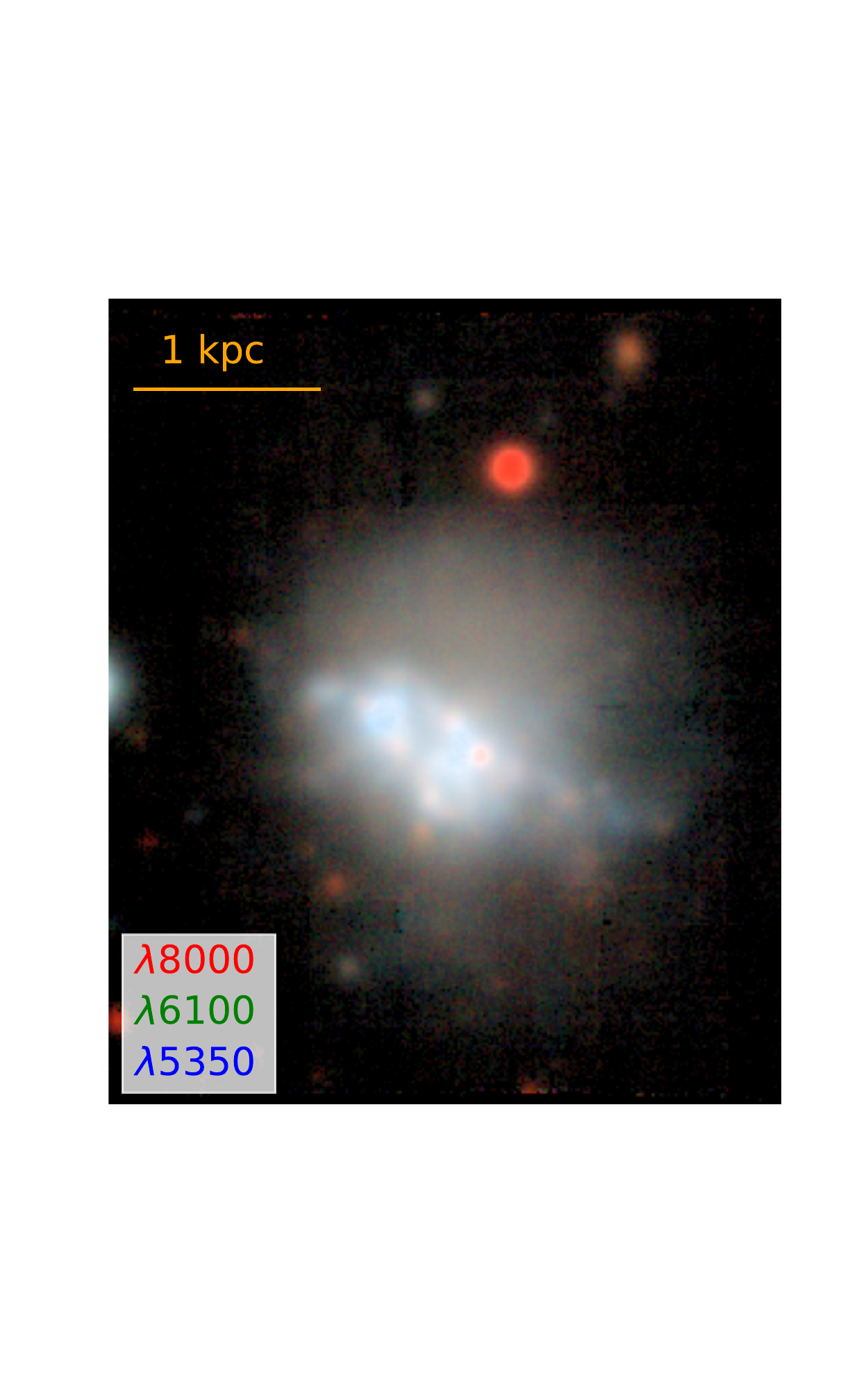} 
\includegraphics[angle=0, trim=40 120 30 115, width=0.24\textwidth, clip=,]{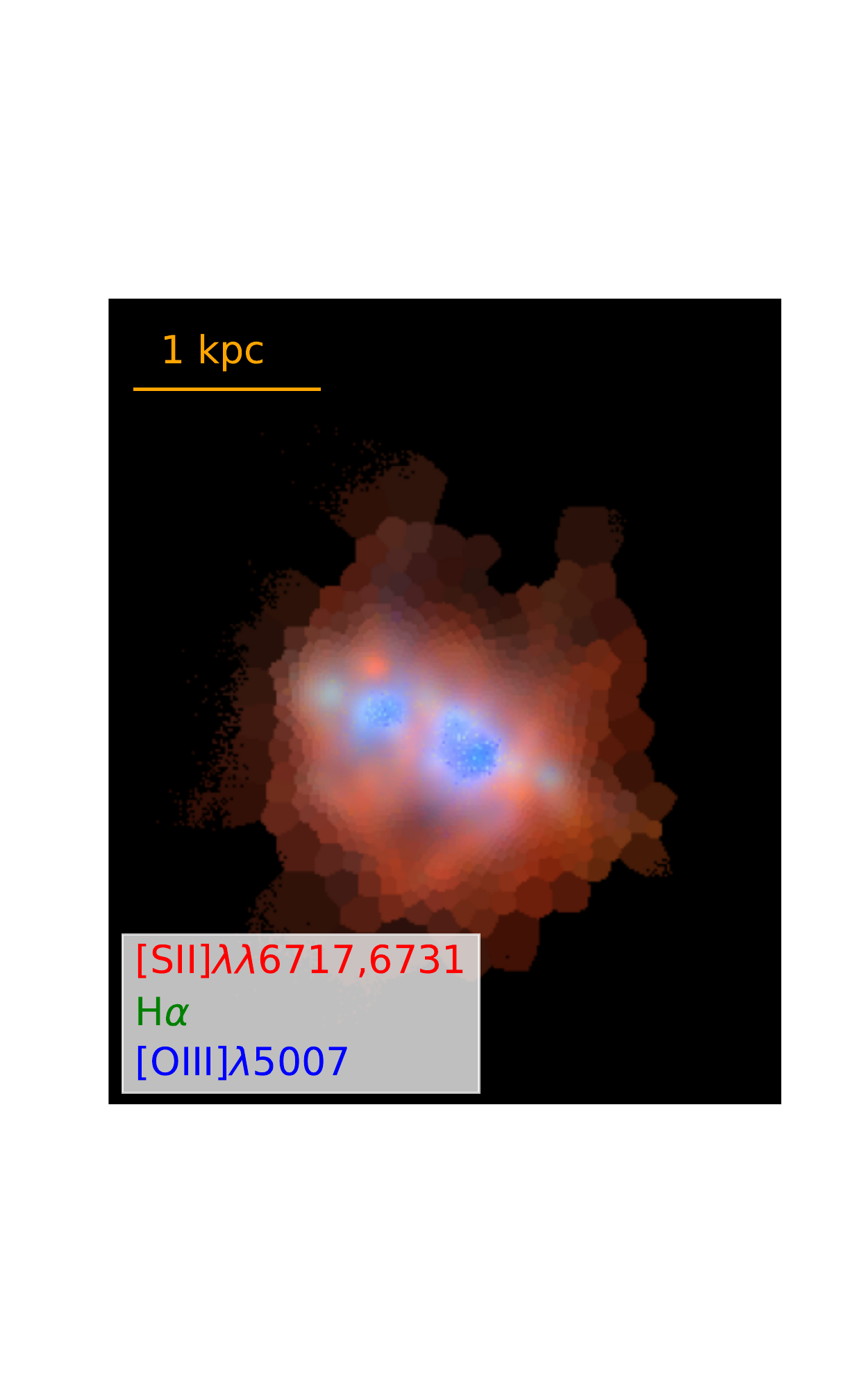}

\includegraphics[angle=0, trim=40 120 30 115, width=0.24\textwidth, clip=,]{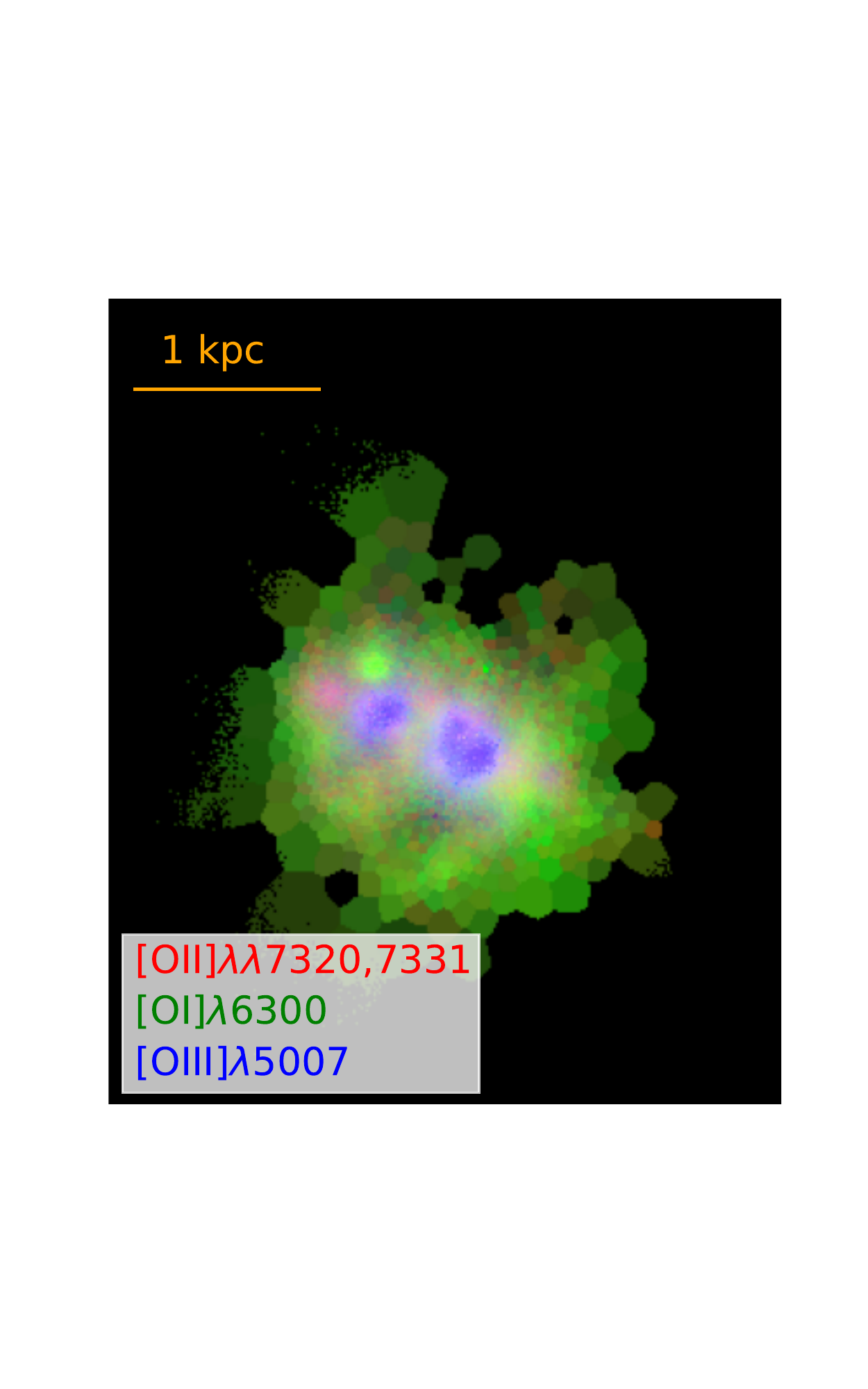}      \includegraphics[angle=0, trim=40 120 30 115, width=0.24\textwidth, clip=,]{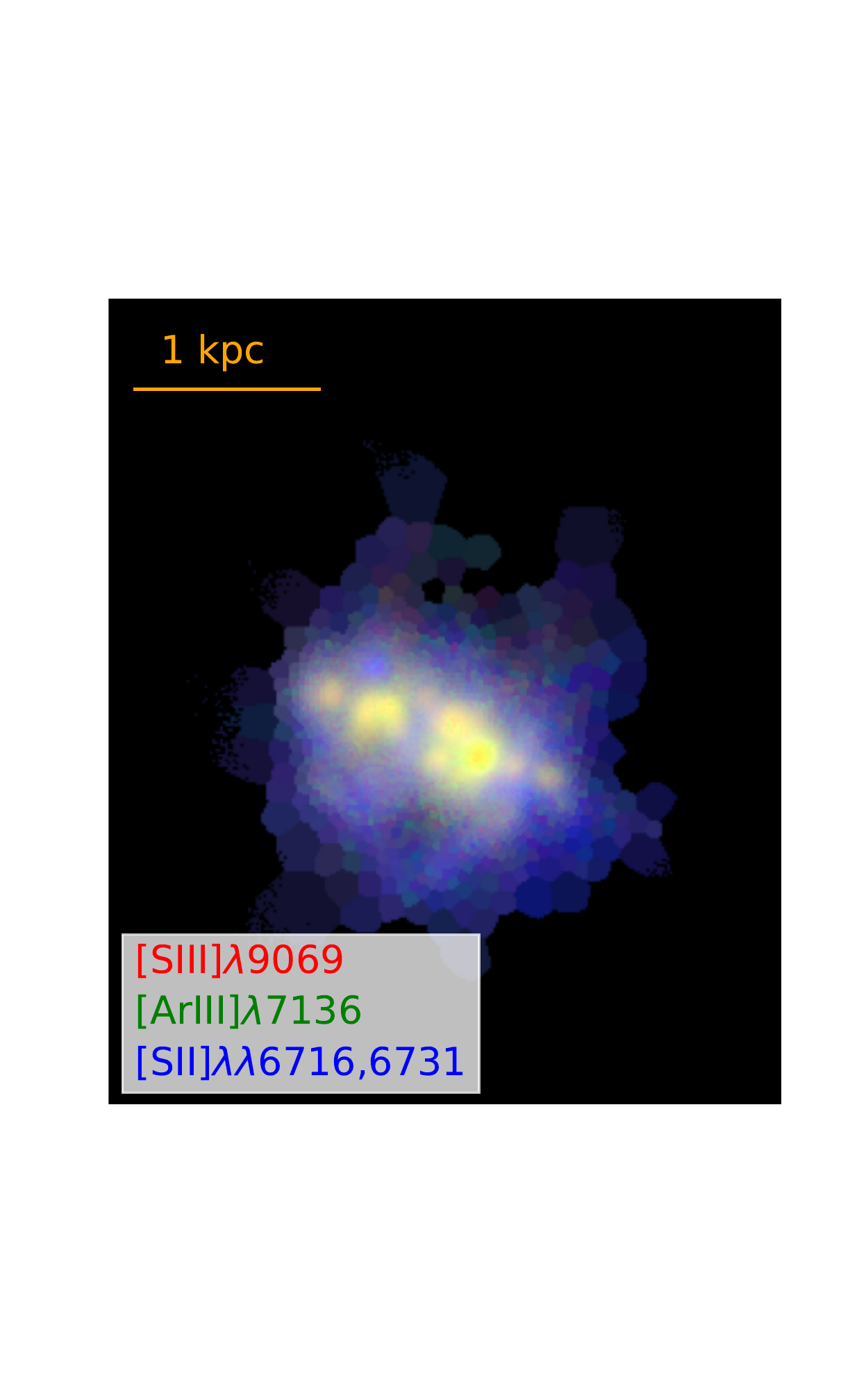}     

   \caption{Set of four RGB images.
    Individual images were allocated in the corresponding channel ordered by wavelength (i.e.  reddest filter in the R  channel, central filter in the G channel and bluest filter in the B channel).
   The first panel was made with three images extracted from the MUSE data in  100 \AA-wide line-free spectral ranges. Central wavelengths are listed.
    The other three panels are RGB images made with different sets of emission line fluxes derived by means of Gaussian line fitting, and after having tessellated the MUSE field of view to increase the S/N of the data. 
   The second panel contains, together with \ha, two of the strongest CELs, usually used for ionisation parameter mapping \citep{Pellegrini12}.
   For the third panel, we used emission lines corresponding to three different ionisation degrees for oxygen: neutral, single ionised and two times ionised.
      The last one has maps for lines tracing the two ionisation states available for sulphur, and a line for two times ionised argon, the ion with the highest ionisation potential after the two times ionised oxygen with an available map.  
 All four panels together summarise in a synthetic manner the structural information for the stars and the ionised gas. 
   North is up and east towards the left. 
   }
   \label{apuntado}
    \end{figure}

The data were reduced with the standard MUSE pipeline \citep[v2.4,][]{Weilbacher20}, called from within the MuseWise framework
\citep{Vriend15}. Both CCD-level reduction -- bias subtraction, flat-fielding, tracing, wavelength calibration, and geometric calibration, and application of these calibrations to the science data -- and processing of on-sky calibrations -- twilight skyflats, standard stars, and astrometric calibration -- used default parameters. The standard star was used for both computation of the response curve and the telluric correction. We employed the default sky subtraction assuming 40\% sky within the science field of view. The four exposures were aligned relative to the first exposure, and combined into a single datacube that is corrected for atmospheric refraction and shifted
to barycentric velocity.

The final cube is sampled $0\farcs2 \times 0\farcs2 \times 1.25$\,\AA, has a wavelength range of 4749.79--9349.79\,\AA, covers approximately 60\arcsec$\times$60\arcsec and has an image quality FWHM$_R\approx1\farcs44$ ($\sim$100~pc, at the distance of \object{UM\,462}), as judged by a \citet{Moffat69} fit to an image integrated over the $R$ band.
The absolute astrometry was adjusted using two stars in the Gaia DR2 catalogue \citep{Lindegren18}. The relative astrometry is correct to better than 0\farcs1.
Before any further analysis, the final cube was corrected for Galactic extinction (see Table \ref{tabbasicdata}) and assuming a total-to-selective extinction ratio $R_V$=3.1, and the \citet{Cardelli89} extinction curve.

Here, we analyse a portion of 52\farcs8 $\times$ 63\farcs2, corresponding to 3.7~kpc$\times$4.4~kpc. That area is presented in Fig. \ref{apunV}, that contains a reconstructed image of the galaxy in the $V$ band with contours tracing the emission in \ha. The figure displays a complex morphology for both stars and  gas in the galaxy. Besides, Fig. \ref{apuntado} heralds in a synthetic manner the complexity inherent to this galaxy and serves as a preview of what we shall present and discuss in this work. The first panel shows how most of the stars are located in a series of partially resolved clusters in the NE-SW direction, but the  extended stellar population is clearly visible. 
Besides, it signals changes on its properties within the galaxy, with bluer colours in the centre and more reddish colours in the outer parts. The data recover well the structure presented by \citet{Micheva13}, although here with a more limited spatial resolution.
The red circle on the northern part of the galaxy is a foreground star, masked in the study presented here.

The other three panels illustrate the ionisation stratification within the galaxy, with a dominance of high ionisation lines (e.g. \oiii$\lambda$5007) in the centre and of low ionisation lines (e.g. \oi$\lambda$6300) in the outer parts. Subtle differences between these three maps exist, however. The second and third maps were made with lines from ions with relatively different ionisation potential. They present a richer structure with the easternmost and westernmost knots in the central structure of the galaxy having lower ionisation degree than the rest of the central structure (lighter versus darker blue in the second map, pink versus violet in the third map). Besides, there is a bright point-like source towards the north dominated by strong low-ionisation lines (red because of the \sii\, lines in the second map, green because of the  \oi\, line in the third map). These subtleties are much more difficult to identify in the last map since two of ions used here (S$^{++}$ and Ar$^{++}$) have pretty similar ionisation potential.

\section{General post-reduction data processing  \label{secdataproc}}

We followed two approaches in this study. On the one hand, we did a very detailed spectroscopic analysis of a small subset of regions.  On the other hand, we took the most of the 2D spatial information at the expense of sacrificing some diagnostics based on extremely faint spectral features. In order to extract the required information, both of them relied on common post-reduction data processing. This is described here. In the first subsection, we describe how we selected the small subset of regions.  

 \begin{figure}[ht]
 \centering
\includegraphics[angle=0,  trim=80 12 60 35, width=0.35\textwidth, clip=,]{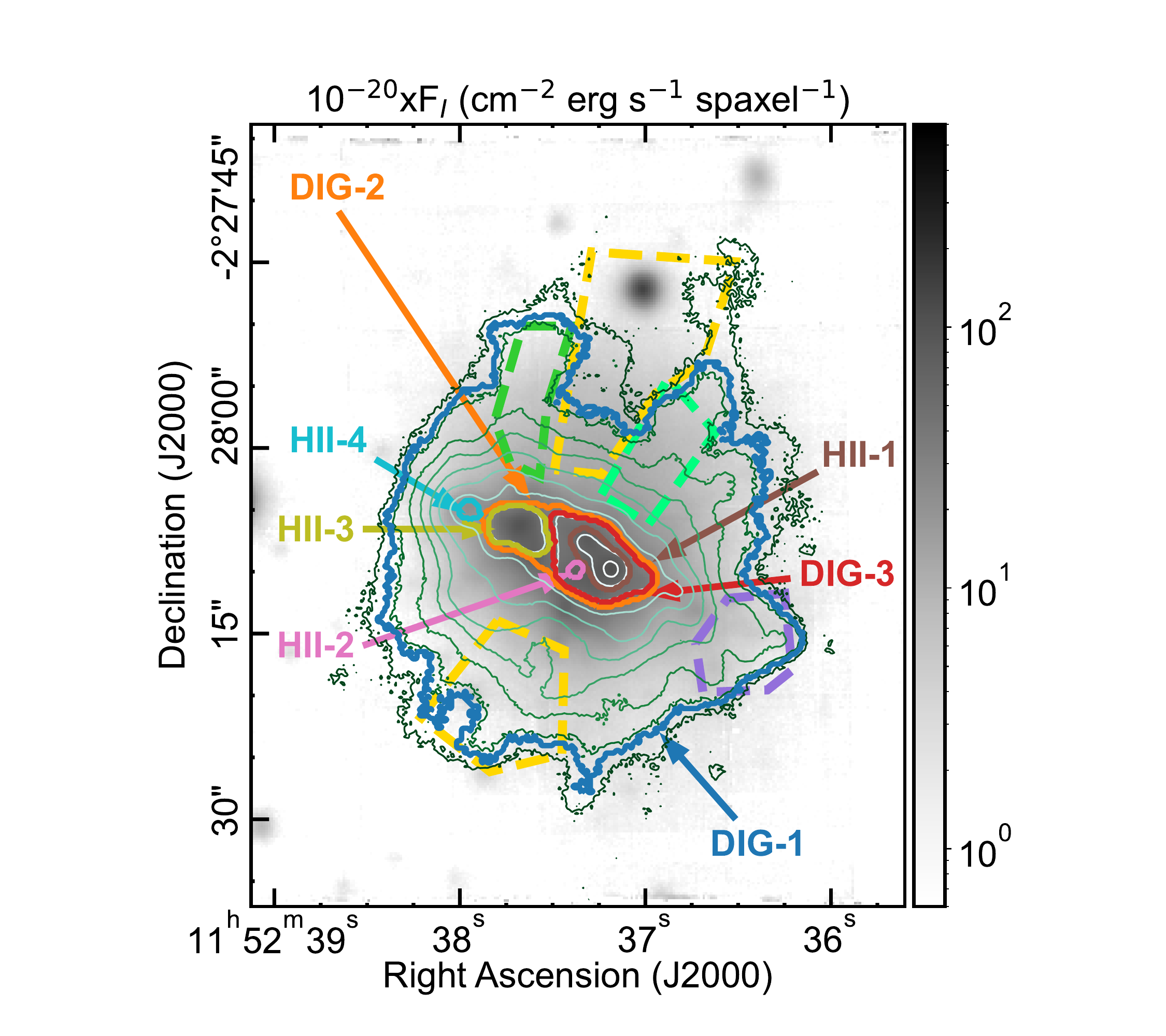}
            \caption{
Reconstructed image of \object{UM~462} made by simulating the action of a broad  filter in the $I$ band with contours showing the 'leaves' and 'branches' according to the hierarchical clustering provided by \texttt{astrodendro}.
Besides, several polygons in dashed lines mark the location of structures within the galaxy of particular interest in terms of ionisation or kinematics.
A map in \ha\, flux made by line fitting on a spaxel-by-spaxel basis is overplotted with ten evenly spaced contours (in logarithmic scale) ranging from 1.26$\times$10$^{-18}$~erg~cm$^{-2}$~s$^{-1}$~spaxel$^{-1}$ to 1.26$\times$10$^{-15}$~erg~cm$^{-2}$~s$^{-1}$~spaxel$^{-1}$.
\label{astrodendrostruc}
}
\end{figure}

   \begin{figure*}[ht]
   \centering
    \includegraphics[angle=0,  trim=5 0 0 0, width=0.92\textwidth, clip=,]{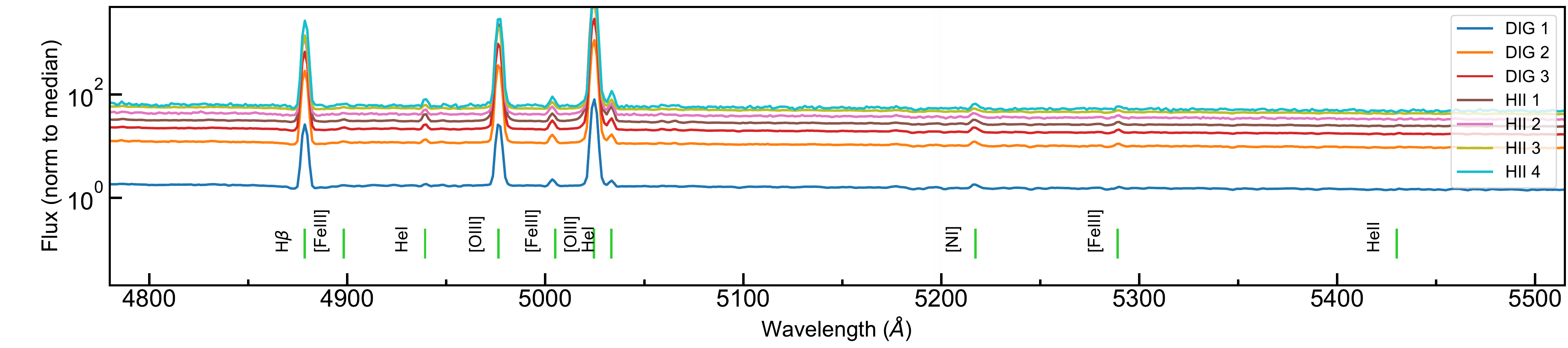}
    \includegraphics[angle=0,  trim=5 0 0 0, width=0.92\textwidth, clip=,]{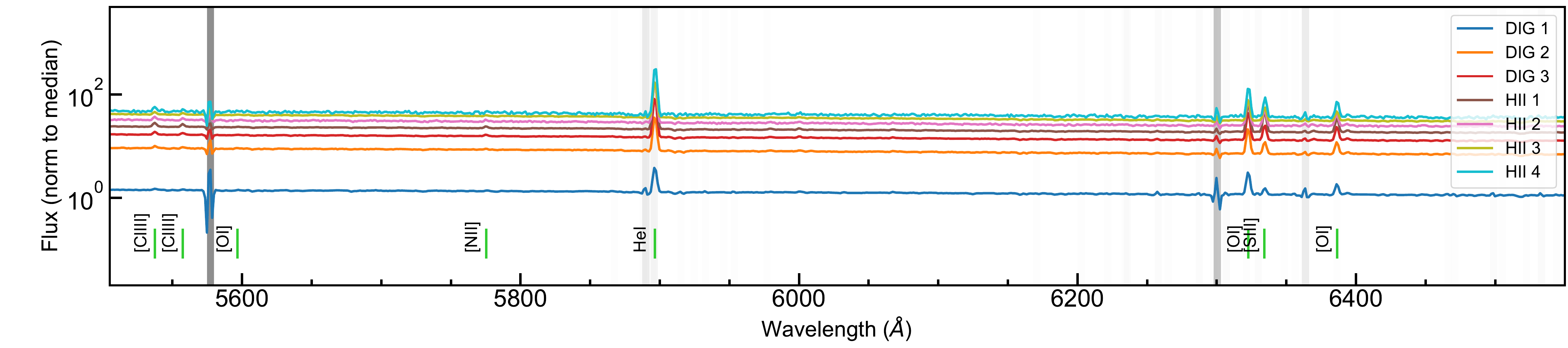}
    \includegraphics[angle=0,  trim=5 0 0 0, width=0.92\textwidth, clip=,]{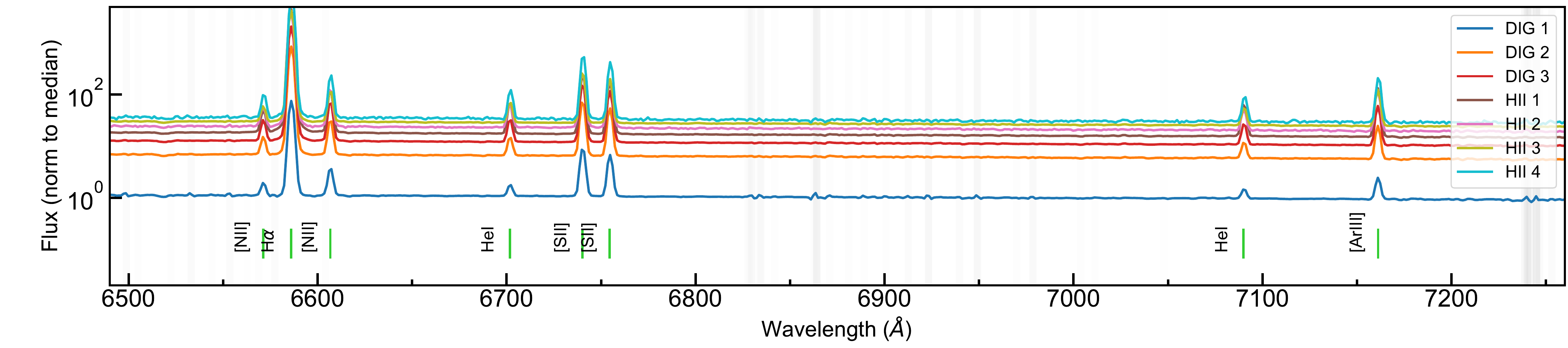}
    \includegraphics[angle=0,  trim=5 0 0 0, width=0.92\textwidth, clip=,]{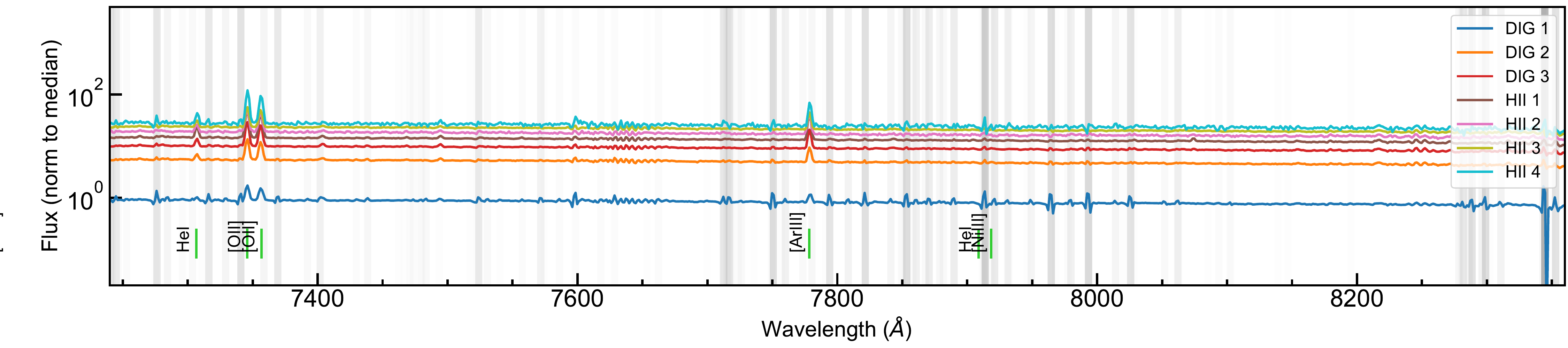}
    \includegraphics[angle=0,  trim=5 0 0 0, width=0.92\textwidth, clip=,]{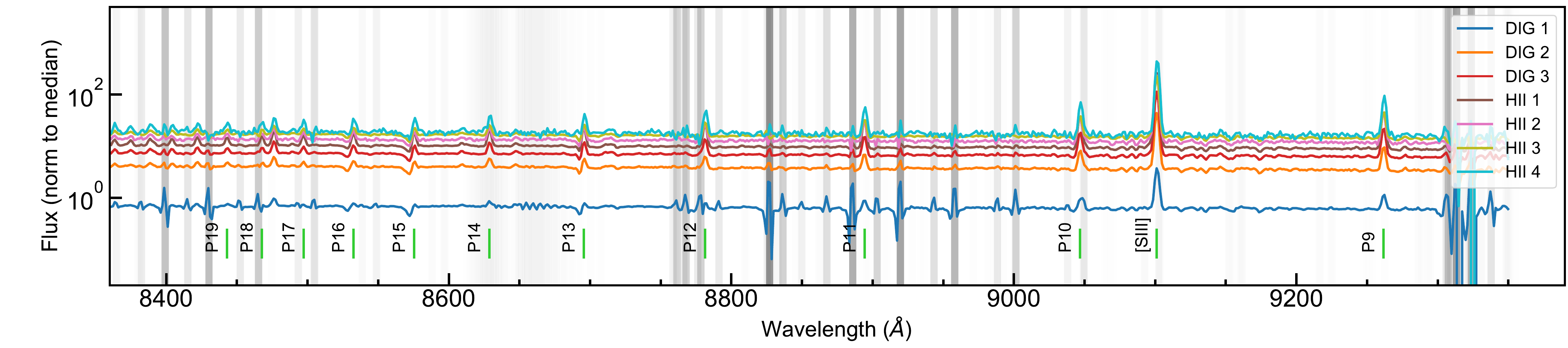}
 \caption{Integrated spectra for the regions defined by \texttt{astrodendro}. Fluxes are presented in logarithmic stretch to optimally visualise at the same time bright and faint features.
 Spectra have been offset by a factor $\times(1+2\,N)$ with $N$ the spectrum number, as indicated by the  alphabetically ordering of the labels, to avoid an overlap of the spectra. Small vertical limegreen lines indicate the expected position of the relevant emission lines. The vertical grey bands mark the position of the brightest skylines, the stronger the line is, the more opaque the band appears.}
 \label{spectraregions}
\end{figure*}

\subsection{Definition of the regions and spectral extraction  \label{secselareas}}

Ideally, one would like to map as many  diagnostic lines as possible to have a complete 2D view of the physical and chemical properties of the gas in the galaxy.
For doing so, we applied the Voronoi binning technique \citep{Cappellari03}. This algorithm divides the datacube in tiles of different shape and size so that the integrated spectrum within a given tile has always a signal-to-noise ratio ($S/N$)  for a given spectral feature of interest comparable or larger than a given threshold. This guarantees a relatively  homogeneous quality of the data. In our particular case, we used the  \ha\, flux map to tesselate the cube, since this traces the morphology of the ionised gas better than any broad band image. We looked for a compromise between having as much spatially resolved information as possible and good detection of the strongest lines, over most of the covered area.
The resulting spectra had a mean and median $S/N\sim8$ in the continuum, measured as the ration between the mean and the standard deviation in 100~\AA\, window centred at 6\,050~\AA. 
With this tesselation, we recovered maps for the 15 brightest lines, some providing redundant information, like \nii$\lambda$6584 and  \nii$\lambda$6548,  \oi$\lambda$6300 and  \oi$\lambda$6363, or \oiii$\lambda$4959 and  \oiii$\lambda$5007. We used a selection  of them to create the three-colour images presented in Fig.~\ref{apuntado}. 

Alternatively, one could extract spectra of  larger areas, to obtain as much spectral information as possible at the expense of some lost of spatial information.
For doing so, we used the \texttt{astrodendro}\footnote{\texttt{http://dendrograms.org}} package that allows us to attach a physical meaning to the extracted regions, at least to a certain extent. This package finds structures in a set of data and groups them hierarchically. The outcome is a set of structures that can be classified as 'leaves', that is to say structures without any sub-structure, and 'branches', i.e. structures with sub-structures within. We use an image in \ha\, made after fitting the spectrum in each spaxel as input for \texttt{astrodendro}. In this way, leaves can be assimilated to the \hii\, regions in the galaxy (up to the spatial resolution of the data), while branches sample areas surrounding the \hii\, regions corresponding to  the diffuse component of the ionised gas, at different levels of surface brightness. Given the seeing, we did not apply any Gaussian filtering to the input image, and requested that a given structure should be larger than eight spaxels.
In total, we identified four 'leaves' (named \hii-$i$ with $i$=1,\ldots,4,  $i$ increasing from west to east) and three 'branches'  (named DIG-$i$ with $i$=1,\ldots,3, with $i$ increasing from the outer to the inner parts).
The identified regions are displayed in Fig. \ref{astrodendrostruc} on top of a reconstructed image in the $I$ band for the galaxy while the corresponding extracted spectra are presented in Fig. \ref{spectraregions}. There, besides the usual strong emission lines, we see multiple \hei\, lines, at least two iron lines, two faint nitrogen features tracing different ionisation states, \niA$\lambda$5199, and \nii$\lambda$5755. This last line   can be used  together with  \nii$\lambda$6584 to estimate the  electron temperature. Besides, the \cliii$\lambda\lambda$5518,5538 doublet, useful to estimate the electron density in high-ionisation regions, is also detected. Finally, the blue stellar continuum, as well as some stellar features (most  notably \hb\, in absorption)  are clearly visible.

\subsection{Subtracting the emission of the stellar population \label{secmodconto}}

The spectrum of  (a portion of) a galaxy is made of the added contribution of a stellar (+nebular) continuum, a set of emission lines tracing the ionised gas, and other atomic and molecular absorption features associated to the ISM. To adequately measure the emission line fluxes, these  should be separated from their underlying continuum. This is particularly important for features like  \hb, existing both in emission (gas) and absorption (stars). For doing so, we used \texttt{FADO}\footnote{\texttt{http://spectralsynthesis.org/fado.html}} \citep{Gomes17}. The code reproduces a given observed spectrum by selecting a linear combination of a subset of $N_\star$ spectral components from a pre-defined set of base spectra. Besides, it includes the contribution expected for nebular continuum.
Given the low metallicity of \object{UM 462} (see Table \ref{tabbasicdata}), this is particularly important here.
\texttt{FADO} allows the user to adjust certain input parameters, and constrains some of the output parameters, and provides default values for both.
In the following, we describe the specific choices that we took that depart from those default values.
Regarding the base spectra, we used  a selection  of single star populations from the set provided by \citet{Bruzual03}. These are based on the Padova 2000 evolutionary tracks \citep{Girardi00} and assume a Salpeter initial mass function between 0.1 and 100~M$_\odot$.
In view of the  stellar metallicity already estimated for this galaxy \citep{Micheva13} we restricted to those with metallicity $Z=0.008, 0.004$, making a total of 40 base spectra, with ages ranging from 1~Myr to 15~Gyr.
The low stellar mass (see Table \ref{tabbasicdata}), also supports our choice of restricting to sub-solar metallicity base spectra. A  dwarf galaxy could have high, meaning about solar, metallicity if, for instance, it were the by-product of a major merger of two spiral galaxies, a so-called tidal dwarf galaxy \citep[e.g.][]{MonrealIbero07,Weilbacher03}. So far, there is no literature supporting this scenario for \object{UM 462}.
The spectral range used for the modelling with \texttt{FADO} was from 4700~\AA\, to 9275~\AA, and included strong stellar features, as \hb\, in absorption in the blue and the calcium triplet in the red. We changed the normalisation to 5500\AA, since the value provided by default with  \texttt{FADO} falls outside of the MUSE spectral range.
Besides, key spectral regions containing spectral features not considered by \texttt{FADO} (to our knowledge) were masked since they may taint the modelling. In particular, the spectral range from 5770~\AA\, to 5920~\AA, was not used since targets with large amount of ISM may display the Diffuse Interstellar Bands at $\lambda$5780 and $\lambda$5797 \citep{MonrealIbero18} and/or an excess in the Sodium doublet  \citep{Poznanski12}.
The stellar velocity was allowed to vary between $-400$ and $+400$~km s$^{-1}$ around the systemic velocity, which was a good compromise between the expected velocity gradients in dwarf galaxies, and leaving room for eventual discovery of high-velocity locations in the galaxy (even if this turned out unnecessary in our case). Finally,
we assumed a single extinction law for all the base spectra that was modelled as a uniform dust screen with the extinction law by \citet{Cardelli89}, and allowed for a stellar attenuation up to $A_V{\star} = 0.9$, since leaving the upper limit as $A_V{\star} =6.0$ (the default in 
\texttt{FADO}) predicted unrealistically highly attenuated areas in the outer parts of the galaxy. 
The best-fit spectra were resampled and subtracted from the original MUSE data.
We used these last spectra to measure the emission line parameters, as explained next.

\begin{sidewaystable*}
\caption{Relative emission line fluxes with respect to 100$\times$f(\hb) for summed spectra in the regions defined by \texttt{astrodendro}. \label{rellinefluxes}} 
 \centering  
\begin{tabular}{lccccccccc}
\hline\hline
              Line ID &                    DIG-1 &                    DIG-2  &                    DIG-3 &                    \hii-1 &                    \hii-2 &                    \hii-3 &                    \hii-4 \\
\hline
       4861 \textsc{H}$\beta$ &     100.000$\pm$ 0.157 &   100.000$\pm$ 0.061 &   100.000$\pm$ 0.051 &   100.000$\pm$ 0.037 &   100.000$\pm$ 0.161 &   100.000$\pm$ 0.046 &   100.000$\pm$ 0.170 \\
 4922 He\textsc{i} &    0.894$\pm$ 0.014 &    0.766$\pm$ 0.034 &    0.825$\pm$ 0.024 &    1.221$\pm$ 0.027 &    0.972$\pm$ 0.124 &    0.891$\pm$ 0.030 &    1.032$\pm$ 0.114 \\
 4959 \textsc{[O iii]} &  106.747$\pm$ 0.024 &  135.933$\pm$ 0.067 &  150.878$\pm$ 0.048 &  186.449$\pm$ 0.057 &  167.988$\pm$ 0.251 &  158.417$\pm$ 0.060 &  122.935$\pm$ 0.189 \\
4987  [Fe\textsc{iii}]  &    2.330$\pm$ 0.014 &    1.901$\pm$ 0.035 &    1.565$\pm$ 0.024 &    1.020$\pm$ 0.027 &    1.554$\pm$ 0.125 &    1.444$\pm$ 0.031 &    1.300$\pm$ 0.095 \\
 5007 \textsc{[O iii]} &  308.500$\pm$ 0.058 &  392.848$\pm$ 0.173 &  436.038$\pm$ 0.124 &  538.836$\pm$ 0.151 &  485.484$\pm$ 0.658 &  457.825$\pm$ 0.157 &  355.283$\pm$ 0.478 \\
    5016 He\textsc{i}&    2.187$\pm$ 0.014 &    2.193$\pm$ 0.036 &    2.278$\pm$ 0.025 &    2.839$\pm$ 0.029 &    2.572$\pm$ 0.131 &    2.373$\pm$ 0.032 &    2.456$\pm$ 0.118 \\
  5267 [Fe\textsc{iii}]  &   0.868$\pm$ 0.008 &     0.690$\pm$ 0.014 &     0.574$\pm$ 0.007 &     0.393$\pm$ 0.006 &     0.478$\pm$ 0.034 &     0.584$\pm$ 0.009 &     0.392$\pm$ 0.044 \\ 
5518 [Cl \textsc{iii}] &    0.249$\pm$ 0.006 &     0.295$\pm$ 0.010 &     0.318$\pm$ 0.005 &     0.328$\pm$ 0.004 &     0.323$\pm$ 0.025 &     0.306$\pm$ 0.007 &     0.380$\pm$ 0.040 \\
5538 [Cl \textsc{iii}]  &     0.191$\pm$ 0.006 &     0.182$\pm$ 0.010 &     0.216$\pm$ 0.005 &     0.222$\pm$ 0.003 &     0.180$\pm$ 0.024 &     0.202$\pm$ 0.006 &     0.154$\pm$ 0.038 \\
5755 [N\textsc{ii}] &   0.194$\pm$ 0.004 &     0.130$\pm$ 0.008 &     0.174$\pm$ 0.004 &     0.143$\pm$ 0.003 &     0.171$\pm$ 0.022 &     0.113$\pm$ 0.005 &     0.129$\pm$ 0.033 \\
5876 He\textsc{i}  &   10.541$\pm$ 0.017 &    10.295$\pm$ 0.017 &    10.488$\pm$ 0.021 &    11.467$\pm$ 0.008 &    11.243$\pm$ 0.045 &    10.389$\pm$ 0.013 &    10.909$\pm$ 0.056 \\
6300  [O\textsc{i}]&   8.043$\pm$ 0.013 &     5.028$\pm$ 0.014 &     4.679$\pm$ 0.008 &     3.457$\pm$ 0.011 &     4.063$\pm$ 0.034 &     3.431$\pm$ 0.009 &     3.754$\pm$ 0.042 \\
6312 [S\textsc{iii}] & 1.505$\pm$ 0.009 &     1.730$\pm$ 0.012 &     1.822$\pm$ 0.007 &     1.751$\pm$ 0.009 &     1.986$\pm$ 0.030 &     1.796$\pm$ 0.008 &     1.758$\pm$ 0.038 \\
6364 [O\textsc{i}] &  2.681$\pm$ 0.006 &     1.676$\pm$ 0.008 &     1.560$\pm$ 0.005 &     1.152$\pm$ 0.006 &     1.354$\pm$ 0.019 &     1.144$\pm$ 0.005 &     1.251$\pm$ 0.024 \\
6548 [N\textsc{ii}]  &   3.483$\pm$ 0.010 &     2.915$\pm$ 0.019 &     2.912$\pm$ 0.013 &     2.344$\pm$ 0.013 &     2.705$\pm$ 0.056 &     2.100$\pm$ 0.015 &     2.372$\pm$ 0.050 \\
6563 \textsc{H}$\alpha$ &   300.000$\pm$ 0.334 &   299.305$\pm$ 0.145 &   310.423$\pm$ 0.122 &   326.461$\pm$ 0.096 &   317.341$\pm$ 0.415 &   306.060$\pm$ 0.114 &   317.428$\pm$ 0.424 \\
6584  [N\textsc{ii}] &    10.449$\pm$ 0.019 &     8.746$\pm$ 0.033 &     8.737$\pm$ 0.023 &     7.033$\pm$ 0.022 &     8.115$\pm$ 0.097 &     6.301$\pm$ 0.026 &     7.116$\pm$ 0.087 \\
6678 He\textsc{i}  &     3.095$\pm$ 0.006 &     3.009$\pm$ 0.010 &     3.182$\pm$ 0.005 &     3.466$\pm$ 0.004 &     3.296$\pm$ 0.025 &     3.096$\pm$ 0.007 &     3.329$\pm$ 0.036 \\
6716 [S\textsc{ii}] &    32.340$\pm$ 0.037 &    23.380$\pm$ 0.019 &    21.420$\pm$ 0.017 &    15.821$\pm$ 0.009 &    19.293$\pm$ 0.049 &    17.185$\pm$ 0.013 &    19.739$\pm$ 0.056 \\
6731 [S\textsc{ii}]  &    22.694$\pm$ 0.026 &    16.511$\pm$ 0.017 &    15.318$\pm$ 0.016 &    11.684$\pm$ 0.008 &    13.876$\pm$ 0.044 &    12.209$\pm$ 0.012 &    14.072$\pm$ 0.050 \\
7065 He\textsc{i}  &     2.221$\pm$ 0.006 &     2.194$\pm$ 0.009 &     2.524$\pm$ 0.005 &     3.676$\pm$ 0.004 &     2.466$\pm$ 0.024 &     2.258$\pm$ 0.006 &     2.379$\pm$ 0.034 \\
7136 [Ar\textsc{iii}] &     5.952$\pm$ 0.009 &     6.631$\pm$ 0.011 &     7.350$\pm$ 0.008 &     8.511$\pm$ 0.007 &     8.058$\pm$ 0.033 &     7.108$\pm$ 0.009 &     6.822$\pm$ 0.035 \\
7281 He\textsc{i}  &     0.460$\pm$ 0.009 &     0.497$\pm$ 0.008 &     0.538$\pm$ 0.005 &     0.647$\pm$ 0.004 &     0.558$\pm$ 0.021 &     0.531$\pm$ 0.007 &     0.574$\pm$ 0.033 \\
7320 [O \textsc{ii}] &     3.731$\pm$ 0.011 &     3.230$\pm$ 0.017 &     3.181$\pm$ 0.011 &     2.812$\pm$ 0.010 &     2.885$\pm$ 0.039 &     2.701$\pm$ 0.013 &     3.610$\pm$ 0.055 \\
7331 [O \textsc{ii}] &     3.135$\pm$ 0.012 &     2.779$\pm$ 0.018 &     2.688$\pm$ 0.012 &     2.392$\pm$ 0.010 &     2.464$\pm$ 0.041 &     2.289$\pm$ 0.014 &     3.000$\pm$ 0.059 \\
8752 Pa 12 &     0.824$\pm$ 0.012 &     1.107$\pm$ 0.012 &     1.201$\pm$ 0.006 &     1.394$\pm$ 0.005 &     1.297$\pm$ 0.033 &     1.140$\pm$ 0.009 &     1.247$\pm$ 0.038 \\
8865 Pa 11 &     2.338$\pm$ 0.020 &     1.433$\pm$ 0.010 &     1.576$\pm$ 0.006 &     1.782$\pm$ 0.004 &     1.553$\pm$ 0.026 &     1.512$\pm$ 0.008 &     1.643$\pm$ 0.040 \\
9017 Pa 10 &     2.603$\pm$ 0.014 &     2.038$\pm$ 0.017 &     2.112$\pm$ 0.010 &     2.420$\pm$ 0.007 &     2.264$\pm$ 0.046 &     2.039$\pm$ 0.012 &     2.185$\pm$ 0.061 \\ 
9069 [S\textsc{iii}] &    13.802$\pm$ 0.019 &    15.269$\pm$ 0.023 &    17.032$\pm$ 0.015 &    19.402$\pm$ 0.013 &    18.839$\pm$ 0.061 &    16.642$\pm$ 0.018 &    17.503$\pm$ 0.061 \\
9231 Pa 9 &     2.773$\pm$ 0.009 &     2.618$\pm$ 0.014 &     2.819$\pm$ 0.008 &     3.226$\pm$ 0.006 &     2.962$\pm$ 0.036 &     2.760$\pm$ 0.010 &     3.066$\pm$ 0.044 \\
\hline
F(\hb) $^{(a)}$ &   147.24$\pm$   26.58 &     28.90$\pm$    0.15 &     66.67$\pm$    0.58 &     79.05$\pm$    0.42 &      4.62$\pm$    0.03 &     44.11$\pm$    0.21 &      3.50$\pm$    0.02 \\
\hline   
\end{tabular}

$^{(a)}$In units of  $10^{-15}$ erg s$^{-1}$ cm$^{-2}$.
\end{sidewaystable*}

 \begin{figure}[!ht]
 \centering
\includegraphics[angle=0,  width=0.48\textwidth, clip=,]{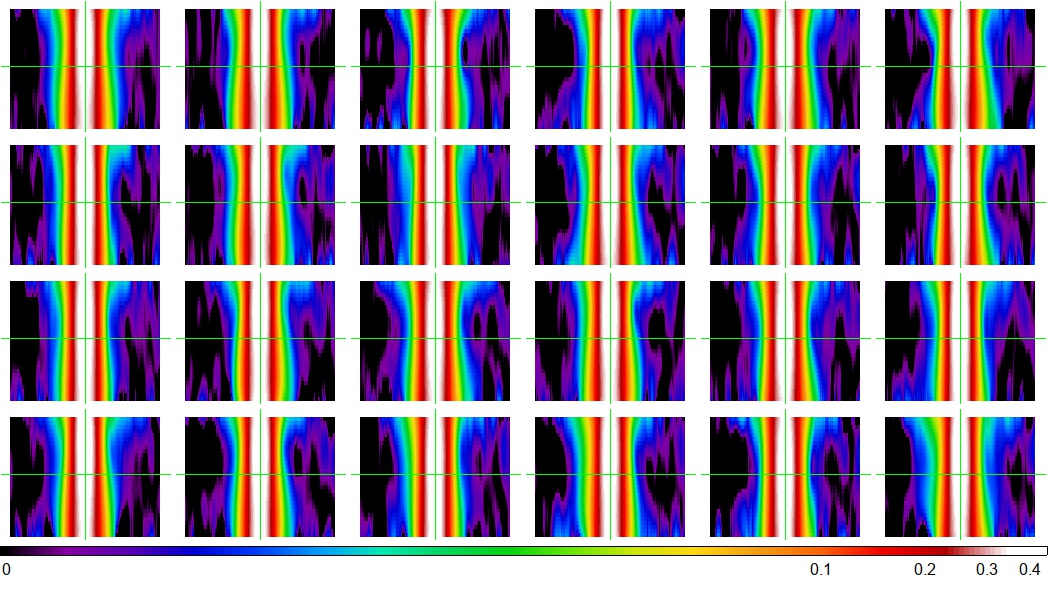}
\includegraphics[angle=0,  width=0.48\textwidth, clip=,]{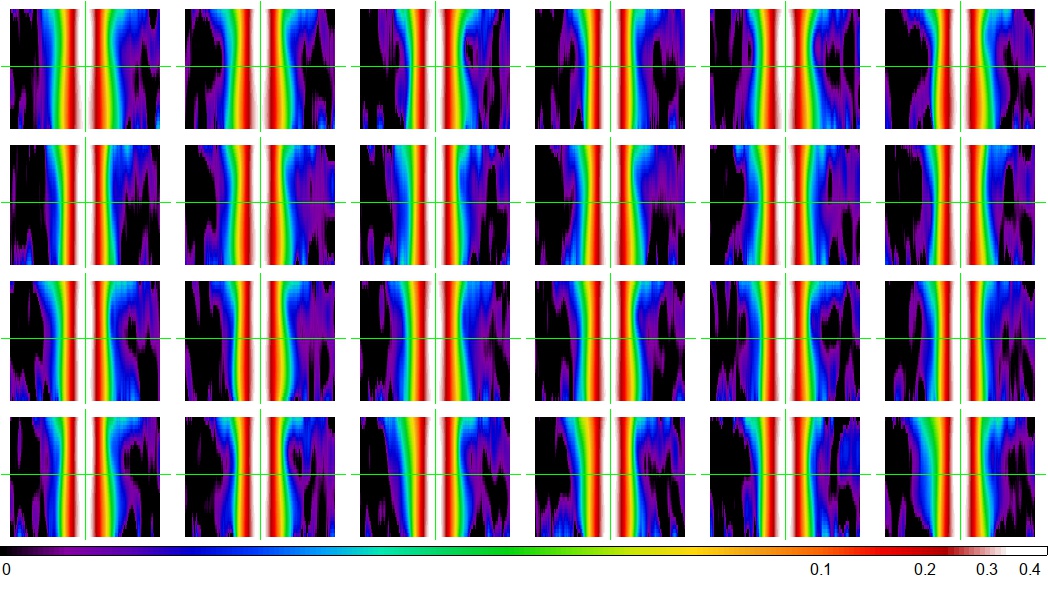}
\caption{LSF derived from arc lines for two slices (\emph{top:} slice 20, \emph{bottom:} slice 23) in all 24 MUSE spectrographs. Each of the 24 panels shows the profile for the given slice in one IFU, with the line wavelength increasing vertically from 4650 to 9300\,\AA. The normalised line profile at each wavelength is distributed horizontally, $\pm7.5$\,\AA\ from the line centre.
}
 \label{LSF}
\end{figure}

\subsection{Line emission measurements \label{seclinefitting}}

We fitted the different spectral features needed for our analysis by Gaussian functions plus a 1-degree polynomial to take into account local residuals in the continuum substraction using the Python package \texttt{LMFIT}\footnote{\texttt{https://lmfit.github.io/lmfit-py/}}. Since we basically followed the same methodology as that throughly described  by \citet{MonrealIbero20}, we refer to that publication for  further details on the line fitting. 
A set of lines not described in that work is the complex
\hei$\lambda\lambda$4922,5016+\oiii$\lambda\lambda$4959,5007+\feiii$\lambda$4987,
Here, the fitting was rather challenging since there are $\sim$2 orders of magnitude of difference between the relative fluxes of the \oiii\, lines and the other lines in the complex. Besides, some of them are strongly blended (e.g. \oiii$\lambda$5007+\hei$\lambda$5016).
Thus, we measure the parameters for the faint lines in the complex only in the selected regions presented in Sect.  \ref{secselareas}. 
For that we imposed, one and the same line width for the two \oiii\, lines, and one and the same line width for the other three lines, we fixed the difference in wavelength between the five lines to the expected one according to their rest-frame wavelength and redshift of the galaxy, and we added an offset to the continuum of $\times$1/1000 the peak in \oiii$\lambda$5007 (that was taken into account in the fit).
A close inspection of this spectral range for the \texttt{astrodendro} spectra revealed a broad component at the base of the \oiii\, lines at the level of $\lesssim1/500$ the peak in \oiii$\lambda$5007. These components cannot directly be attributed to the target since they are at a level where the line spread function (LSF) for MUSE is  not very well defined. These wings can clearly be seen in other publicly available MUSE data cubes with strong emission lines at high S/N \citep[e.g.][]{MonrealIbero20b} and also in the LSF measured for the sky subtraction during the data reduction. As illustration, Fig. \ref{LSF} displays those used for  slice 20 (top) and 23 (bottom) in all 24 IFUs, in logarithmic stretch. It is clear from the figure that the LSF changes quite a bit on rather small scales (the slices are adjacent on the sky), especially in the wings. Elucidating whether there is information in these broad components beyond the LSF and attributable to the target is a delicate exercise that will not be discussed in this contribution. However it suggests a nice avenue for future research.
Finally, derived line widths were corrected by the MUSE LSF for all the lines. This was estimated assuming a Gaussian shape - thus not including the wings mentioned above - and using a polynomial originally derived using the MUSE pipeline and convolved with the 1.25 \AA~pix$^{-1}$ binning of the cube and had a value of $\sigma \sim1.18$~\AA\, around \oiii$\lambda$5007 and $\sigma\sim1.06$~\AA\, around \ha.

Fluxes for detected emission lines relative to \hb\, in the areas extracted with \texttt{astrodendro} are presented in Tab.~\ref{rellinefluxes}. They were used to derive the physical and chemical properties presented in Sect. \ref{secionigas}.

 \begin{figure}[!ht]
 \centering
\includegraphics[angle=0,  trim=80 12 60 35, width=0.35\textwidth, clip=,]{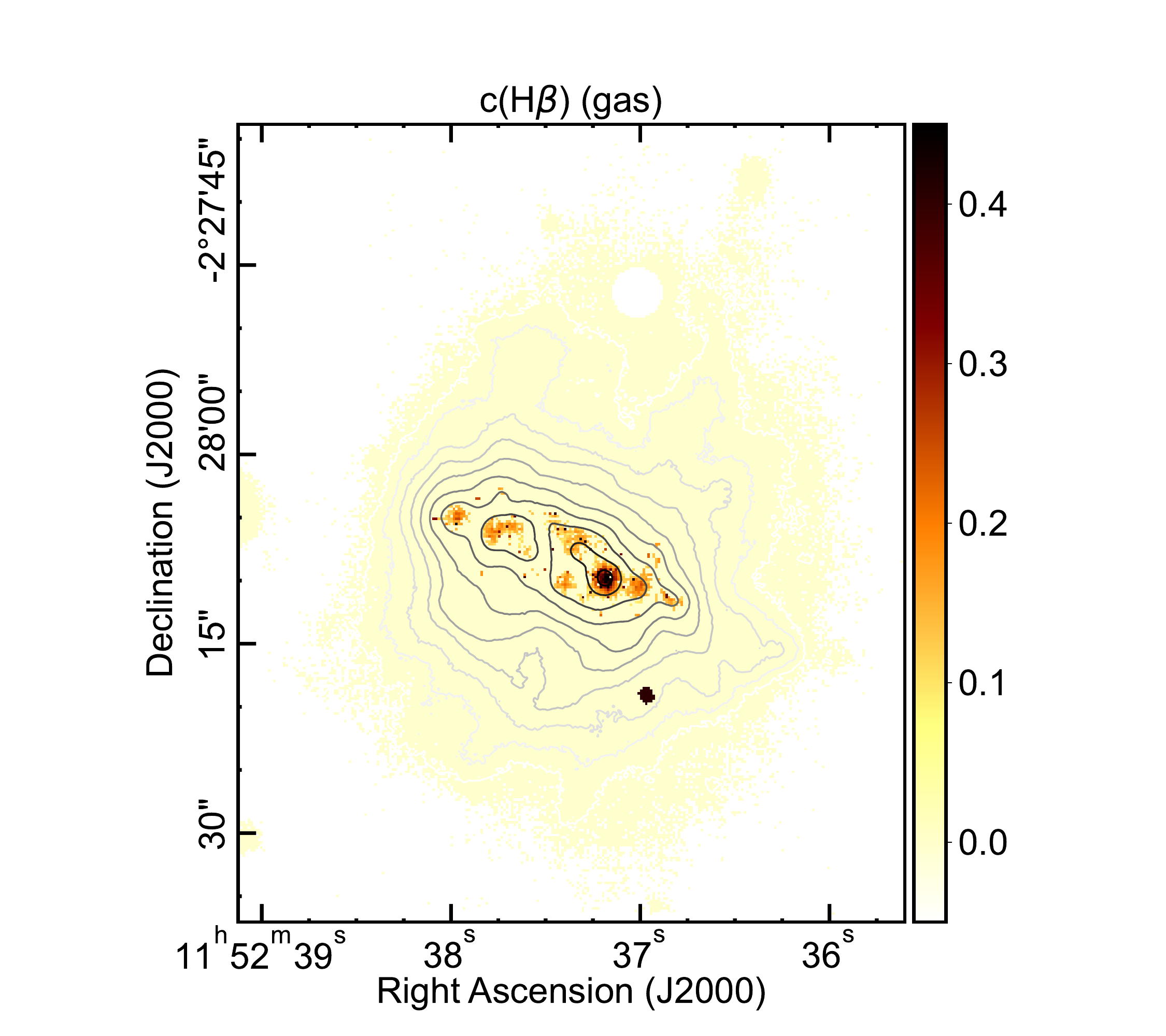}

\includegraphics[angle=0,  trim=80 12 60 35, width=0.35\textwidth, clip=,]{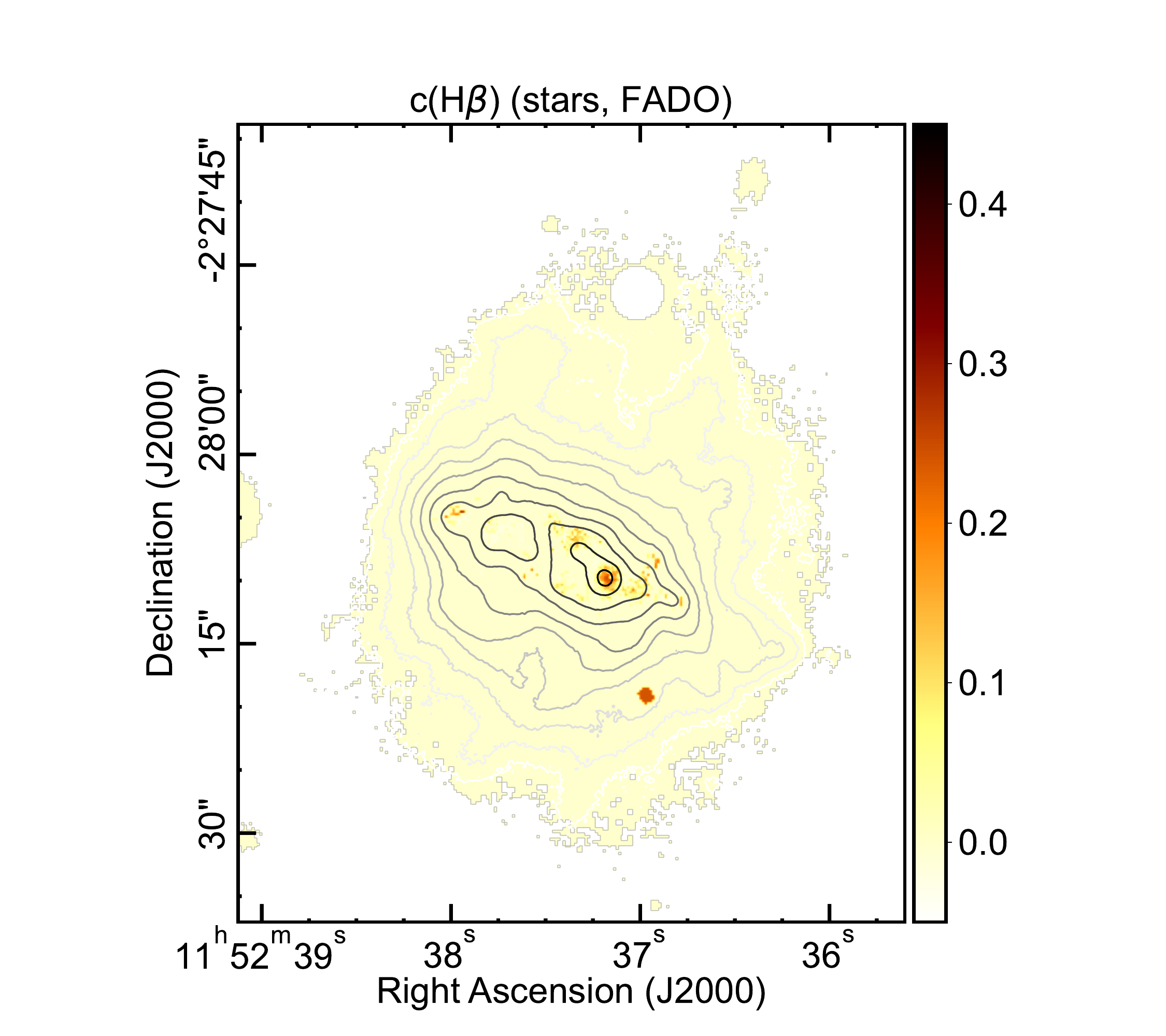}
\caption{Mapping the extinction in \object{UM\,462}. \emph{Top:} Log extinction at \hb, c(\hb), assuming \ha/\hb=2.89, and the \citet{Cardelli89} extinction law. The median c(\hb) is 0.04. The estimated median uncertainty for the presented map was $\sim$0.03, with 90\% of the uncertainties ranging between 0.02 and 0.04.
\emph{Bottom:} Similar information for the stellar populations as derived from \texttt{FADO}.
}
 \label{mapchb}
\end{figure}

\section{Results \label{secresults}}

\subsection{Extinction structure \label{secextin}}

Extinction for the gas was derived from the \ha\, and \hb\, emission lines and the \texttt{RedCorr()} class in \texttt{PyNeb}. We assumed an intrinsic Balmer emission line ratio of \ha/\hb=2.89 \citep{Osterbrock06}, and the extinction law provided by \citet{Cardelli89} with $R_V=3.1$. The derived c(\hb) map is presented in the upper panel of Fig. \ref{mapchb}. Uncertainties, estimated by means of Monte Carlo simulations with 100 realisations, were $\sim$0.02-0.04. The map is consistent with no extinction over most of the galaxy. The most outstanding exception is an area with a size comparable to that of the seeing disc associated with the peak of emission in \ha, and within \hii-1, where c(\hb)$\sim$0.4-0.5. The extinction map suggests other locations with a more moderate amount of extinction (c(\hb)$\sim$0.2) coinciding with other secondary peaks of SF (e.g. \hii-4 and part of \hii-3), unresolved at the spatial resolution of these data. For comparison, \citet{James10} report c(\hb)=0.19 while  \citet{Nicholls14a} reports c(\hb)=0.128, 0.021 for the two locations analysed in there, roughly corresponding to our   \hii-1 and   \hii-3.

Among many other physical quantities, \texttt{FADO} also estimates the stellar attenuation needed to reproduce, together with the age and metallicity of the stars, the observed stellar continuum. The corresponding c(\hb) is presented in the lower panel of Fig. \ref{mapchb}. The extinction for the stars is overall smaller than for the gas, as expected. 

\subsection{The ionised gas \label{secionigas}}

\subsubsection{Electron temperatures and densities  \label{secphysprop}}

Thanks to the high S/N of the extracted spectra, we could derive electron density and temperatures for the seven defined areas based from the extinction corrected fluxes of the collisionally excited lines (CEL) ratios \nii\,5755/6584, \sii\,6731/6716 (lower ionisation plasma), \siii\,6312/9069, and \cliii\,5538/5518 (higher ionisation plasma). We used the \texttt{Atom.getTemDen()} method in the  \texttt{PyNeb} package  \citep{Luridiana15}, version 1.1.13 with the default set of atomic data (transition probabilities and collision cross sections). We assumed $T_e$=12\,000~K for the $n_e$ and $n_e=100$~cm$^{-3}$ for the $T_e$. 
Uncertainties were estimated by means of Monte carlo simulations with 100 realisations. The reported uncertainties were calculated as half the difference between the 14 and 86 percentile of the distribution of a given outcome of the MC simulation.

   \begin{figure}[h]
   \centering
 \includegraphics[angle=0, trim=0 0 20 0, width=0.45\textwidth, angle=0, clip=,]{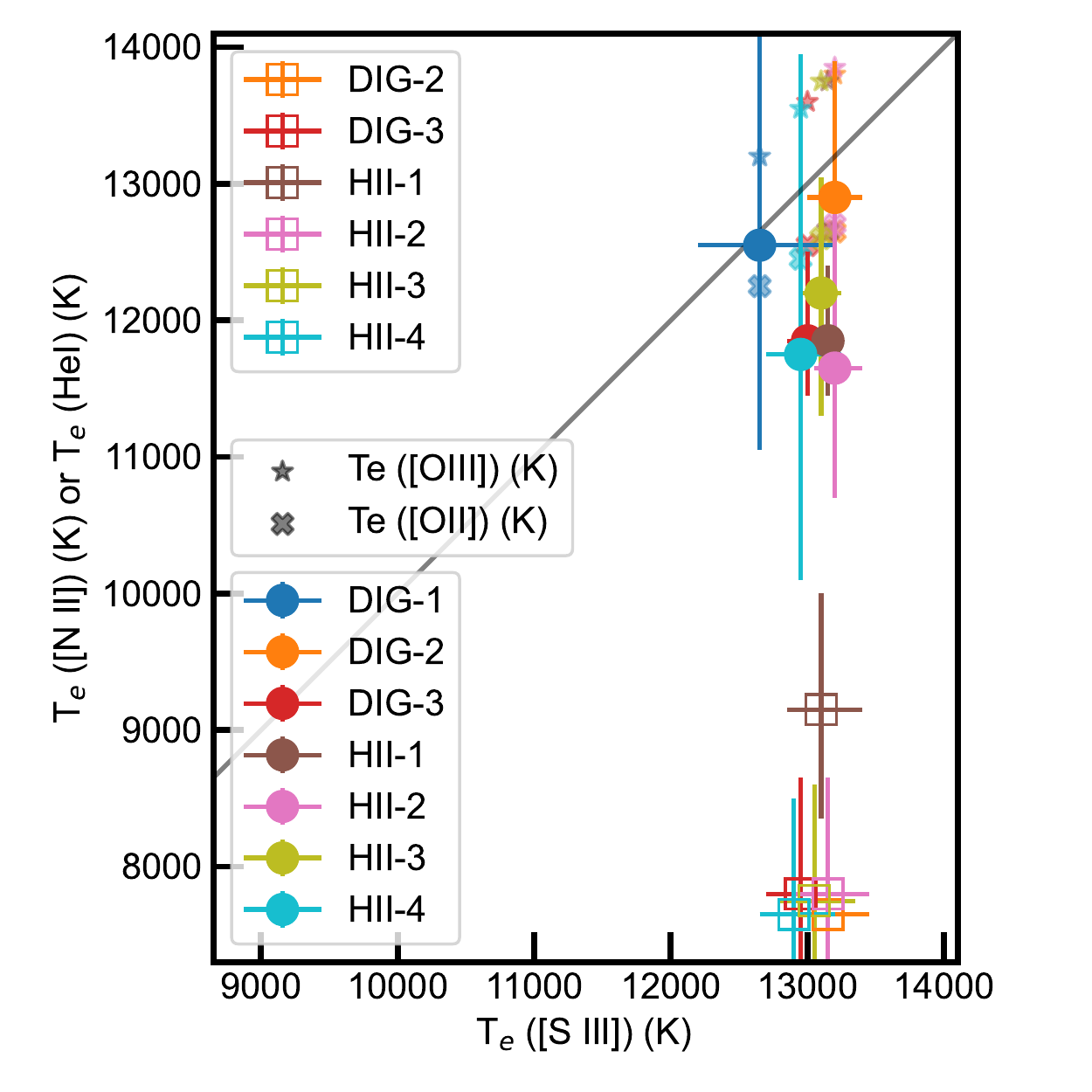}
   \caption{Comparison of the different derived $T_e$ values. The grey diagonal signals the locus of equal temperatures. Data points with error bars represent our measurements for $T_e$(\nii) (\emph{solid circles}) and $T_e$(\hei) (\emph{open squares}) vs $T_e$(\siii). Lighter smaller data points without error bars represent the predicted  $T_e$(\oiii) (\emph{stars}) and  $T_e$(\oii)  (\emph{crosses}) as a function of $T_e$(\siii) according to the relations proposed by \citet{Garnett92}.}
   \label{CompaTes}
    \end{figure}

Results for the seven selected regions are presented in Table~\ref{physprop}. Densities for low ionisation plasma are always  in the low density regime. The only exception is \hii-1, were they reach values of $\sim$100~cm$^{-3}$. Densities for high ionisation plasma seem somewhat higher ($\sim$200~cm$^{-3}$), although again in the low density regime for that diagnostic. Uncertainties are large for this diagnostic due to the faintness of the chlorine lines.

\begin{table*}[th!]
\caption{Physical properties derived from the line fluxes presented in Table \ref{rellinefluxes}. 
\label{physprop}} 
\centering  
\small
\begin{tabular}{lccccccccc}
\hline\hline
Property     &                    DIG-1 &                    DIG-2  &                    DIG-3 &                    \hii-1 &                    \hii-2 &                    \hii-3 &                    \hii-4 \\
\hline
\smallskip
$E(B-V)$               &     0.01$\pm^{   0.01}_{   0.01}$ &     0.02$\pm^{   0.01}_{   0.01}$ &     0.06$\pm^{   0.01}_{   0.01}$ &     0.12$\pm^{   0.01}_{   0.01}$ &     0.10$\pm^{   0.01}_{   0.01}$ &     0.06$\pm^{   0.01}_{   0.01}$ &     0.10$\pm^{   0.01}_{   0.01}$ \\
\smallskip
$n_e$(\sii)  (cm$^{-3}$) &           50$\pm^{60}_{40}$ &                 70$\pm^{80}_{50}$ &                70$\pm^{120}_{50}$ &               110$\pm^{110}_{70}$ &                 60$\pm^{10}_{10}$ &                 70$\pm^{80}_{50}$ &                 40$\pm^{10}_{10}$ \\
\smallskip
$n_e$(\cliii) (cm$^{-3}$)  &      1530$\pm^{1480}_{1130}$ &              280$\pm^{490}_{230}$ &              330$\pm^{360}_{270}$ &              210$\pm^{190}_{160}$ &              280$\pm^{290}_{170}$ &              250$\pm^{320}_{180}$ &                  500$\pm^{140}_{140}$ \\       
\smallskip
$T_e$(\nii)   (K) &                 12550$\pm^{1650}_{1500}$ &          12900$\pm^{1000}_{1000}$ &            11850$\pm^{650}_{400}$ &            11850$\pm^{550}_{400}$ &           11650$\pm^{1200}_{950}$ &            12200$\pm^{850}_{900}$ &          11750$\pm^{2200}_{1650}$ \\
\smallskip
$T_e$(\siii)  (K) &           12650$\pm^{550}_{450}$ &            13200$\pm^{200}_{200}$ &            13000$\pm^{150}_{150}$ &            13150$\pm^{100}_{100}$ &            13200$\pm^{200}_{150}$ &            13100$\pm^{150}_{150}$ &            12950$\pm^{300}_{250}$ \\
\smallskip
$T_e$(\hei)  (K)             &                  \ldots &   7650$\pm^{700}_{650}$ &             7800$\pm^{600}_{500}$ &             9150$\pm^{400}_{350}$ &             7800$\pm^{900}_{800}$ &             7750$\pm^{750}_{750}$ &             7650$\pm^{850}_{800}$ \\
\smallskip
$T_e$(\oiii)$^{\rm{(a)}}$  (K)     &           13200$\pm^{650}_{550}$ &            13800$\pm^{250}_{250}$ &            13600$\pm^{150}_{200}$ &            13750$\pm^{100}_{100}$ &            13850$\pm^{250}_{150}$ &            13750$\pm^{200}_{200}$ &            13550$\pm^{350}_{300}$ \\
\smallskip
$T_e$(\oii)$^{\rm{(a)}}$   (K)         &               12250$\pm^{450}_{400}$ &            12650$\pm^{200}_{150}$ &            12550$\pm^{100}_{150}$ &              12650$\pm^{50}_{50}$ &            12700$\pm^{150}_{100}$ &            12600$\pm^{150}_{150}$ &            12450$\pm^{250}_{200}$ \\
\hline
\end{tabular}

(a): Using relations provided by \citet{Garnett92}.
\end{table*}

 \begin{figure*}[!ht]
 \centering
\includegraphics[angle=0,  trim=80 12 60 35, width=0.35\textwidth, clip=,]{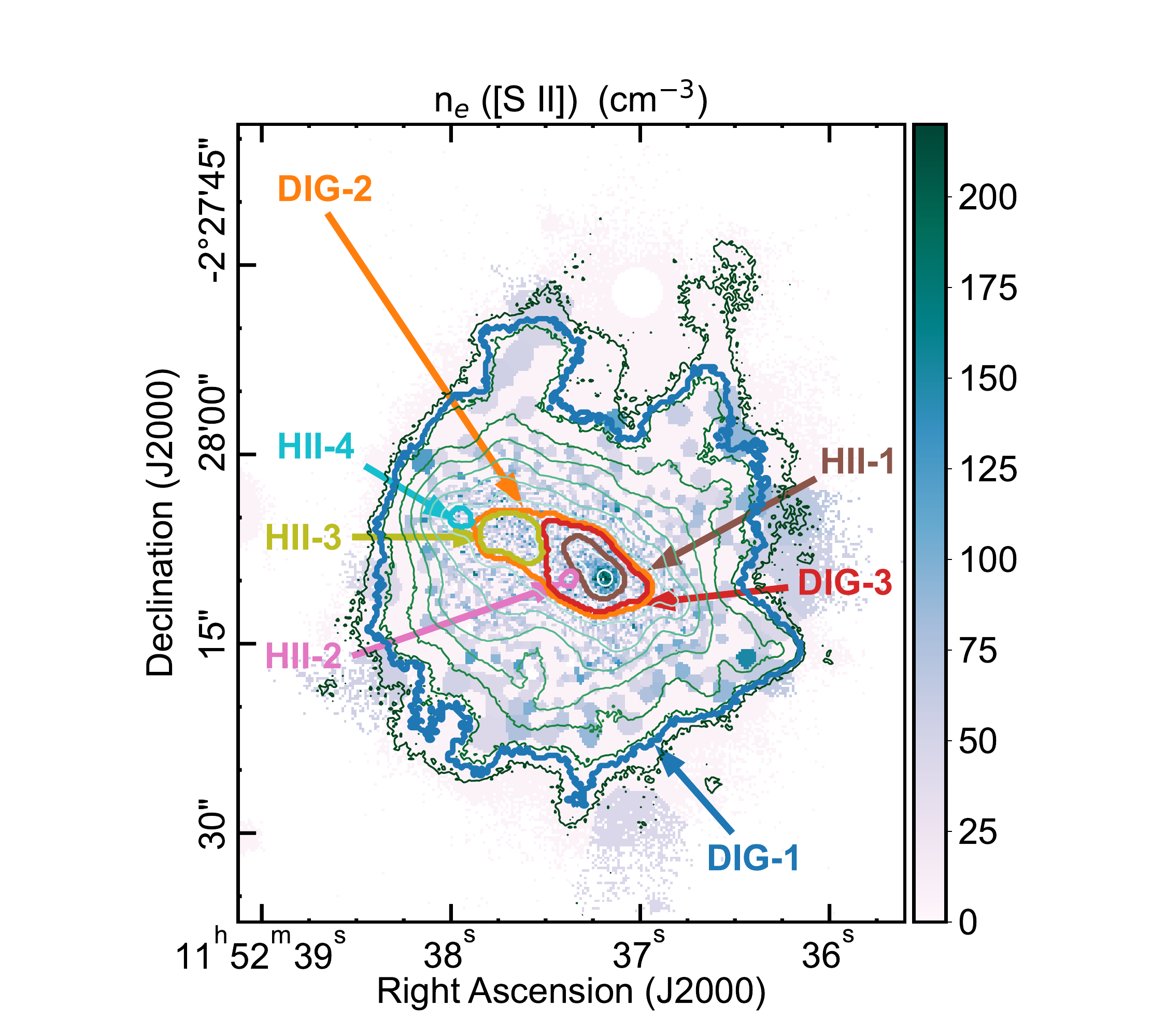}
\includegraphics[angle=0,  trim=80 12 60 35, width=0.35\textwidth, clip=,]{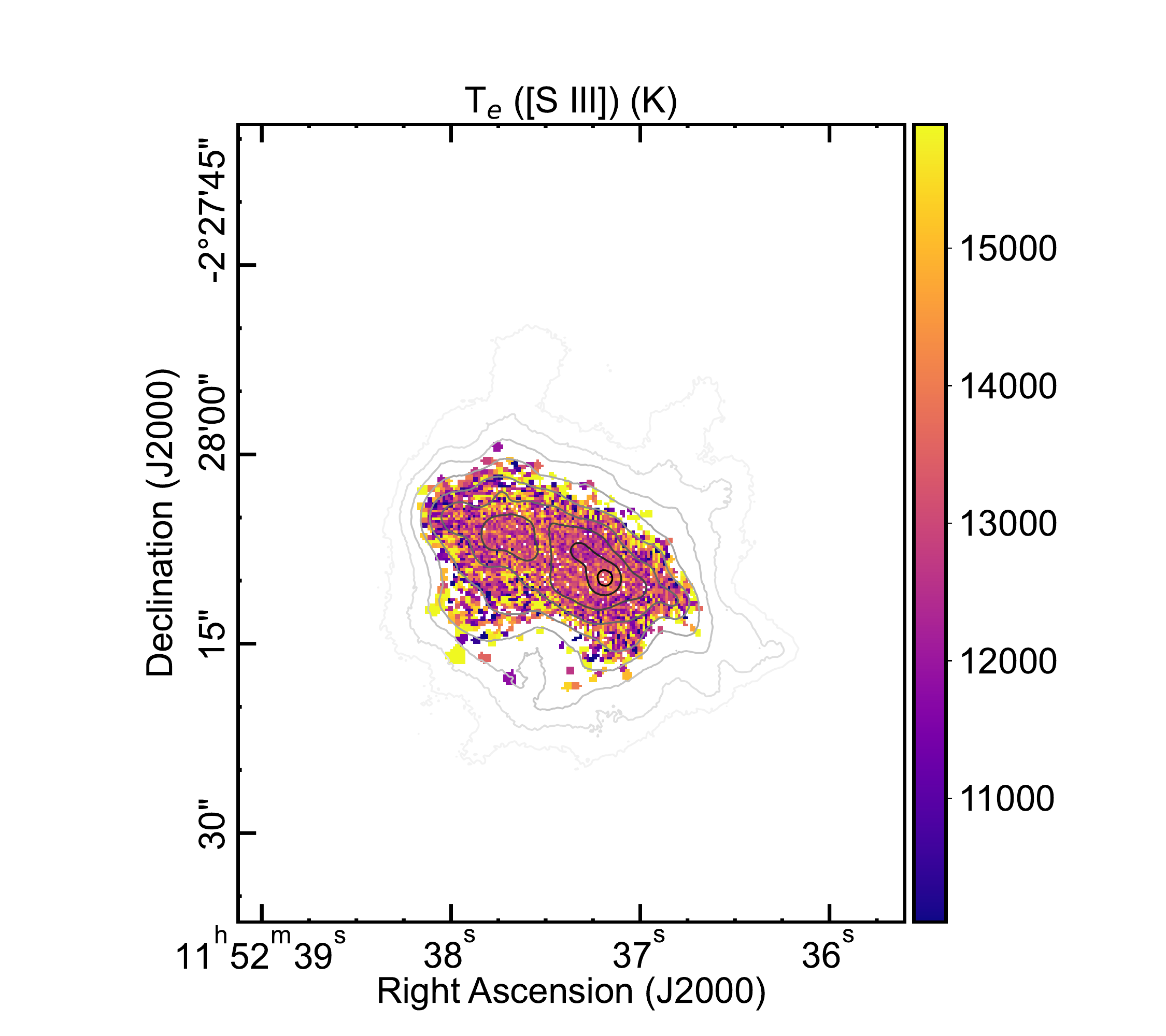}
\caption{Mapping the physical properties of the ionised gas in \object{UM\,462}.
\emph{Left:} Map of $n_e$ determined from the \sii\, emission lines and assuming a $T_e$=12\,000~K. 
\emph{Right:} Map of $T_e$ determined from the \siii\, emission line and assuming a $n_e$=100~cm$^{-3}$.
}
 \label{mapsTeNe}
\end{figure*}

A widely used description of  the thermal structure in a ionised nebula is the three-zone model  \citep{Garnett92}. Here, an ionised area is divided in three zones corresponding to three ranges of ionisation degree. The high-ionisation zone corresponds to ions such as \element[++][]{O}, or \element[+++][]{Ar}, while the intermediate zone to ions such as \element[++][]{Ar}  or \element[++][]{S} , and  the low-ionisation zone to species such as \element[+][]{O}  or \element[+][]{N} . The usual approximation is that temperature in each zone is represented by the temperature derived from emission lines of species in that zone. We could estimate the temperatures for the intermediate zone (by means of the \siii\, lines) and the low-ionisation zone (by means of the \nii\, lines).
These, $T_e$(\nii) and $T_e$(\siii) are displayed in Fig. \ref{CompaTes} with filled circles. 
Derived $T_e$(\siii) values are consistent with a uniform temperature of $\sim$13\,000~K all over the galaxy.
$T_e$(\nii) is comparable or systematically lower than $T_e$(\siii) by up to $\sim$1\,200~K. 
Uncertainties are large, specially in the $T_e$(\nii) since the flux of the \nii$\lambda$5755 was especially difficult to measure and is subject to large uncertainties  but Fig. \ref{CompaTes}  suggests a relatively homogeneous electron temperature all over the galaxy (i.e. with temperature differences,  $\Delta\,T_e \lesssim$1\,000~K) by means of both diagnostics.
We could not measure a diagnostic for the $T_e$ in the high-ionisation zone since the commonly used \oiii$\lambda$4363 auroral line is not covered with MUSE. Instead, we used an expression provided by \citet{Garnett92}, that relates $T_e$(\oiii) with $T_e$(\siii).  The derived values in relation with $T_e$(\siii) are displayed with with stars in Fig. \ref{CompaTes}).
$T_e$(\oiii) are somewhat larger than $T_e$(\siii) but never beyond 14\,000~K. For comparison, \citep{Guseva07} reports a $T_e$(\oiii)=16\,600~K and 13\,900~K  for two locations in the galaxy that roughly corresponds to our \hii\,1+ \hii\,2 and \hii\,3.

The expressions provided by \citet{Garnett92} also allow us to put in context our derived $T_e$(\nii) values. In principle, they should be equivalent to  $T_e$(\oii), since both ions belong to the same ionisation zone.
The crosses in Fig. \ref{CompaTes} display $T_e$(\oii) as derived using those  expressions. $T_e$(\nii) and $T_e$(\oii) differ by an amount that can range between $\sim-1\,000$~K and $\sim300$~K.
This difference can perfectly be understood simply by taking into account the uncertainty intrinsic to  the adopted relation between  $T_e$(\oii)  and $T_e$(\siii), actually a two-step relation, since  \citet{Garnett92} relates   $T_e$(\oii)  and $T_e$(\siii) through  $T_e$(\oiii).
In a recent paper, \citet{ArellanoCordova20} made a detailed comparison of the different  $T_e$(\nii)-$T_e$(\oiii) temperature relation in \hii\, regions published so far, and concluded that, together with the relation provided in their work, the one provided by \citet{Garnett92} was among the most reliable ones.  To our knowledge, there is not a similar work discussing relations involving the $T_e$(\siii).
For the range in temperatures covered here, the uncertainty in the relation for each of the steps would be $\sim$500~K.
Considering additional sources of uncertainties, in particular, that associated to the measurement of the flux for the  \nii$\lambda$5755 emission line, an absolute difference between the measured $T_e$(\nii) and derived $T_e$(\oii) of $\sim$500~K seems reasonable and consistent with both temperatures being in agreement.
Anyway, the comparison above stresses the difficulty of deriving reliable temperatures for each ionisation zone in extragalactic (and not particularly distant) objects. 
For the remaining of the study, when a temperature for a low-ionisation zone was needed, we used $T_e$(\nii).

\begin{table*}[h!]
\caption{Ionic and total abundances for helium, oxygen, nitrogen, sulphur, argon  and chlorine in selected apertures. \label{tabchemprop}} 
\centering  
\begin{tabular}{lccccccccc}
\hline\hline
Property     &                    DIG 1 &                    DIG 2  &                    DIG 3 &                    \hii\, 1 &                    \hii\, 2 &                    \hii\, 3 &                    \hii\, 4 \\
\hline
He$^+$/H$^+$($\lambda$4922)$\times10^2$&                  7.20$\pm$   2.27 &                  6.11$\pm$   2.78 &                  6.10$\pm$   3.27 &                  9.15$\pm$   4.53 &                  7.10$\pm$   5.78 &                  7.03$\pm$   3.37 &                  7.93$\pm$   3.62 \\
He$^+$/H$^+$($\lambda$5016)$\times10^2$&                  7.29$\pm$   1.58 &                  6.72$\pm$   2.18 &                  7.61$\pm$   2.61 &                  9.30$\pm$   3.24 &                  8.35$\pm$   4.14 &                  7.51$\pm$   2.31 &                  8.08$\pm$   2.40 \\
He$^+$/H$^+$($\lambda$5876)$\times10^2$  &                  7.80$\pm$   1.31 &                  7.82$\pm$  11.47 &                  7.81$\pm$   1.18 &                  8.13$\pm$   1.20 &                  8.17$\pm$   2.01 &                  7.74$\pm$   1.12 &                  7.85$\pm$   1.18 \\
He$^+$/H$^+$($\lambda$6678)$\times10^2$&                  7.95$\pm$   0.92 &                  7.90$\pm$   7.71 &                  7.90$\pm$   0.99 &                  8.17$\pm$   0.79 &                  8.15$\pm$   1.35 &                  7.81$\pm$   1.01 &                  7.96$\pm$   1.02 \\
He$^+$/H$^+$($\lambda$7065)$\times10^2$         &                  7.95$\pm$   0.74 &                  7.82$\pm$   5.39 &                  7.97$\pm$   1.00 &                  8.34$\pm$   2.14 &                  8.13$\pm$   0.94 &                  7.82$\pm$   0.87 &                  7.91$\pm$   0.77 \\
\rowcolor{light-gray}  He$^+$/H$^+$ $\times10^2$&                  7.86$\pm$   0.50 &                  7.84$\pm$   4.25 &                  7.86$\pm$   0.49 &                  8.18$\pm$   0.58 &                  8.16$\pm$   0.74 &                  7.77$\pm$   0.47 &                  7.89$\pm$   0.48 \\
\smallskip
O$^0$/H$^+$ ($\lambda$6300)$\times10^5$         &  0.67$\pm$   0.25 &                  0.39$\pm$   0.11 &                  0.44$\pm$   0.07 &                  0.32$\pm$   0.04 &                  0.40$\pm$   0.11 &                  0.36$\pm$   0.11 &                  0.40$\pm$   0.26 \\
O$^0$/H$^+$ ($\lambda$6364)$\times10^5$         &  0.67$\pm$   0.28 &                  0.40$\pm$   0.10 &                  0.49$\pm$   0.07 &                  0.32$\pm$   0.03 &                  0.46$\pm$   0.14 &                  0.38$\pm$   0.11 &                  0.36$\pm$   0.17 \\ 
O$^+$/H$^+$ ($\lambda$7320)$\times10^5$        &  3.56$\pm$   2.17 &                  3.02$\pm$   1.17 &                  4.47$\pm$   1.05 &                  3.16$\pm$   0.63 &                  4.15$\pm$   2.14 &                  3.90$\pm$   1.65 &                  4.92$\pm$   2.93 \\
O$^+$/H$^+$ ($\lambda$7331)$\times10^5$        &  4.12$\pm$   3.14 &                  3.12$\pm$   1.12 &                  4.32$\pm$   1.03 &                  3.40$\pm$   0.76 &                  4.19$\pm$   1.57 &                  3.99$\pm$   1.25 &                  5.16$\pm$   4.49 \\
O$^{++}$/H$^+$ ($\lambda$4959)$\times10^5$   &   5.25$\pm$   0.69 &                  6.21$\pm$   0.26 &                  6.97$\pm$   0.22 &                  8.39$\pm$   0.19 &                  7.26$\pm$   0.25 &                  7.14$\pm$   0.20 &                  5.80$\pm$   0.33 \\
O$^{++}$/H$^+$ ($\lambda$5007)$\times10^5$     &  5.16$\pm$   0.61 &                  5.97$\pm$   0.22 &                  6.81$\pm$   0.22 &                  8.11$\pm$   0.18 &                  7.02$\pm$   0.25 &                  6.84$\pm$   0.23 &                  5.57$\pm$   0.31 \\
O /H$\times10^5$                 &       9.18$\pm$   2.44 &                  9.57$\pm$   0.98 &                 11.92$\pm$   0.84 &                 11.72$\pm$   0.61 &                 11.25$\pm$   1.48 &                 10.50$\pm$   0.90 &                 11.99$\pm$   6.96 \\
\rowcolor{light-gray} 12+$\log$(O/H)            &    7.96$\pm^{   0.10}_{   0.13}$ &     7.98$\pm^{   0.04}_{   0.05}$ &     8.08$\pm^{   0.03}_{   0.03}$ &     8.07$\pm^{   0.02}_{   0.02}$ &     8.05$\pm^{   0.05}_{   0.06}$ &     8.02$\pm^{   0.04}_{   0.04}$ &     8.08$\pm^{   0.20}_{   0.38}$ \\
N$^0$/H$^+$ ($\lambda$5199)$\times10^6$       &  0.35$\pm$   0.16 &                  0.22$\pm$   0.05 &                  0.27$\pm$   0.04 &                  0.18$\pm$   0.02 &                  0.26$\pm$   0.08 &                  0.18$\pm$   0.04 &                  0.19$\pm$   0.10 \\
N$^+$/H$^+$ ($\lambda$6584)$\times10^6$     &   1.15$\pm$   0.26 &                  0.96$\pm$   0.17 &                  1.09$\pm$   0.11 &                  0.85$\pm$   0.09 &                  1.05$\pm$   0.20 &                  0.83$\pm$   0.16 &                  0.88$\pm$   0.35 \\
ICF(N)$_{\rm I06}^{\rm{(a)}}$          &   2.06$\pm$   1.05 &                  2.83$\pm$   1.05 &                  2.34$\pm$   0.44 &                  3.22$\pm$   0.56 &                  2.45$\pm$   1.65 &                  2.46$\pm$   1.08 &                  1.96$\pm$   1.08 \\
N/H$\times10^6_{\rm I06}\,^{\rm{(a)}} $          &     1.49$\pm$   0.42 &                  1.17$\pm$   0.22 &                  1.36$\pm$   0.16 &                  1.03$\pm$   0.11 &                  1.32$\pm$   0.28 &                  1.01$\pm$   0.21 &                  1.07$\pm$   0.44 \\
12+$\log$(N/H)$^{\rm{(a)}}$           &      6.37$\pm^{   0.22}_{   0.49}$ &     6.43$\pm^{   0.20}_{   0.37}$ &     6.39$\pm^{   0.12}_{   0.17}$ &     6.44$\pm^{   0.10}_{   0.13}$ &     6.39$\pm^{   0.26}_{   0.72}$ &     6.29$\pm^{   0.21}_{   0.44}$ &     6.24$\pm^{   0.32}_{   1.15}$ \\
\rowcolor{light-gray}  $\log$(N/O)$^{\rm{(a)}}$          &    -1.48$\pm^{   0.27}_{   0.83}$ &    -1.51$\pm^{   0.16}_{   0.25}$ &    -1.68$\pm^{   0.09}_{   0.11}$ &    -1.66$\pm^{   0.08}_{   0.10}$ &    -1.67$\pm^{   0.17}_{   0.28}$ &    -1.67$\pm^{   0.12}_{   0.17}$ &    -1.97$\pm^{   0.34}_{ 0.40}$ \\
S$^+$/H$^+$ ($\lambda$6716)$\times10^6$       & 0.79$\pm$   0.15 &                  0.49$\pm$   0.08 &                  0.54$\pm$   0.06 &                  0.38$\pm$   0.04 &                  0.52$\pm$   0.09 &                  0.41$\pm$   0.05 &                  0.57$\pm$   0.20 \\
S$^+$/H$^+$ ($\lambda$6731)$\times10^6$      & 0.75$\pm$   0.19 &                  0.49$\pm$   0.11 &                  0.51$\pm$   0.06 &                  0.37$\pm$   0.04 &                  0.52$\pm$   0.10 &                  0.41$\pm$   0.07 &                  0.55$\pm$   0.18 \\
S$^{++}$/H$^+$ ($\lambda$9069)$\times10^6$    &   1.25$\pm$   0.11 &                  1.32$\pm$   0.04 &                  1.40$\pm$   0.03 &                  1.40$\pm$   0.02 &                  1.40$\pm$   0.03 &                  1.36$\pm$   0.03 &                  1.35$\pm$   0.04 \\
S$^{++}$/H$^+$ ($\lambda$6312)$\times10^6$     &  1.31$\pm$   0.25 &                  1.32$\pm$   0.07 &                  1.41$\pm$   0.05 &                  1.38$\pm$   0.05 &                  1.41$\pm$   0.07 &                  1.35$\pm$   0.05 &                  1.35$\pm$   0.10 \\
ICF(S)$_{\rm I06}^{\rm{(a)}}$     &   1.06$\pm$   0.04 &                  1.08$\pm$   0.04 &                  1.06$\pm$   0.02 &                  1.09$\pm$   0.03 &                  1.08$\pm$   0.07 &                  1.07$\pm$   0.04 &                  1.05$\pm$   0.08 \\
S/H$\times10^6_{\rm I06}\,^{\rm{(a)}} $          &   2.17$\pm$   0.55 &                  1.95$\pm$   0.22 &                  2.03$\pm$   0.15 &                  1.94$\pm$   0.11 &                  2.09$\pm$   0.34 &                  1.90$\pm$   0.16 &                  2.04$\pm$   0.60 \\
12+$\log$(S/H)      &  6.34$\pm^{   0.10}_{   0.13}$ &     6.29$\pm^{   0.05}_{   0.05}$ &     6.31$\pm^{   0.03}_{   0.03}$ &     6.29$\pm^{   0.02}_{   0.02}$ &     6.32$\pm^{   0.07}_{   0.08}$ &     6.28$\pm^{   0.03}_{   0.04}$ &     6.31$\pm^{   0.11}_{   0.15}$ \\
\rowcolor{light-gray}  $\log$(S/O)            &     -1.61$\pm^{   0.13}_{   0.19}$ &    -1.69$\pm^{   0.07}_{   0.08}$ &    -1.76$\pm^{   0.04}_{   0.05}$ &    -1.77$\pm^{   0.04}_{   0.04}$ &    -1.73$\pm^{   0.07}_{   0.09}$ &    -1.73$\pm^{   0.05}_{   0.06}$ &    -1.77$\pm^{   0.23}_{   0.49}$ \\
Ar$^{++}$/H$^+$ ($\lambda$7136)$\times10^7$   &   2.77$\pm$   0.25 &                  2.90$\pm$   0.09 &                  3.16$\pm$   0.07 &                  3.35$\pm$   0.07 &                  3.23$\pm$   0.07 &                  3.03$\pm$   0.07 &                  2.85$\pm$   0.09 \\
ICF(Ar)$_{\rm I06}^{\rm{(a)}}$      &                1.08$\pm$   0.02 &                  1.08$\pm$   0.03 &                  1.07$\pm$   0.00 &                  1.08$\pm$   0.01 &                  1.08$\pm$   0.04 &                  1.08$\pm$   0.02 &                  1.08$\pm$   0.04 \\
Ar/H$\times10^7_{\rm I06}\,^{\rm{(a)}} $      &                3.01$\pm$   0.55 &                  3.14$\pm$   0.22 &                  3.41$\pm$   0.16 &                  3.63$\pm$   0.15 &                  3.51$\pm$   0.21 &                  3.27$\pm$   0.17 &                  3.09$\pm$   0.26 \\
12+$\log$(Ar/H)        &     5.48$\pm^{   0.07}_{   0.09}$ &     5.50$\pm^{   0.03}_{   0.03}$ &     5.53$\pm^{   0.02}_{   0.02}$ &     5.56$\pm^{   0.02}_{   0.02}$ &     5.54$\pm^{   0.03}_{   0.03}$ &     5.51$\pm^{   0.02}_{   0.02}$ &     5.49$\pm^{   0.04}_{   0.04}$ \\
\rowcolor{light-gray} $\log$(Ar/O)     &  -2.48$\pm^{   0.14}_{   0.20}$ &    -2.48$\pm^{   0.05}_{   0.06}$ &    -2.54$\pm^{   0.04}_{   0.04}$ &    -2.50$\pm^{   0.03}_{   0.03}$ &    -2.51$\pm^{   0.06}_{   0.07}$ &    -2.50$\pm^{   0.04}_{   0.05}$ &    -2.58$\pm^{   0.20}_{   0.39}$ \\
Cl$^{++}$/H$^+$ ($\lambda$5518)$\times10^8$     &   1.33$\pm$   0.37 &                  1.46$\pm$   0.15 &                  1.56$\pm$   0.10 &                  1.55$\pm$   0.05 &                  1.55$\pm$   0.16 &                  1.52$\pm$   0.11 &                  1.77$\pm$   0.29 \\
Cl$^{++}$/H$^+$ ($\lambda$5538)$\times10^8$  &    1.35$\pm$   0.50 &                  1.22$\pm$   0.21 &                  1.48$\pm$   0.12 &                  1.46$\pm$   0.07 &                  1.15$\pm$   0.22 &                  1.40$\pm$   0.14 &                  1.06$\pm$   0.40 \\
ICF(Cl)$_{\rm I06}^{\rm{(a)}}$       &   1.27$\pm$   0.04 &                  1.27$\pm$   0.05 &                  1.27$\pm$   0.01 &                  1.28$\pm$   0.03 &                  1.27$\pm$   0.03 &                  1.27$\pm$   0.02 &                  1.31$\pm$   0.11 \\
Cl/H$\times10^8_{\rm I06}\,^{\rm{(a)}} $            &       1.70$\pm$   0.96 &                  1.71$\pm$   0.33 &                  1.93$\pm$   0.18 &                  1.93$\pm$   0.13 &                  1.72$\pm$   0.35 &                  1.86$\pm$   0.26 &                  1.93$\pm$   0.71 \\12+$\log$(Cl/H)           &    4.23$\pm^{   0.19}_{   0.36}$ &     4.23$\pm^{   0.08}_{   0.09}$ &     4.29$\pm^{   0.04}_{   0.04}$ &     4.29$\pm^{   0.03}_{   0.03}$ &     4.24$\pm^{   0.08}_{   0.10}$ &     4.27$\pm^{   0.06}_{   0.07}$ &     4.29$\pm^{   0.14}_{   0.20}$ \\
\rowcolor{light-gray}  $\log$(Cl/O)               &    -3.73$\pm^{   0.20}_{   0.39}$ &    -3.74$\pm^{   0.09}_{   0.12}$ &    -3.79$\pm^{   0.06}_{   0.06}$ &    -3.78$\pm^{   0.04}_{   0.04}$ &    -3.82$\pm^{   0.09}_{   0.12}$ &    -3.75$\pm^{   0.07}_{   0.09}$ &    -3.78$\pm^{   0.37}_{   0.43}$ \\
\hline
\end{tabular}

(a): ICFs in Eq. 18b, and 20b, 21b, and 22b by \citet{Izotov06}. 
\end{table*}

The temperature diagnostics discussed so far rely on measurements of CELs. But to prove the physical properties of the optical recombination lines  (ORL)  emitting regions, line ratios of certain recombination lines can be used. In particular, the \hei\,$\lambda$7281 and  \hei\,$\lambda$6678 recombination line ratio constitutes a suitable diagnostic for the ORL temperature \citep{Zhang05}. We used these lines together with  the analytic fits presented by \citet{Benjamin99} and the most recent  emissivities for these lines \citep{Porter13} to get an estimation of this temperature, $T_e$(\hei). This ratio does not depend very much on electron density \citep{Zhang05}. Thus, we simply assumed $n_e$=100~cm$^{-3}$.
The derived temperatures are in the range of $\sim$7\,600-9\,500~K, in all the cases but DIG-1, where we could not derive a reliable $T_e$(\hei). 
A comparison of these temperatures with $T_e$(\siii) is presented in Fig. \ref{CompaTes} with open squares.
$T_e$(\hei) is systematically lower than any CEL-based temperature by a factor $\sim0.6$. As a comparison, \citet{LopezSanchez07} found in \object{NGC\,5253}, a BCD galaxy with similar metallicity as \object{UM\,462} that $T_e$(\hei) was lower than  $T_e$(\siii) by a factor $\sim0.8$.

Finally, now we shall look to the 2D distribution of the electron density and temperature.
The \cliii\, doublet,  \nii$\lambda$5755 auroral line, and  \hei$\lambda$7281 recombination line were too faint to be detected over a substantial area of the galaxy in individual tiles. Thus, we obtained 2D maps for $n_e$ only as derived from the \sii\, lines and $T_e$ as derived from the \siii\, lines. The electron density map is presented on the left panel of Fig. \ref{mapsTeNe}. As suggested by the values derived on selected apertures, \object{UM\,462} has values in low density regime almost everywhere but in the peak of SF, and coincident with the location with the highest reddening.
It is already about a decade since pioneer works provided the community with the first $n_e$ density maps in BCDs derived using IFUs \citep[e.g.][]{MonrealIbero10a,MonrealIbero12a,Telles14}. In the meantime, the pool of examples has been slowly but steadily increasing. Nowadays, thanks to MUSE, maps covering an important area of the galaxy under study are routinely derived \citep[e.g.][]{Menacho21,Fernandez23}.  A recurrent finding in these examples, it is the presence of (more or less) central, and extremely young knots of star formation with larger extinction, and high $n_e$. These could be interpreted as the observational footprint of star clusters so young that the feedback has not yet not kicked in and the environment around has not yet been fully cleared out.

The electron temperature map is displayed on the right panel.  It includes only those spaxels with estimated uncertainties of $\lesssim$2\,000~K (median $\sim$900~K)
and covers an area of $\sim20^{\prime\prime}\times8^{\prime\prime}$ ($\sim$1.4~kpc$\times$0.6~kpc). 
The 2D electron temperature map does not reveal any particular $T_e$ gradient and simply supports the result for the extracted apertures on the brightest part of the galaxy. 
This is one of the few  $T_e$ maps in BCDs.  \citet{MonrealIbero12a} discussed maps for $T_e$(\oiii) and $T_e$(\sii)  for  \object{NGC\,5253}, a BCD with metallicity comparable to  \object{UM\,462}, but covering a smaller area  ($\sim8^{\prime\prime}\times4^{\prime\prime}$ or $\sim$140~pc$\times$70~pc). $T_e$(\oiii) ranged between 10\,000~K and 12\,500~K.
Besides, \citet{Kumari18} map $T_e$(\oiii) in a $\sim3^{\prime\prime}\times5^{\prime\prime}$ ($\sim$60~pc$\times$100~pc) area in \object{NGC\,4670}.
They found values comparable to those found in the inner regions of \object{UM\,462}.
Likewise, \citet{Menacho21} presented a $T_e$(\siii) map over an area of  $\sim$10~kpc$\times$10~kpc in Haro~11. As it happens here, $T_e$ was relatively uniform over most of the covered area, with the exception of an area of  about $\sim$7~kpc$\times$3~kpc  in the halo with relatively higher temperatures, where fast shocks have been detected.

The measured electron temperatures are comparable to available electron temperature for GP galaxies. For example, \citet{Amorin12} report $T_e$(\siii) and $T_e$(\nii) $\sim$12\.000-14\,000~K for three GP galaxies. This similarity strengthens the role of \object{UM\,462} as local GP analogue.
Moreover, obtaining spatially resolved information of the electron temperature in GP galaxies is clearly a challenge (because of the faintness of the auroral line). The relative homogeneous  $T_e$(\siii) found here  -- including the map in Fig. \ref{mapsTeNe} but also the value measured for DIG-1 -- supports the idea that a single value representative of the GP galaxy would be enough to characterise a galaxy where such a high spatially resolution as here is not attainable. This would need to be backed with similar maps from additional GP analogues.

 \begin{figure}[ht]
 \centering
\includegraphics[angle=0,  trim=80 12 60 35, width=0.35\textwidth, clip=,]{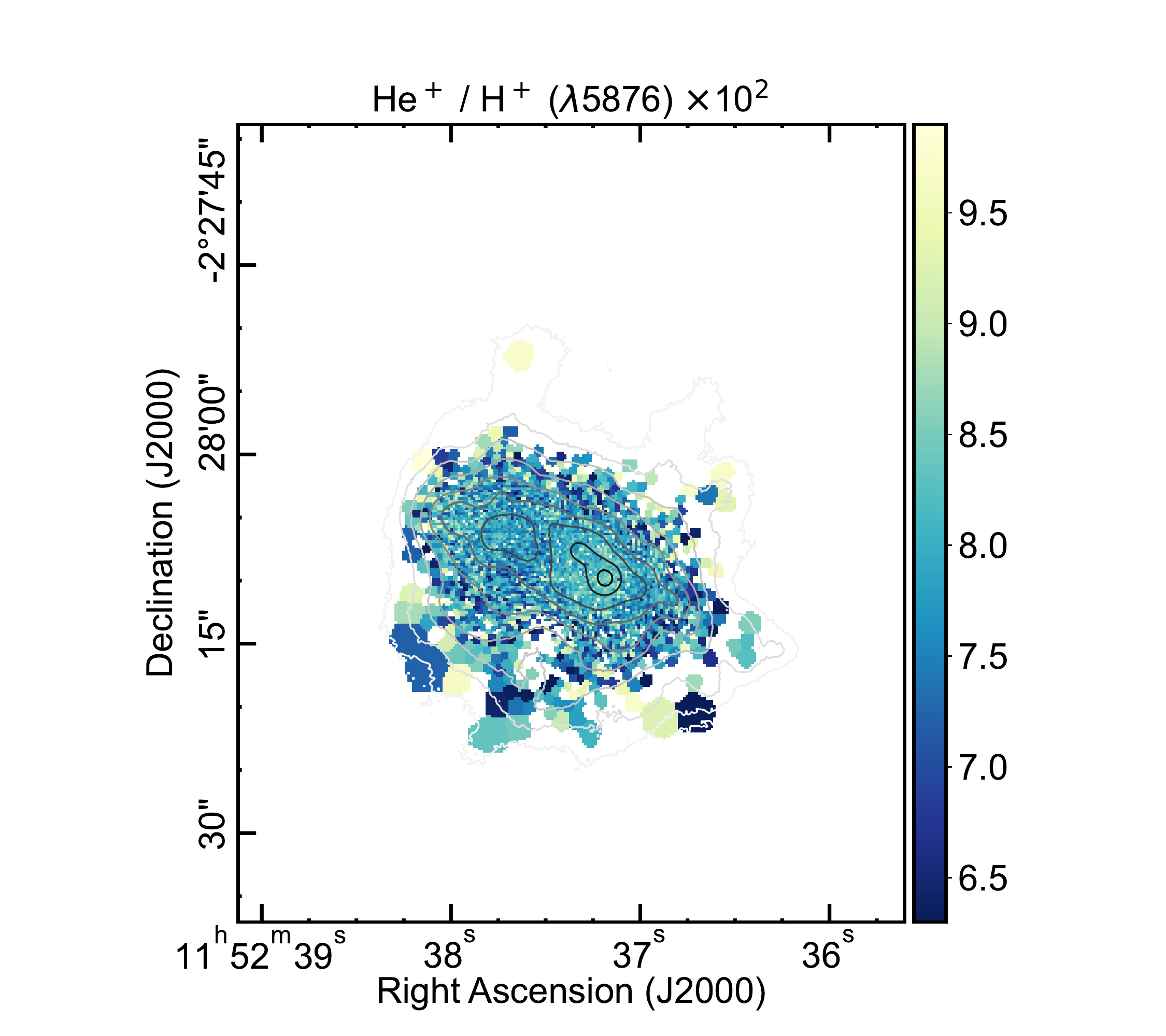}
\caption{
Map for the He$^+$/H$^+$ abundance derived from the \hei$\lambda$5876 emission line.
North is up and east towards the left. 
\label{mapHeI}
}
\end{figure}

\subsubsection{Chemical properties  \label{secchemprop}}

We derived several ionic and total abundances using the set of lines presented in Table \ref{rellinefluxes}. In all the cases, we assumed an $n_e$(\sii)=50 cm$^{-3}$  but \hii-1, where we used the value reported in Table \ref{physprop}.  We did not consider $n_e$(\cliii)  because of the large uncertainty.
Ionic abundances from recombination lines - here only for helium – were calculated using the method \texttt{RecAtom.getEmissivity()}  in \texttt{PyNeb}. Ionic abundances from
CELs were determined using the  \texttt{Atom.getIonAbundance()} method in \texttt{PyNeb}. To derive total abundance, we used the \texttt{ICF} class. Uncertainties were determined by means of Monte Carlo simulations with 100 realisations, and assuming a normal distribution for the errors of the involved lines.
For a given ion, we used as many emission lines as possible to asses the consistency of the results in the apertures defined by \texttt{astrodendro}. Besides, for helium, oxygen and nitrogen we derived 2D abundance maps using the strongest lines for their ions. 
Derived ionic and total abundances  as well as used ionisation correction factors (ICFs) for the selected apertures are listed in Table \ref{tabchemprop}. There, rows shaded in grey mark the adopted total abundance for a given element. The reported uncertainties were calculated as half the difference between the 14 and 86 percentile of the distribution of a given outcome of the MC simulation. 

\paragraph{Helium: \label{sechelium}}

There are several singlet and triplet recombination He$^+$ lines within the MUSE spectral coverage. Here, we measured fluxes for  three singlet ($\lambda$4922, $\lambda$5016, and $\lambda$6678), and two triplet ($\lambda$5876, and $\lambda$7065) lines. While lines in the singlet cascade are insensitive to radiative transfer effects, these can have an impact in lines belonging to the triplet cascade, and in particular, in $\lambda$7065  (see \citet{MonrealIbero13} and references therein for a discussion on these effects).
A choice should be made also about the assumed temperature to calculate the helium abundance. Many works in the literature use $T_e$(\oiii). However, in our case, this was not directly measured from our data, but an estimate that relies on other $T_e$ measurements.
Abundances reported were derived using $T_e$(\siii). To assess the robustness of the derived values, we also calculated the helium abundances using the helium emissivities at the measured $T_e$(\hei). Differences were $\lesssim$0.1\%, much smaller than the estimated uncertainty.
We used the ratio between the $\lambda$6678 and $\lambda$7065 lines to evaluate the relative importance of radiative transfer effects. We found these basically non-existent in every aperture but in \hii-1 which were negligible ($\tau(\lambda3889)\sim$0.020, for a ratio between radial and thermal velocity of $\omega\sim$2).
This is within the expectation given the derived low electron density.

Abundances by means of the different helium lines are reported in  Table \ref{tabchemprop}. For a given aperture, all the lines give consistent results. Since the measurement of the fluxes for the faint and (and heavily blended) lines $\lambda$4922 and $\lambda$5016 were of considerably lower quality, we used the abundances derived from lines $\lambda$5876:$\lambda$6678:$\lambda$7065, with a weight 3:1:1 to derive the single ionised helium abundance.
We also derived a 2D map of He$^+$/H$^+$, this time only using the strongest line (see Fig. \ref{mapHeI}).
He$^+$/H$^+$ is the highest at \hii-1 and the lowest in DIG-1, suggesting a larger contribution of neutral helium when going towards the outer parts of the galaxy. 
As for comparison, \citet{James10} report an He/H = $8.73\pm0.72\times10^{-2}$.
This is comparable within the uncertainties but somewhat higher with the values measured here. 

Regarding He$^{++}$, MUSE in normal mode do not cover the strongest optical line (\heii$\lambda$4686), and we did not detect the much fainter \heii\, line at $\lambda$5411 in any or our apertures.
\citet{James10} report that, if \heii\, were present, then it presents an abundance $\lesssim 1/50^{th}$ that of \hei. Thus, we consider its contribution here as to be negligible. 

 \begin{figure*}[ht]
 \centering
\includegraphics[angle=0,  trim=80 12 60 35, width=0.35\textwidth, clip=,]{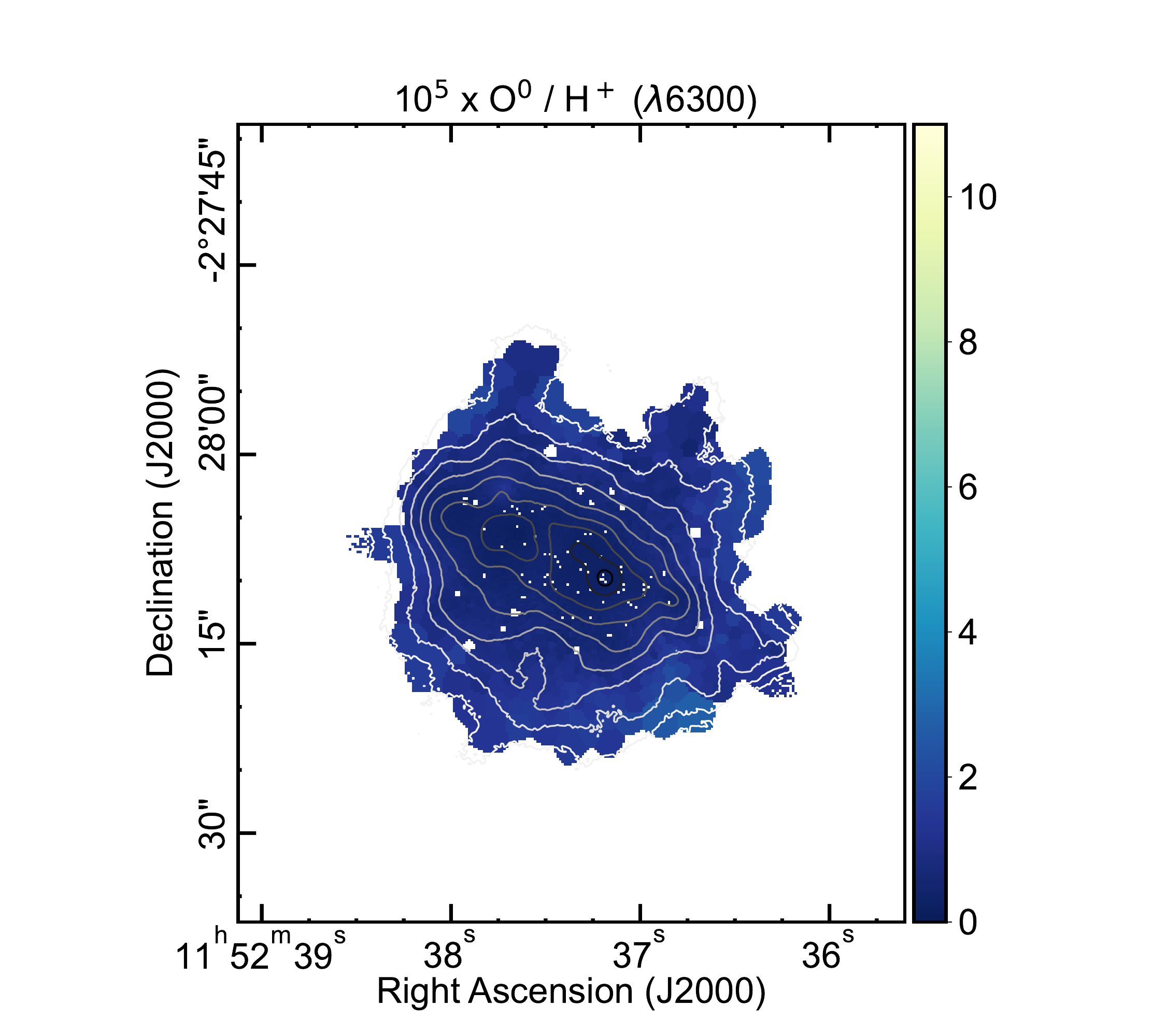}
\includegraphics[angle=0,  trim=80 12 60 35, width=0.35\textwidth, clip=,]{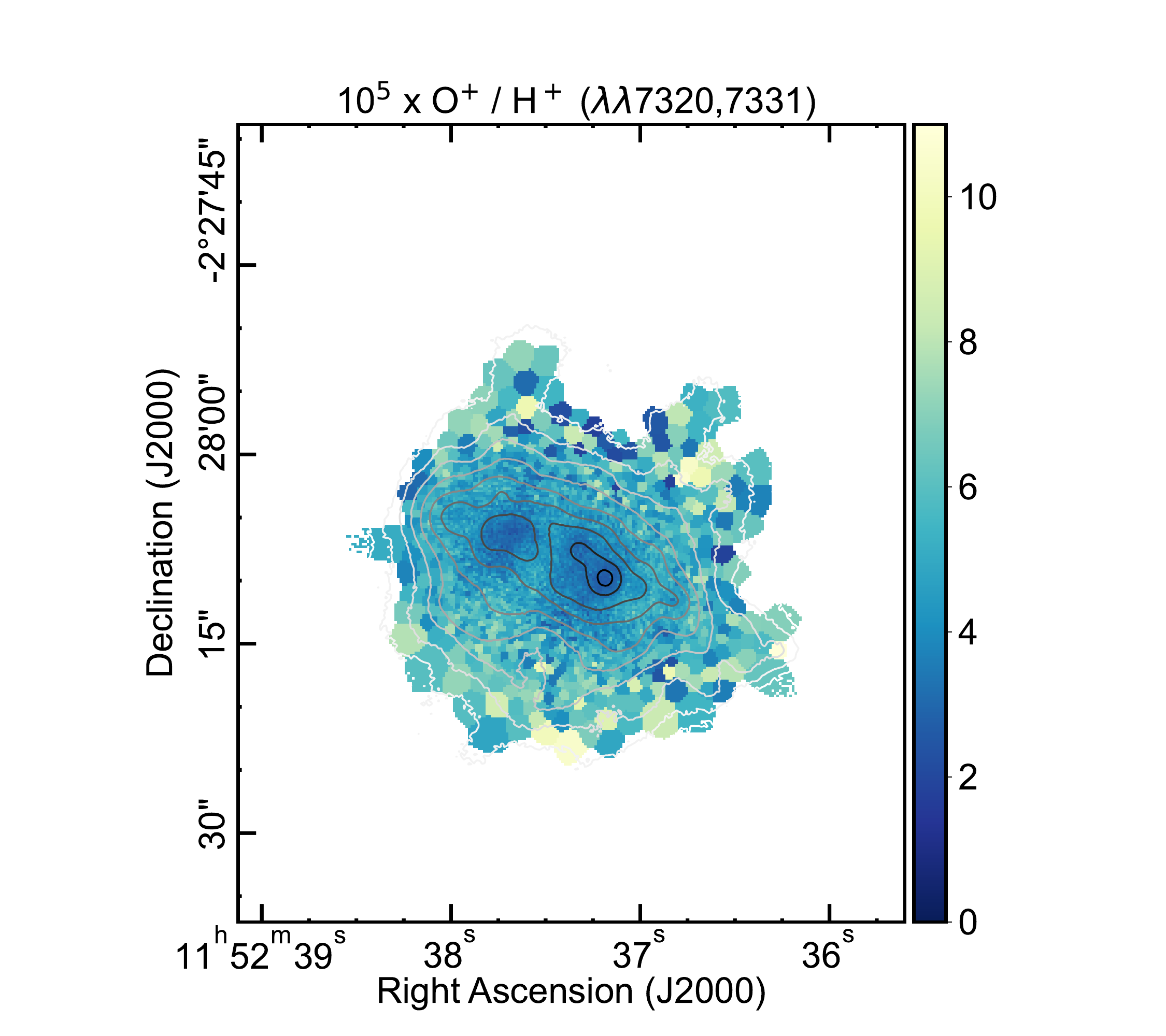}
\includegraphics[angle=0,  trim=80 12 60 35, width=0.35\textwidth, clip=,]{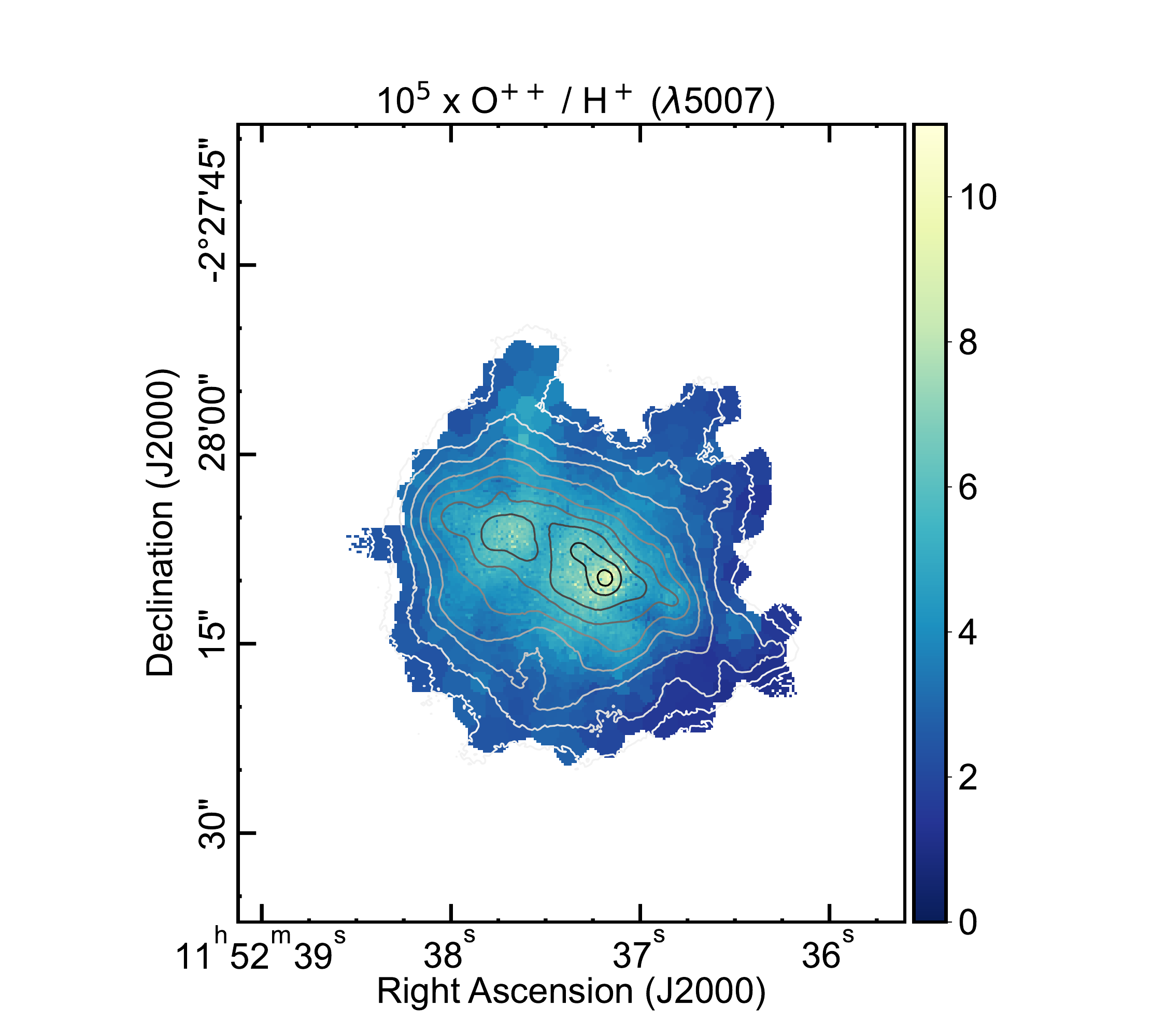}
 \includegraphics[angle=0,  trim=80 12 60 35, width=0.35\textwidth, clip=,]{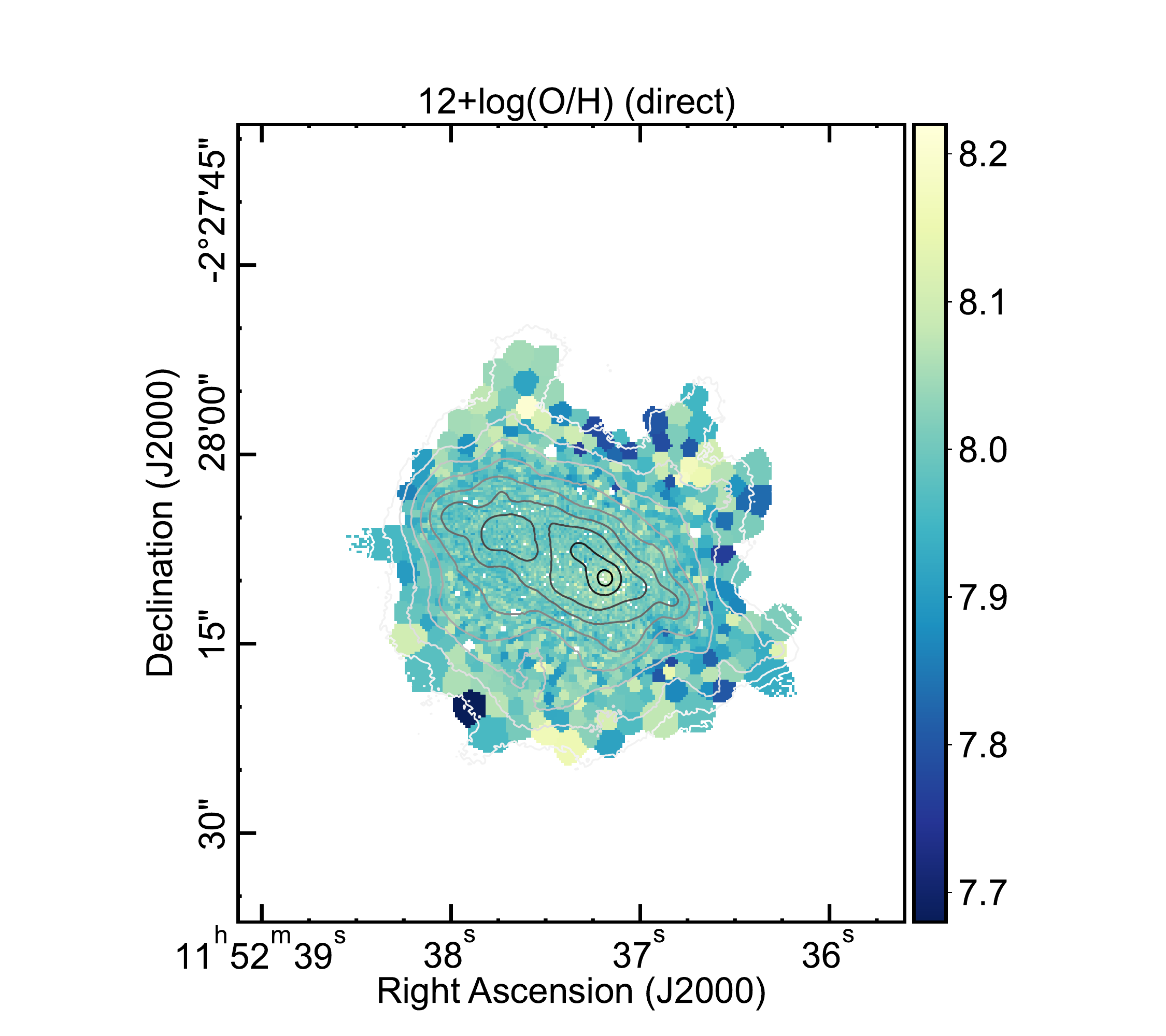}
\caption{
Maps for the $O^{++}$ (\emph{top left}), $O^+$, (\emph{top right}), and $O^0$ (\emph{bottom left}) ionic abundances, derived using the direct method as described in the text. All the three maps display the same range in abundance in order to emphasise the relative contribution of each ion in the different parts of the galaxy. The last map in this figure (\emph{bottom right}) contains the total oxygen abundance map, $12+\log(O/H)$.
North is up and east towards the left.
}
\label{mapOionic}
\end{figure*}

 \begin{figure}[ht]
 \centering
\includegraphics[angle=0,  trim=80 12 60 35, width=0.35\textwidth, clip=,]{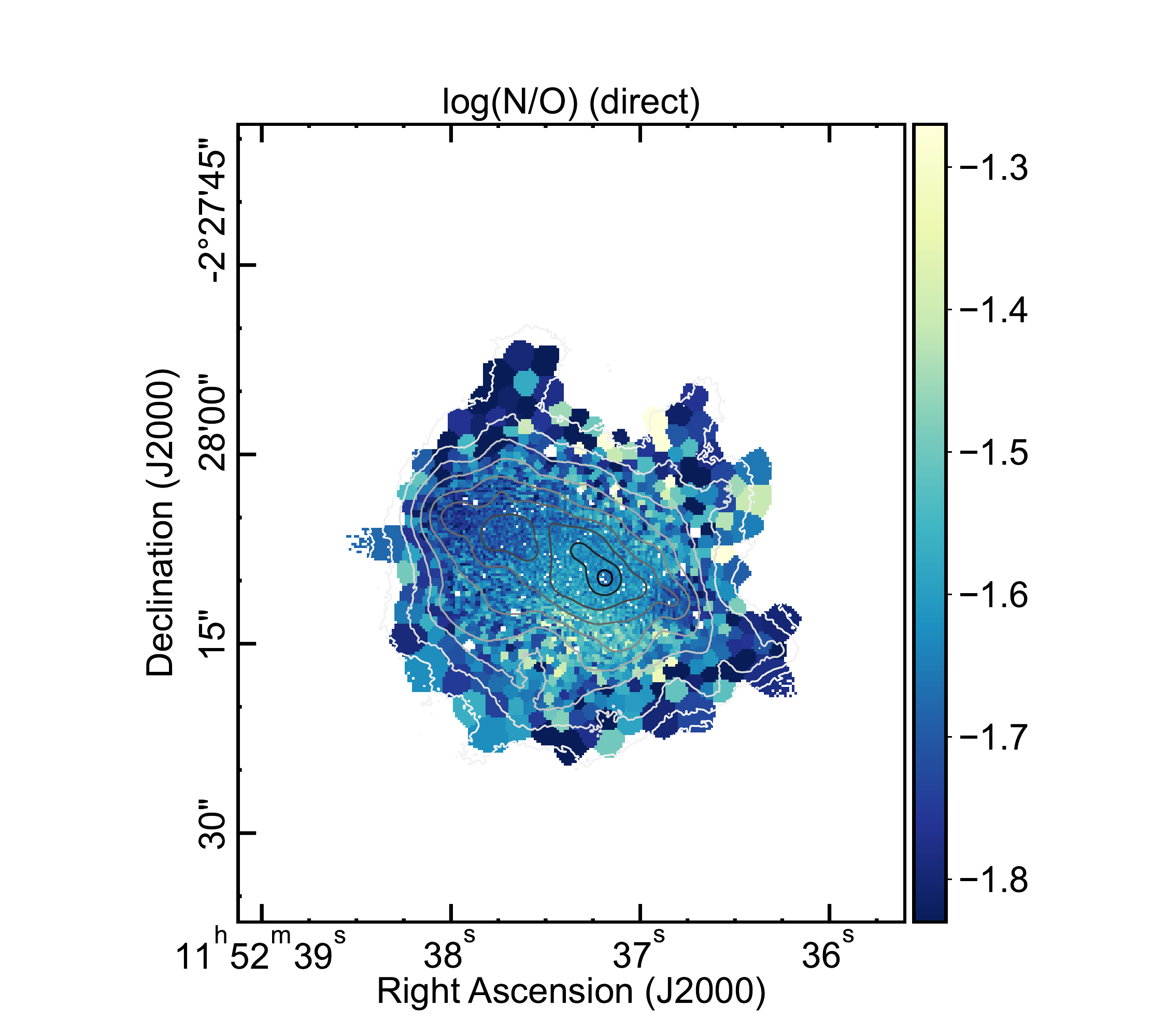}
\caption{
Map for the relative nitrogen-to-oxygen abundance derived using the direct method as described in the text. 
North is up and east towards the left. 
\label{mapN}
}
\end{figure}

\paragraph{Oxygen:}

The MUSE spectral range covers optical lines tracing neutral, singly ionised and doubly ionised oxygen. %
For this last ion, we used both the \oiii$\lambda$4959 and \oiii$\lambda$5007 lines together with the values derived for  $T_e$(\oiii).  
For singly ionised and neutral oxygen, we adopted a temperature of $T_e$= 12\,100~K, the average of the $T_e$(\nii) values. Regarding singly ionised oxygen, the \oii$\lambda$7320 and  \oii$\lambda$7331 lines (actually, four lines seen as a doublet at the MUSE spectral resolution) may be affected by a certain O$^{++}$ recombination contribution. Following the correction scheme proposed by \citet{Liu00}, we estimated a negligible contribution here. Finally, neutral oxygen abundance was derived using the \oi$\lambda$6300 and \oi$\lambda$6363 lines.
Total abundances were calculated by addition of ionic abundances. For the selected apertures, the two \oii\, lines were equally weighted, and the two \oiii\, lines were weighted 3:1, as the two \oi\, lines.  
Results for the selected apertures are presented in Table \ref{secchemprop}, while maps for the ionic and total abundances appear in  Fig. \ref{mapOionic}.

The comparison with the results by \citet{James10}, when possible (our regions \hii-1-- their regions 1+3, \hii-2[+ DIG3] -- approximately, their region 2, and \hii~3 -- approximately, their region 4), shows an agreement within the uncertainties. The relative abundances of neutral, single and double ionised oxygen in the different apertures already delineate the variation in ionisation conditions from higher ionisation degree (in \hii-1, where the proportion of \element[++][]{O}  is the highest and that of \element[0][]{O} is the lowest) to lower ionisation degree (DIG-1, where the situation is reversed). 
This is much better seen in the 2D abundance maps (see Fig. \ref{mapOionic}) where all the ions have been presented with the same scale to make  the comparison of the relative ionic abundances easier.
Regarding total oxygen abundance,  within the reported uncertainties, this is uniform all over the galaxy, and indicates a metallicity for the galaxy of $Z\sim 0.25 Z_\odot$, well within the range of metallicities derived for GP galaxies \citep{Izotov11,Amorin10}.

The direct method is considered the gold standard in terms of determining element abundances in extragalactic astronomy. In that sense, the map presented in Fig. \ref{mapOionic} is, probably, among the largest oxygen abundance maps, in terms of covered area, determined by means of this method (see also \citet{Menacho21} for a similar map in \object{Haro 11}). However, given the difficulty to measure the auroral line(s), strong line methods are more routinely used instead.
Because of the importance of identifying metallicity gradients or inhomogeneities  in dwarf galaxies, we discuss in Sect. \ref{secstronlinemethods} how the map derived here compares to the results by some of the most widely used strong line methods.

\paragraph{Nitrogen:}

We used the \niA$\lambda$5199 emission line (actually a blended doublet at the MUSE spectral resolution) to determine the abundance in neutral nitrogen, and the \nii$\lambda$6584 line for the single ionised nitrogen.  For both, \element[+][]{N} and \element[0][]{N}, we employed  the $T_e$(\nii) values. Derived ionic abundances are listed in Table \ref{tabchemprop}. To our knowledge, this is the first time that neutral nitrogen abundances are reported for \object{UM~462}. Regarding ionised nitrogen, we find an abundance about $\times2$ larger than those reported by \citet{James10}. 

The sum of these two ionic abundances is  clearly higher in the areas far from the main sites of star formation (e.g. DIG-1, DIG-2) than in the brightest \hii\, regions (e.g. \hii-1). 
The results of the ionic abundances for oxygen suggests unseen further (i.e. two times or more) ionised nitrogen in the main sites of SF but does not reject an  inhomogeneity in nitrogen abundance with lower total nitrogen at those locations.
When the abundance of one or several ions of a given element is not available, it is customary the use of different ionisation correction factor (ICF) schemes that rely on
the available abundances of ions of other elements with comparable ionisation potentials to those of the ionic species of the element of interest.
Here, we adopted those presented by \citet{Izotov06} for intermediate oxygen abundances, derived using emission-line galaxies observed in the frame of the SDSS.
The derived total nitrogen abundance and relative to oxygen abundances are displayed in Table \ref{tabchemprop}.  Results using the ICF provided by \citet[][not shown]{TorresPeimbert77} were the same within the uncertainties. 
The map in Fig. \ref{mapN} suggests a  mild ($\lesssim$0.20 gradient of $\log$(N/O) crossing the galaxy in the north-east to south-west direction (i.e. \hii-4 with $\log$(N/O)$\sim-1.8$$\rightarrow$\hii-1 with $\log$(N/O)$\sim-1.6$. This is of the order of the estimated uncertainty and thus, compatible with an homogeneity in relative nitrogen-to-oxygen abundance within the galaxy. For the metallicity derived here, GP galaxies have a relative nitrogen-to-oxygen abundance of $-1.6 < \log$(N/O)$\sim-0.9$ \citep{Amorin10}. The value found for \object{UM 462} puts this galaxy just at the lower limit of the typical nitrogen-to-oxygen abundance measured in GP galaxies.

\paragraph{Sulphur, argon and chlorine:}

We could derive also ionic and total abundances in the selected apertures for some additional elements with somewhat more uncertain ICFs. They are all presented here.

Regarding sulphur, the MUSE spectral range offers information to  derive ionic abundances for  \element[+][]{S} and  \element[++][]{S}. We used either $T_e$(\nii) or $T_e$(\siii), depending on whether we wanted to determine single or double ionised sulphur abundance. As with nitrogen,  the sum of these two ionic abundances is lower in the main sites of SF than in the diffuse component of the ionised gas, suggesting the presence of three times ionised sulphur in there.
Total abundances using the ICF provided by \citet{Izotov06}  are also listed in Tab. \ref{tabchemprop}. Values are compatible with a uniform abundance in sulphur. They are systematically higher than those reported by  \citet{James10}, although in some apertures they are compatible within the uncertainties. Differences may be attributed to the different $T_e$ values used in both works.
 
Finally, we could derive ionic abundances for two times ionised argon and chlorine (\element[++][]{Ar} and \element[++][]{Cl}). In both cases, we used  $T_e$(\siii). These are both ions with relatively high  ionisation potentials (27.63~eV and 23.81~eV, respectively), comparable to that of \element[++][]{S} (23.34~eV). As a consequence, higher ionic abundances are seen in the main site of SF, in particular \hii-1, than in the rest of apertures under consideration.  Again, total abundances were estimated by means of the ICF schemes provided by \citet{Izotov06}  and are consistent with uniform abundance over the galaxy.

\subsubsection{Emission line ratio in the BPT diagnostic diagrams \label{secbpt}}

A widely used tool to gain insight into the ionisation mechanisms playing a role in the ISM is the use of diagnostic diagrams, where different areas of a given diagram are
occupied by gas excited via different mechanisms. Probably, the most popular ones are those proposed by \citet{Baldwin81} and later reviewed by \citet{Veilleux87}: the so called BPT diagrams.  In their origin, the different areas in the diagrams were defined using the nuclear or integrated information of a set of astrophysical objects, including \hii\, regions (and starburst galaxies), planetary nebulae, and different types of active galactic nuclei. Then, the different areas in the diagrams were named after the kind of objects that could be found in there, and when information for a new object was obtained, the position in those diagrams was used to identify the main ionisation mechanism at play. 
Nowadays, the same information is routinely derived for portions (in an ample variety of scales) of astrophysical objects, including anything from a planetary nebulae to high redshift galaxies, and the one-to-one association between the position of a given set of data in the diagram and the original label associated with the objects used to define these diagrams is not so straightforward any more. Ionisation mechanisms not considered originally like shocks \citep[e.g.][]{MonrealIbero06,MonrealIbero10b}, AGB stars \citep[e.g.][]{FloresFajardo11,Kehrig12,Zhang17}, or leaking photons from an \hii\, region \citep[e.g.][]{DellaBruna21,Weilbacher18,Papaderos13} produce spectra with line ratios falling in areas in the BPT diagrams that overlap those originally defined. However, as long as one has this in mind, these diagrams are still a useful tool to explore the ionisation structure within, for example, a given galaxy and learn about the causes of the observed line ratios.

 \begin{figure*}[ht]
 \centering
\includegraphics[angle=0,  trim=80 12 60 35, width=0.35\textwidth, clip=,]{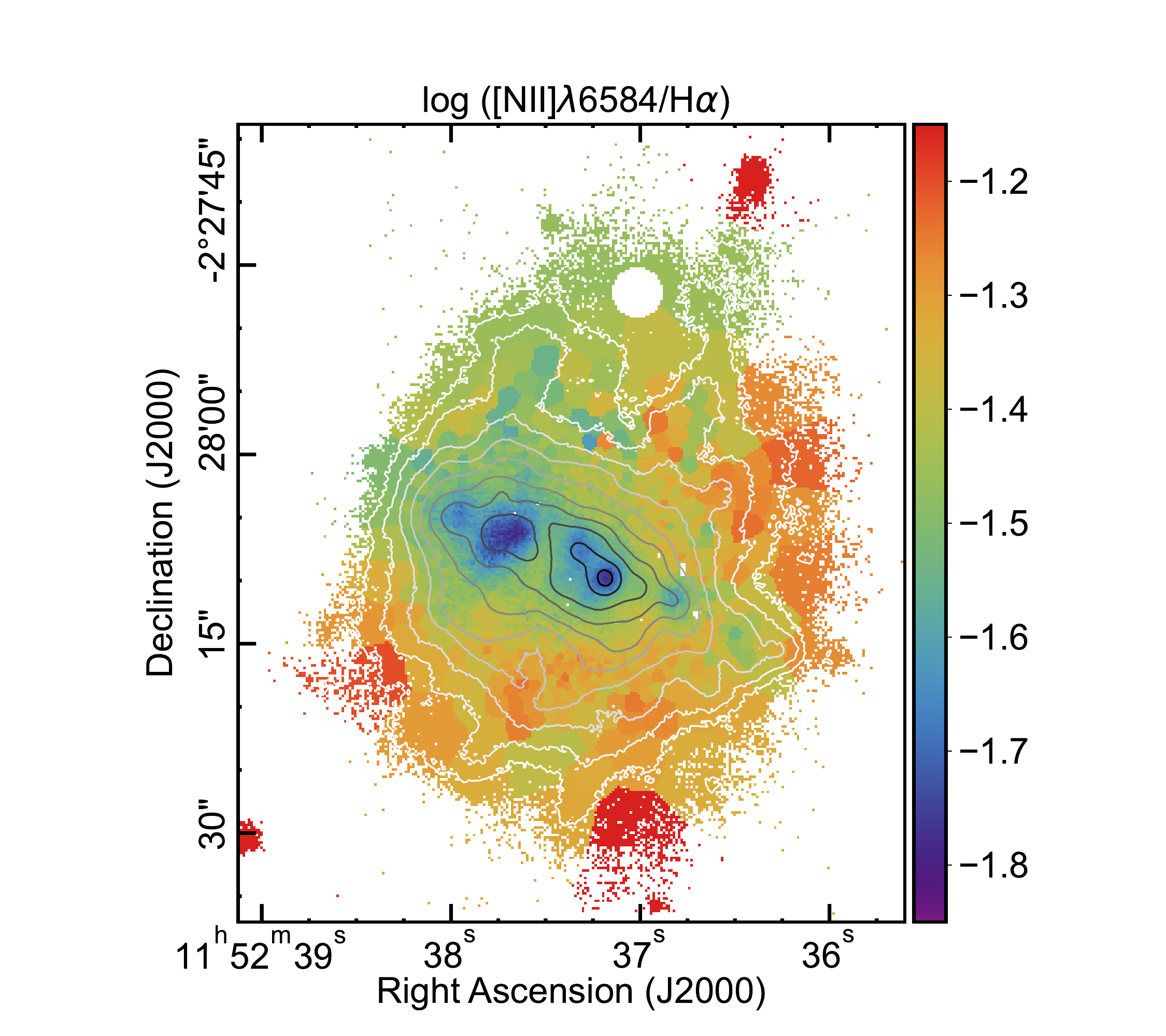}
\includegraphics[angle=0,  trim=80 12 60 35, width=0.35\textwidth, clip=,]{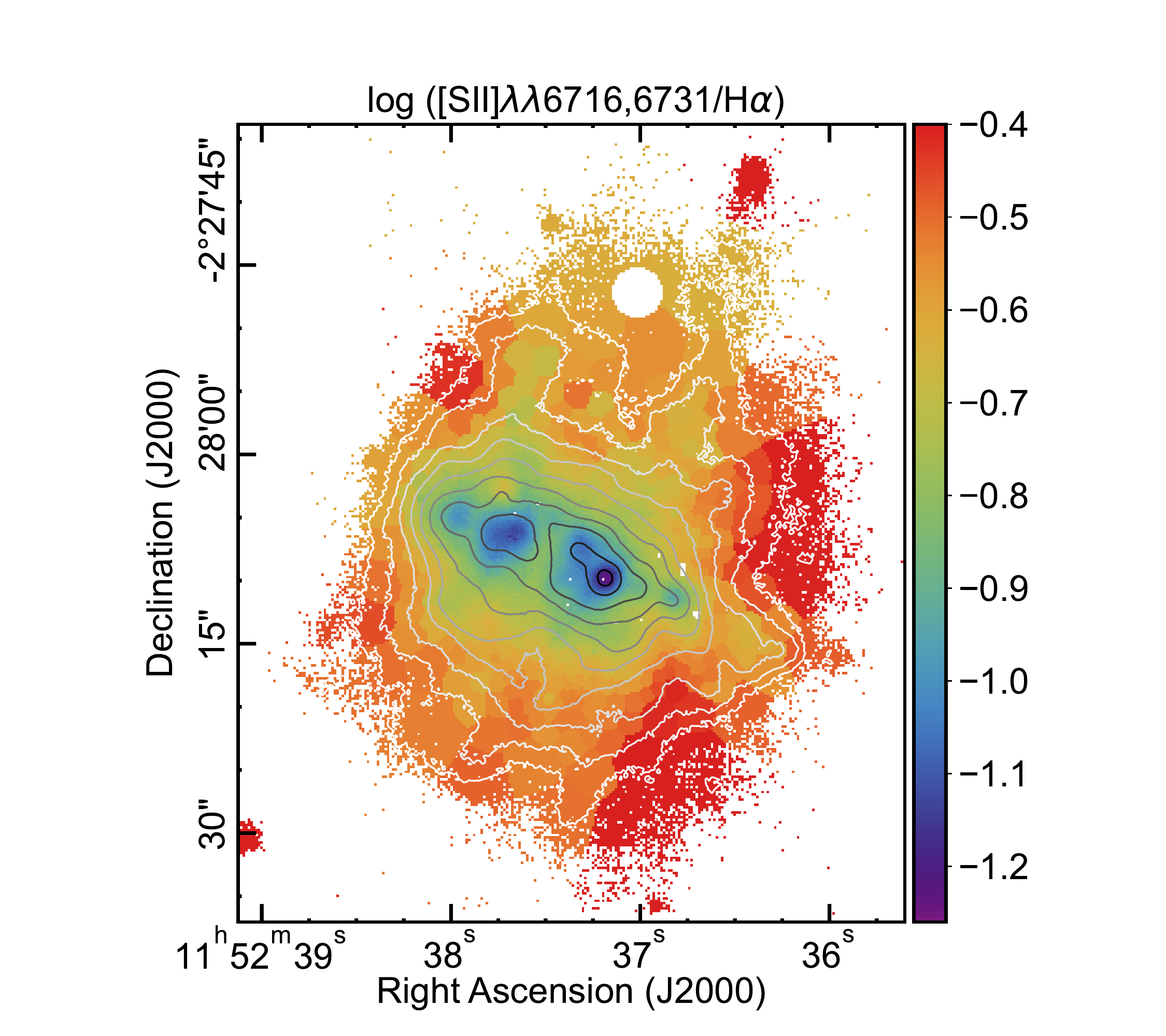}
\includegraphics[angle=0,  trim=80 12 60 35, width=0.35\textwidth, clip=,]{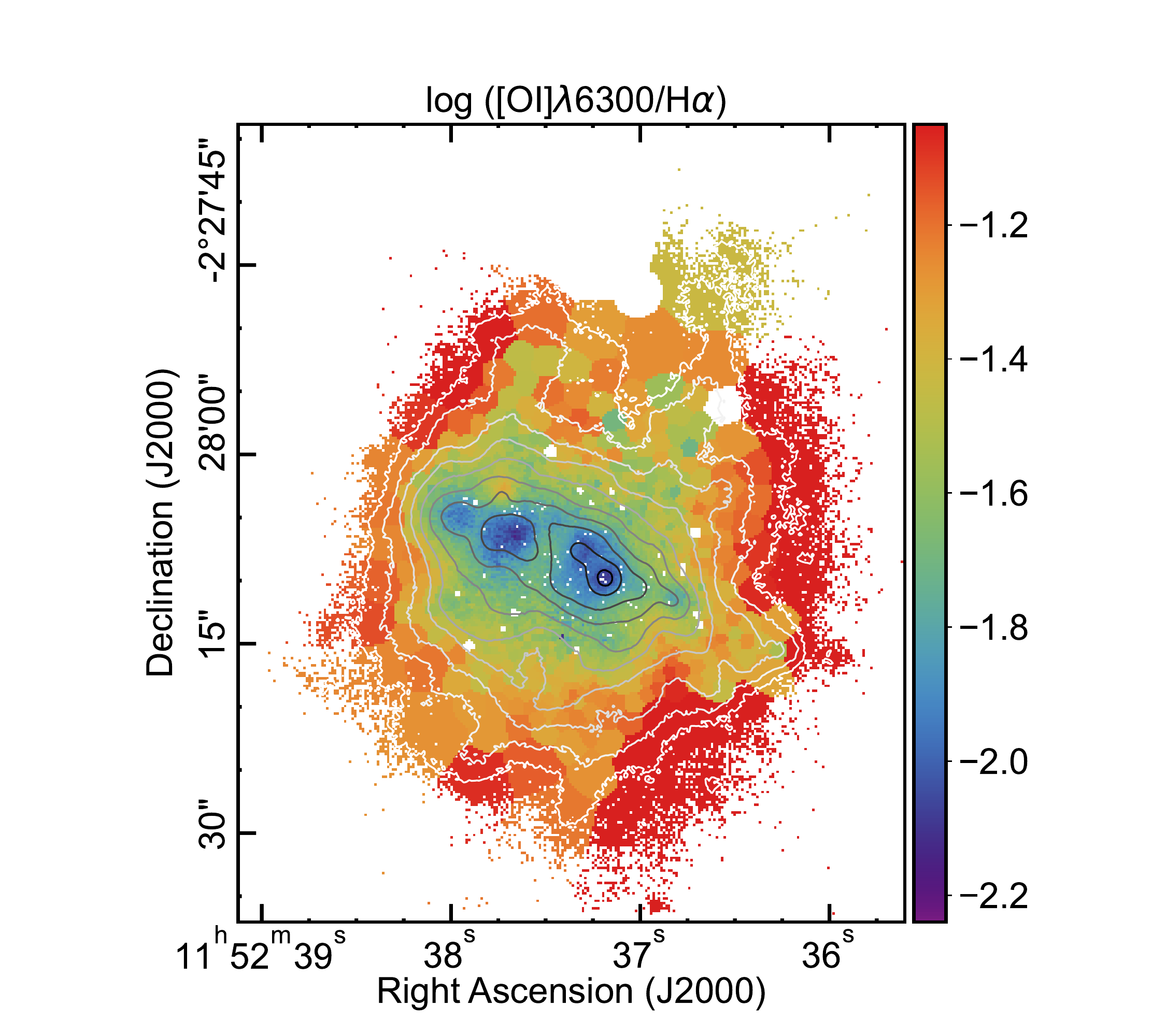}
\includegraphics[angle=0,  trim=80 12 60 35, width=0.35\textwidth, clip=,]{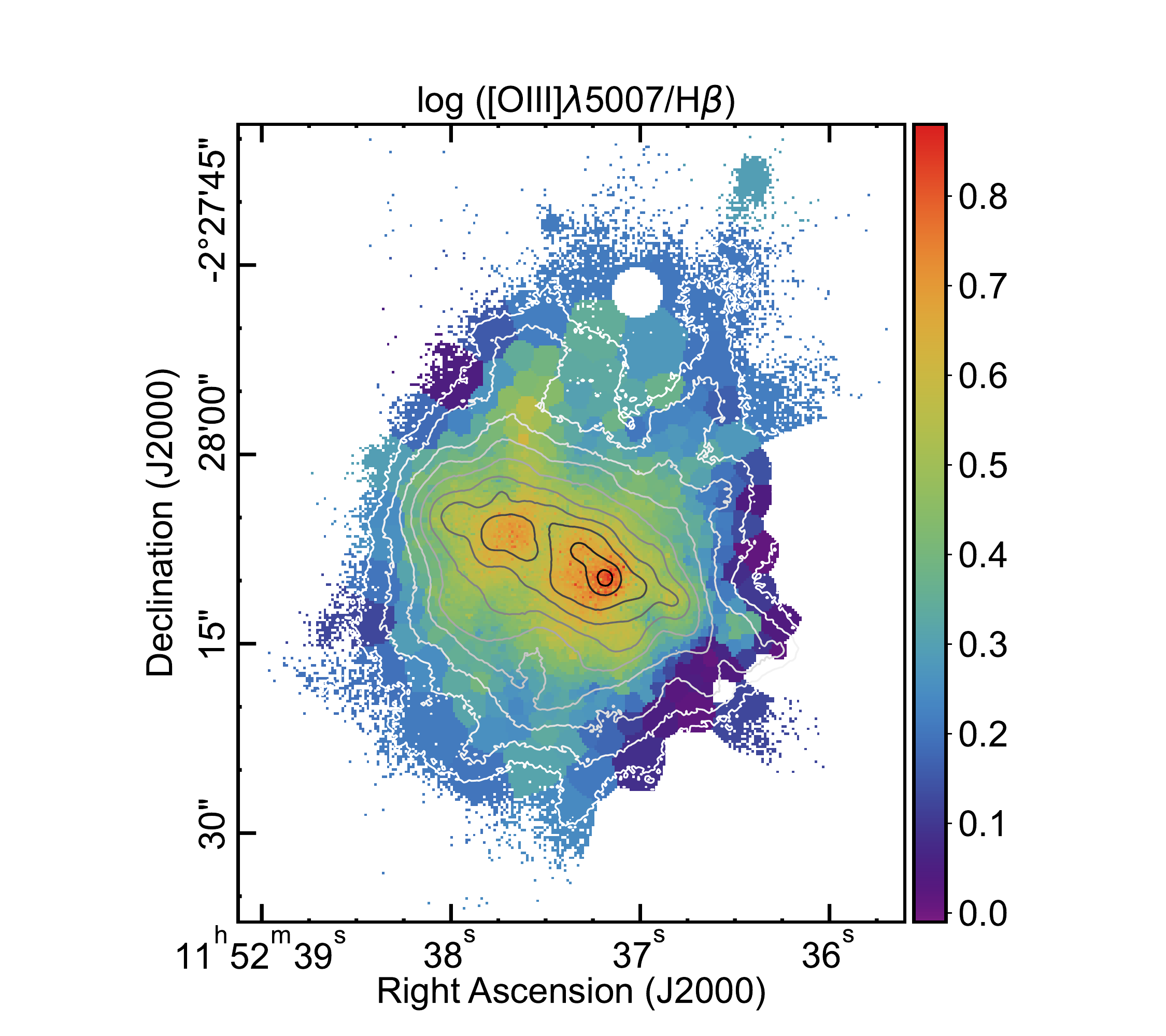}
\caption{Maps for the line ratio involved in the so-called BPT diagrams \citep{Baldwin81}.
  North is up and east towards the left. 
}
 \label{mapBPT}
\end{figure*}

   \begin{figure*}[h]
   \centering
 \includegraphics[angle=0, trim=10 0 0 0, width=0.99\textwidth, angle=0, clip=,]{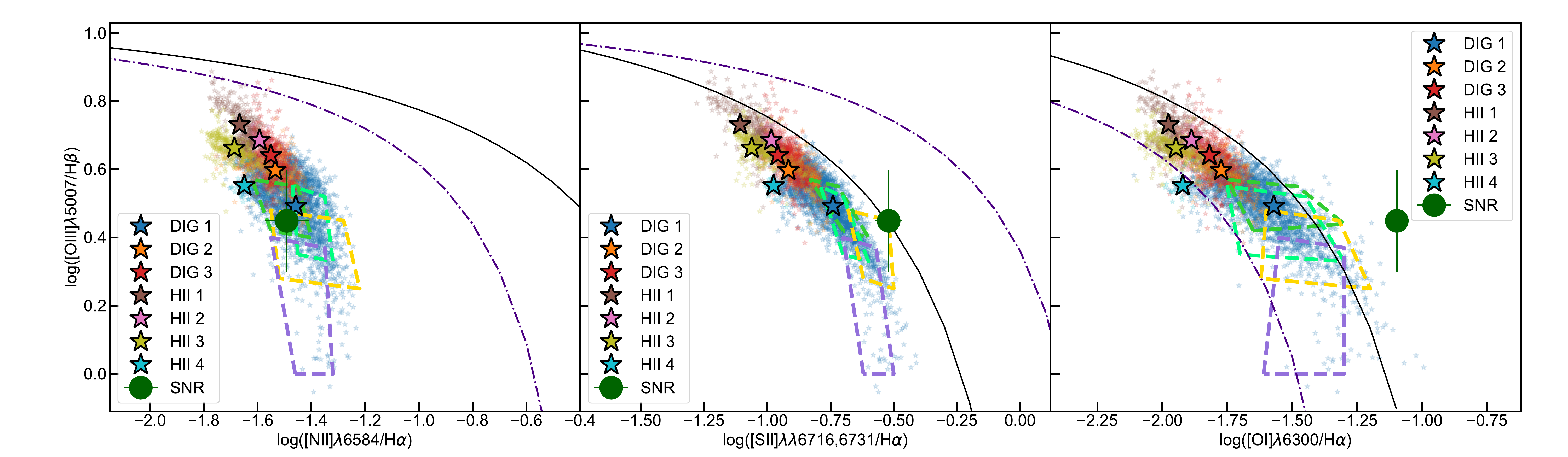}
   \caption{Position of the individual tiles (\emph{small translucent stars}) and the regions defined by \texttt{astrodendro} (\emph{large solid stars}) in the so called BPT diagnostic diagrams \citep{Baldwin81}.  Black continuous lines show the theoretical borders proposed by \citet{Kewley01a}  to delimit the area where the line ratios can be explained by star formation, while indigo dashed-dotted lines mark the maximal starburst prediction for an object with metallicities $Z=0.004$,
   as derived by \citet{Xiao18}.
   The big green circle in each diagram represents the line ratios for the unresolved source at RA(J2000) = 11:52:37.8 and DEC(J2000)=-02:28:03.0, once fluxes have been decontaminated by the diffuse gas emission of the galaxy.
   The four irregular polygons with dashed lines aproximately delimit the locii of line ratios measured in the regions marked in Fig. \ref{astrodendrostruc}.
}
   \label{BPT}
    \end{figure*}

 \begin{figure*}[ht]
 \centering
\includegraphics[angle=0,  trim=80 12 60 35, width=0.35\textwidth, clip=,]{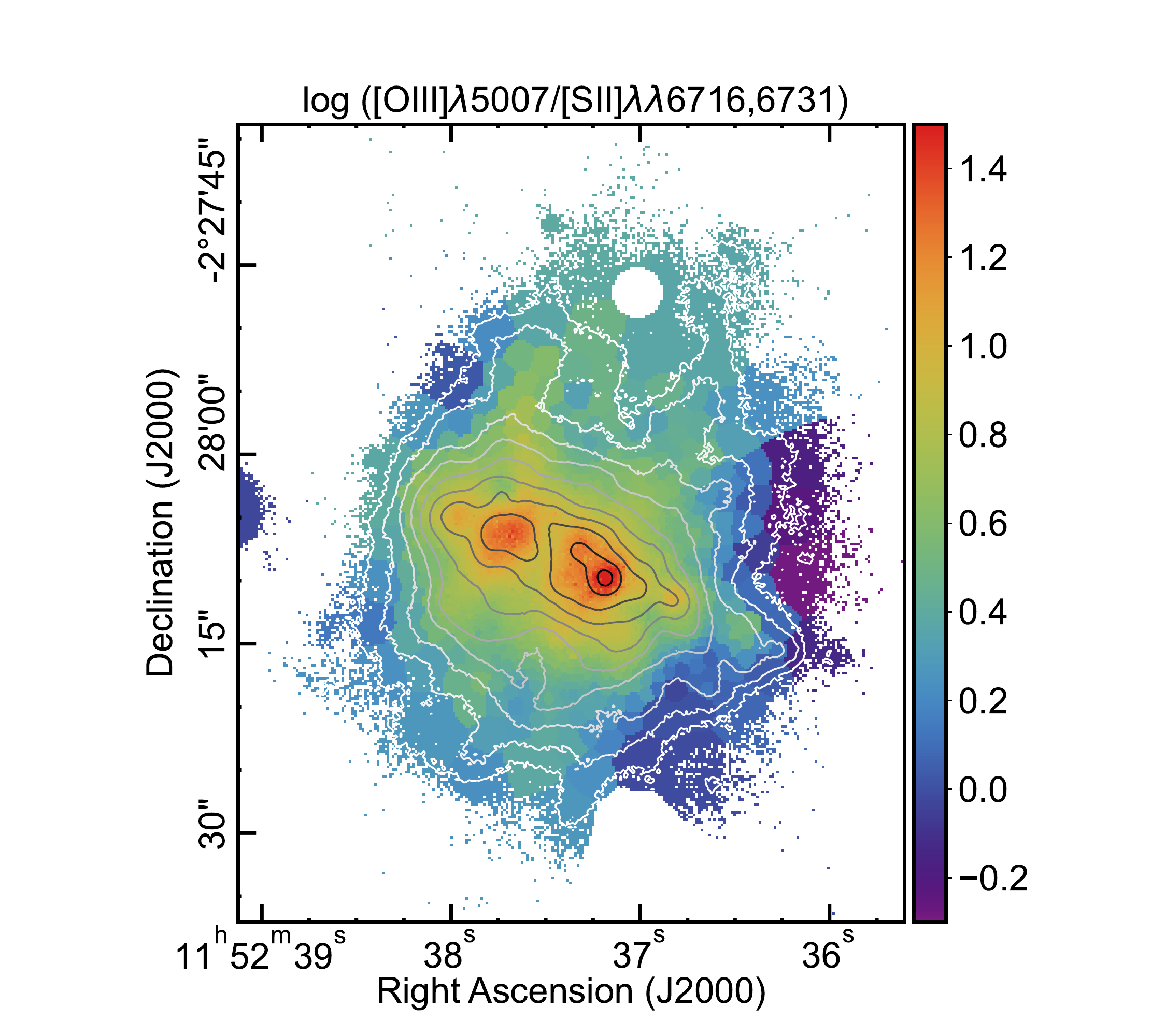}
\includegraphics[angle=0,  trim=80 12 60 35, width=0.35\textwidth, clip=,]{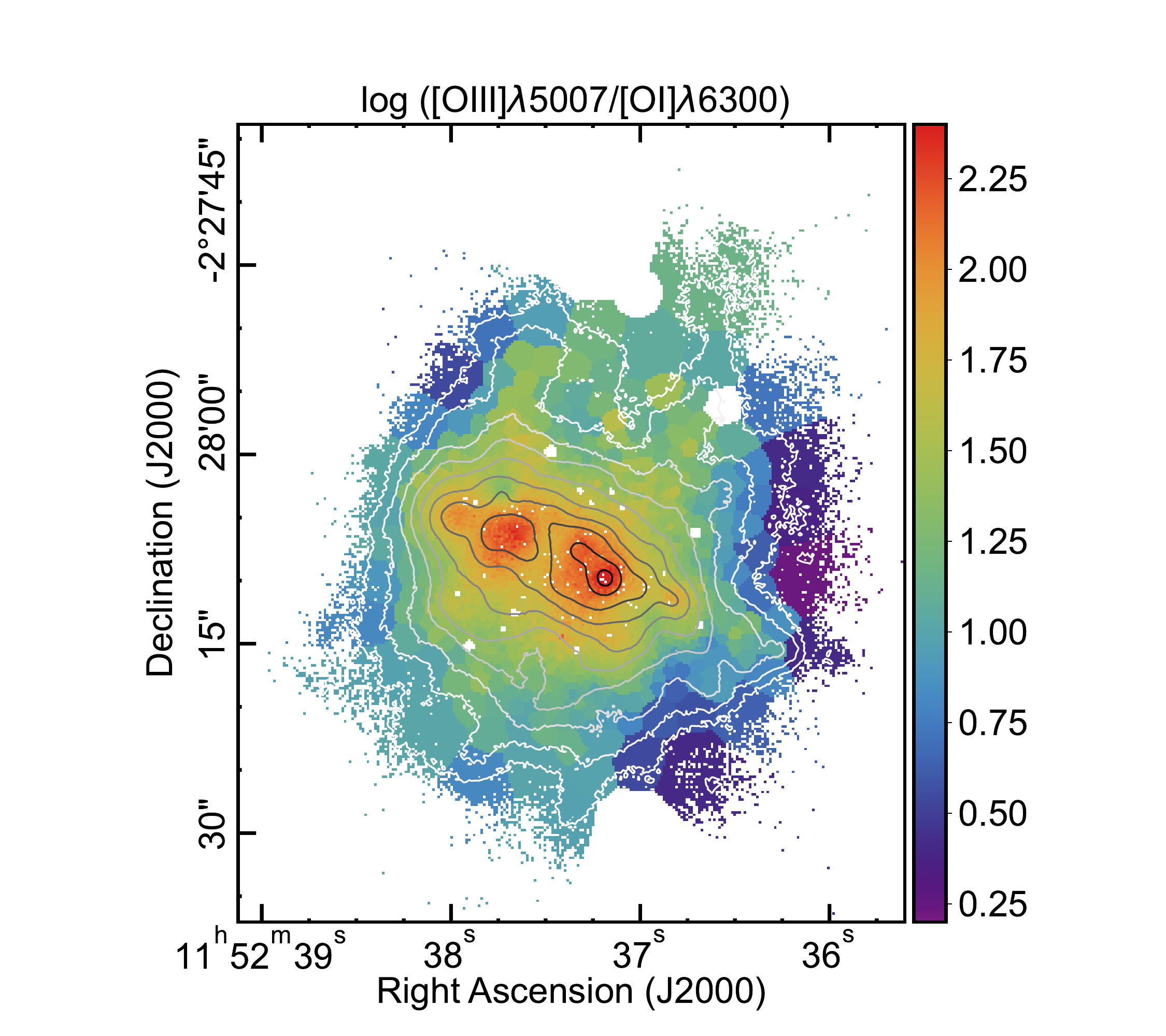}
\includegraphics[angle=0,  trim=80 12 60 35, width=0.35\textwidth, clip=,]{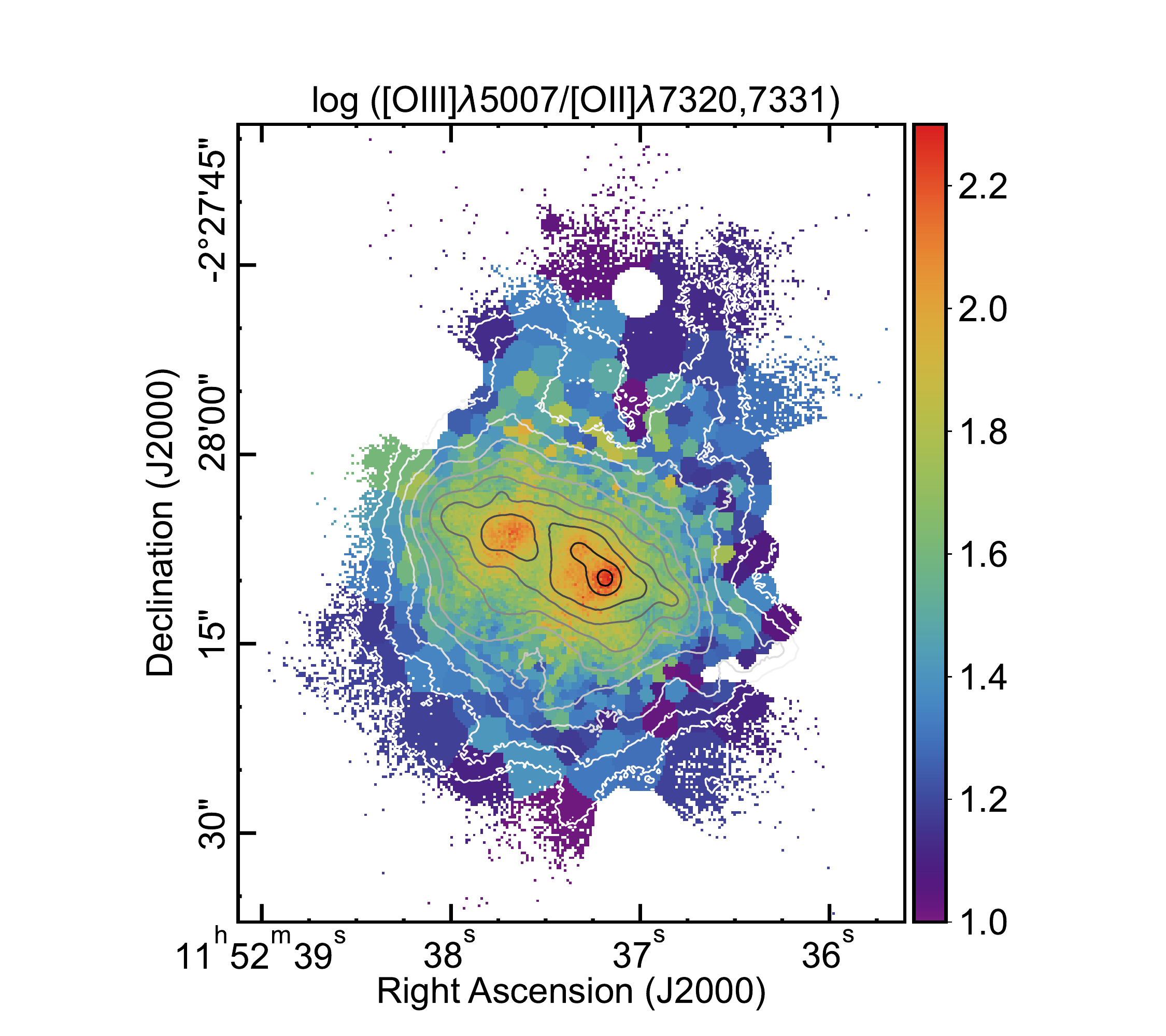}
\includegraphics[angle=0,  trim=80 12 60 35, width=0.35\textwidth, clip=,]{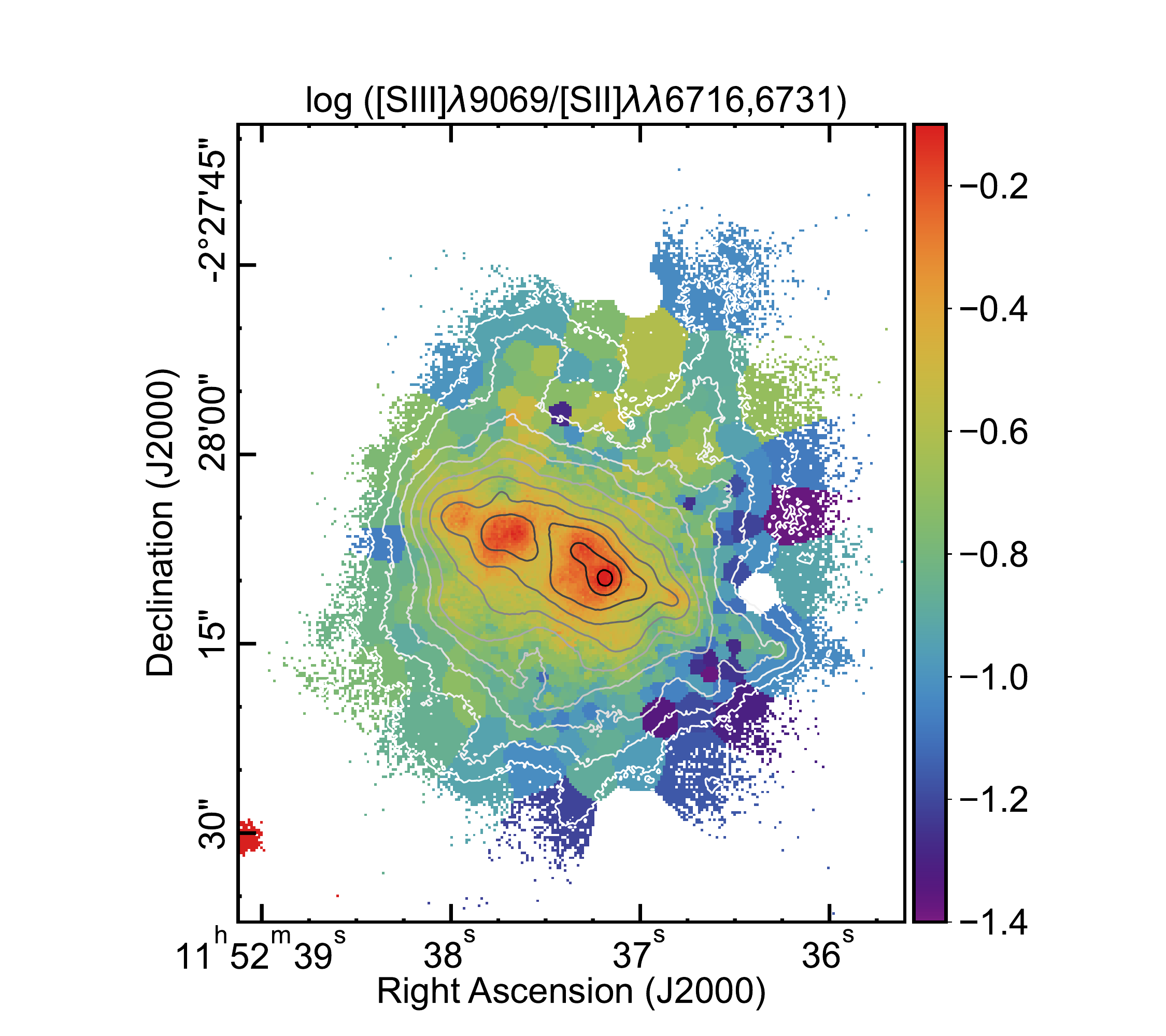}

\caption{Maps for the line ratios tracing the ionisation parameter. \oiii/\sii\, was introduced by \citet{Pellegrini12}, while the other ratios were used in the diagnostic diagrams presented by \citet{Ramambason20}.
  North is up and east towards the left. 
}
 \label{mapU}
\end{figure*}

   \begin{figure}[h]
   \centering
 \includegraphics[angle=0, trim=0 10 50 58, width=0.24\textwidth, angle=0, clip=,]{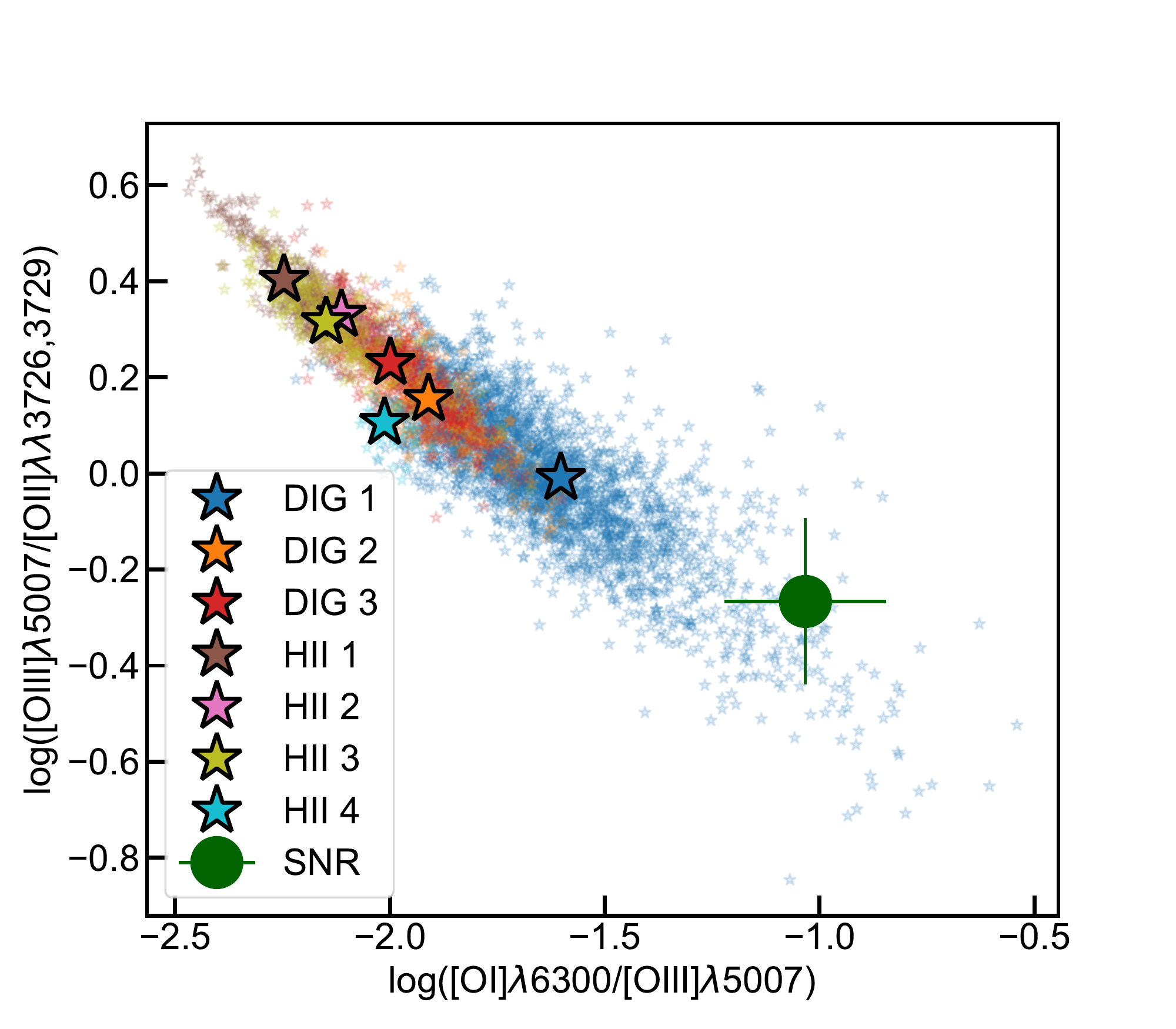}
 \includegraphics[angle=0, trim=0 10 50 58, width=0.24\textwidth, angle=0, clip=,]{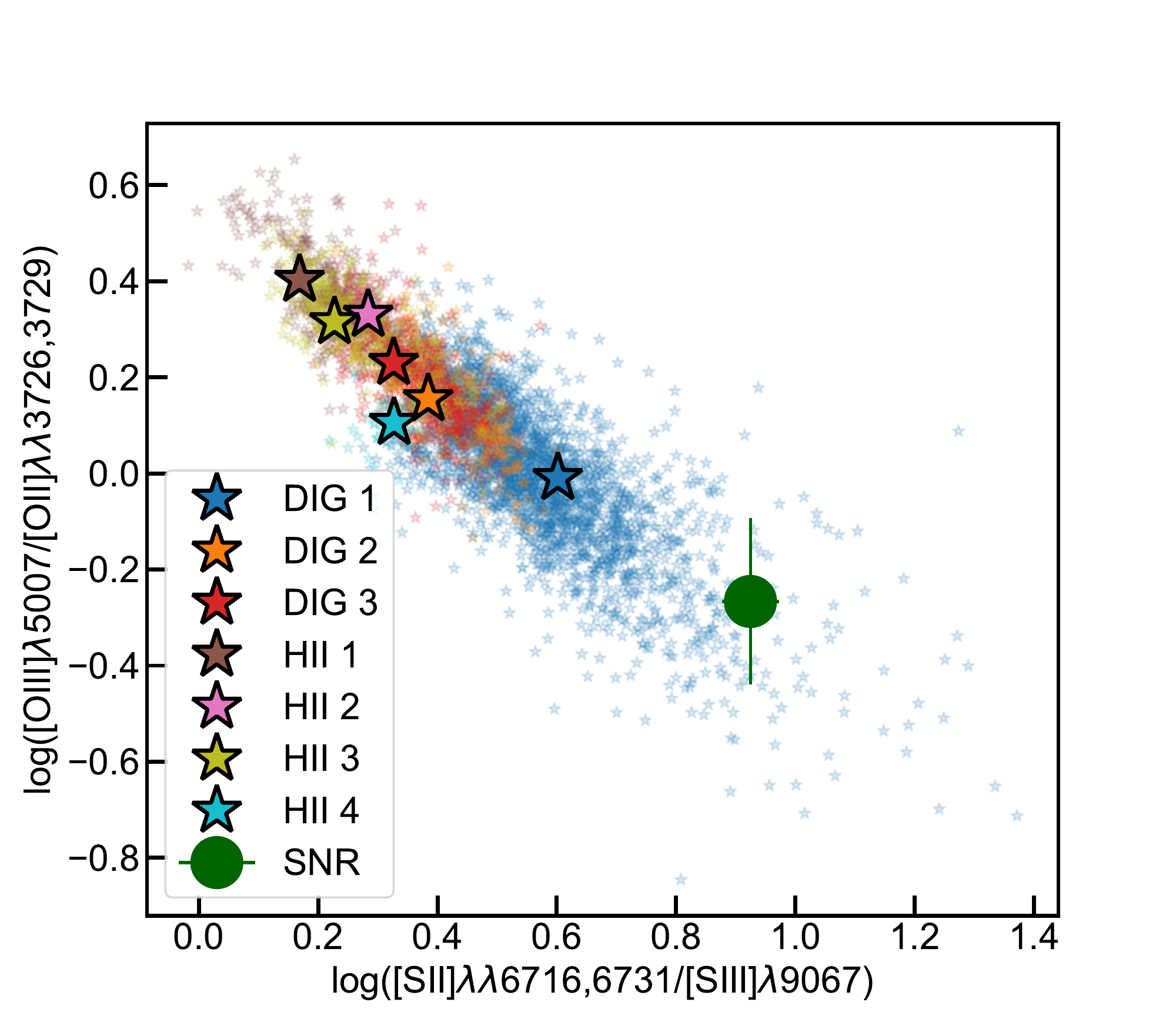}

 \includegraphics[angle=0, trim=0 10 50 58, width=0.24\textwidth, angle=0, clip=,]{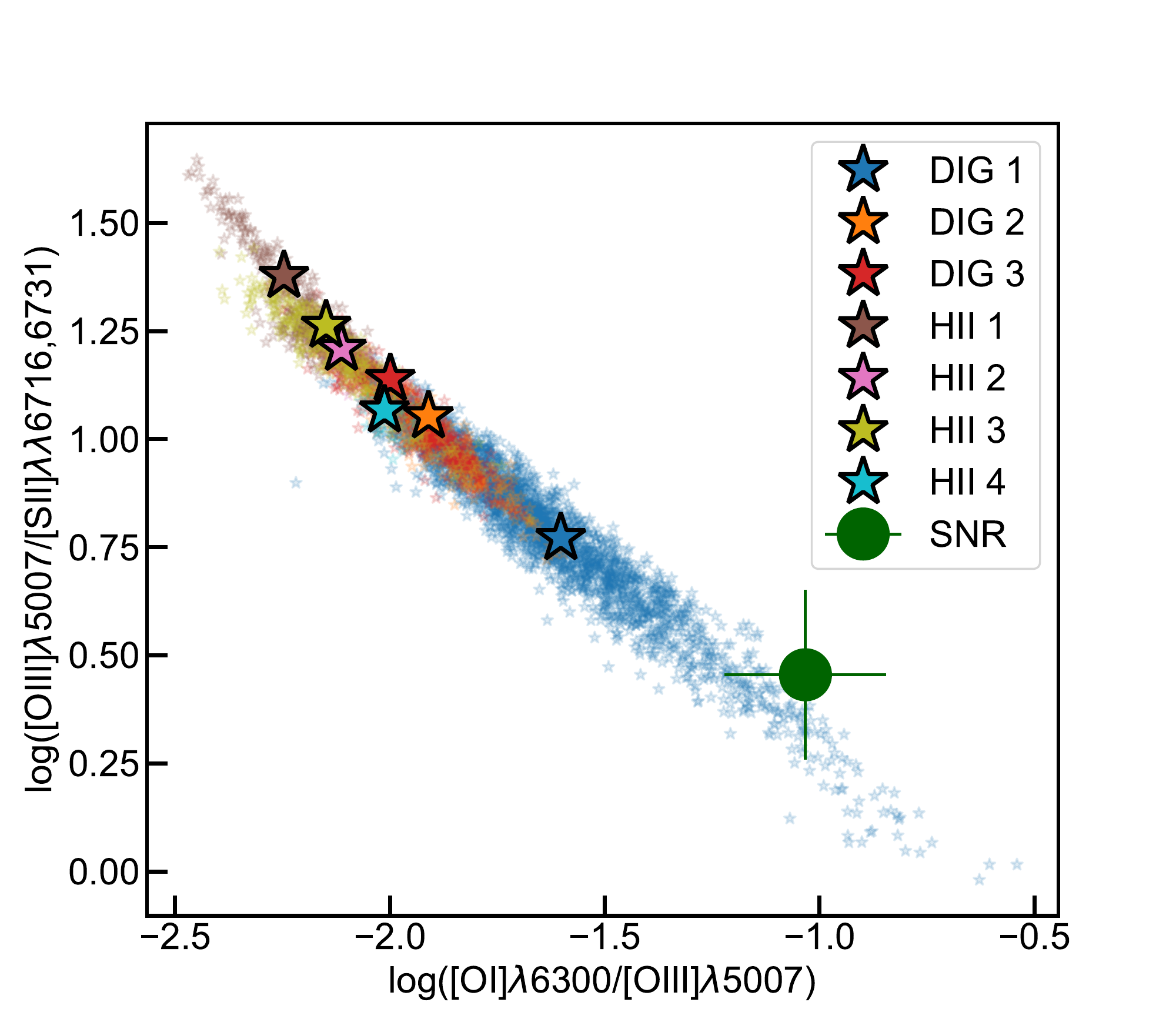}
 \includegraphics[angle=0, trim=0 10 50 58, width=0.24\textwidth, angle=0, clip=,]{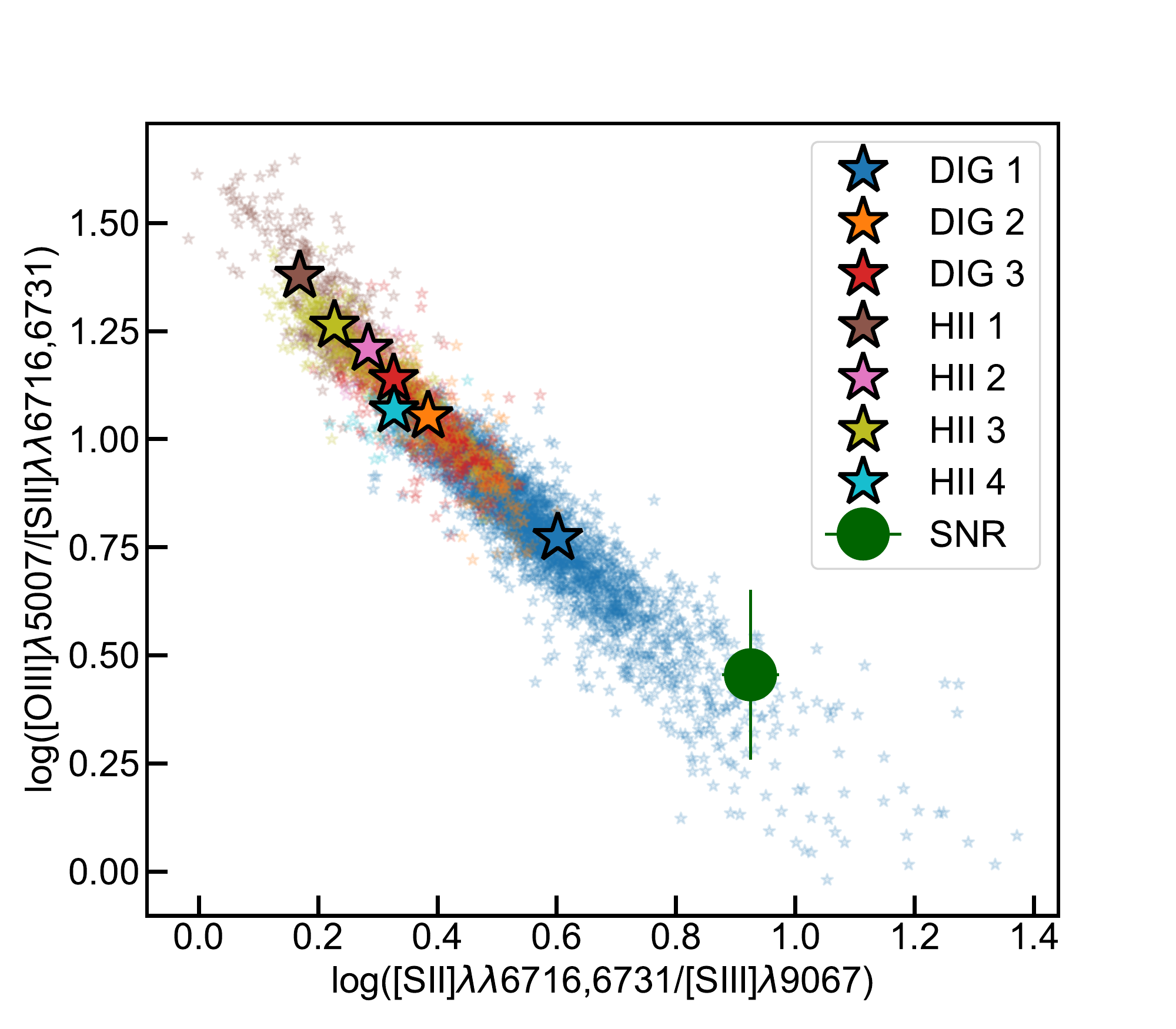}
   \caption {Position of the individual tiles (\emph{small translucent stars}) and the regions defined by \texttt{astrodendro} (\emph{large solid stars}) in some additional diagnostic diagrams.  \emph{Upper row:} The so called O32-O13 and O32-S23 diagnostic diagrams presented by \citet{Ramambason20}.  The \oii$\lambda\lambda$3726,3229 doublet is not covered by MUSE. Instead, we scaled the fluxes for  \oii$\lambda\lambda$7320,7321 by $\times$48. The scaling factor was obtained by estimating the needed \oii$\lambda\lambda$3726,3229 fluxes to obtain the same ionic $O^+/H^+$ abundances as those reported in the apertures defined in Sect. \ref{secselareas}. \emph{Lower row:} Similar to the upper row but the y-axis has been replaced by the \oiii$\lambda$5007/\sii$\lambda\lambda$6761,6731 line ratio. Colour code is the same as in Fig. \ref{BPT}.
}
   \label{diagRamambason20}
    \end{figure}

Maps for the line ratios involved in the three BPT diagrams are presented in Fig. \ref{mapBPT} and the positions of each individual tile in the diagrams themselves are presented in Fig.  \ref{BPT}. Clearly, the ionisation structure is quite complex and the joint information of the maps and the diagrams allow us to identify some well defined substructures within the system.  

The \nii/\ha\footnote{We use the following nomenclature: \nii/\ha\, for \nii$\lambda$6584/\ha,  \sii/\ha\, for  \sii$\lambda\lambda$6716,6731/\ha, \oi/\ha\, for  \oi$\lambda$6300/\ha,  \oiii/\hb\,  for \oiii$\lambda$5007/\hb. When other lines are used, these will be explicitly mentioned.} map displays the most regular distribution of the four. Ratios are the lowest in the main sites of SF, in particular \hii-1 and \hii-3, and  increase in a relatively smooth and unstructured manner when going towards the external parts of the galaxy.
Overall, \sii/\ha\, and \oi/\ha\, present a similar distribution as \nii/\ha\, (i.e lowest values in \hii-1 and \hii-3, then increasing outwards). However,  both maps are more structured and more closely following the \ha\, flux maps (contours in all the presented maps). On first order, \oiii/\hb\, has a reversed distribution  (i.e a structure following the  \ha\, flux maps but with higher values in \hii-1 and \hii-3, then decreasing outwards). There are further subtle differences between these maps and the one for \nii/\ha. First, there are two horn-like structures in the northern part of the galaxy, hereafter called 'the horns'. Their borders are marked in Fig. \ref{astrodendrostruc} with two green dashed polygons. The eastern one roots at \hii-3 and while the western at  \hii-1. Both extend towards the north for $\sim$900~pc. Their typical ratios in the BPT diagrams are represented with green polygons in Fig. \ref{BPT}.  
Excluding the main sites of SF, they are the locations with the highest \oiii/\hb\, line ratios, with somewhat larger ratios in the eastern horn. Both present moderate and basically indistinguishable  \sii/\ha\, and \oi/\ha\, line ratios but eastern one has lower  \nii/\ha\, ratios by $\sim$0.15~dex. 

The southern half of the galaxy presents at least one  structure in \ha, reminiscent of the northern horns (and actually brighter). It roots at the complex \hii-1/\hii-2/DIG-3, extending towards the south first, then bending towards the east. However, all the southern half is much less structured in terms of line ratios. These simply increase (or decrease for \oiii/\hb) with distance from the main site of SF.  

There is an additional resolved structure that stands out in the maps: an ear-like region towards the east, hereafter called 'the ear', marked in Fig. \ref{astrodendrostruc} with a purple dashed polygon. It has relatively high \sii/\ha\, and \oi/\ha\, even if not the highest, and covers a relatively large range in \oiii/\hb, including the lowest values measured within the galaxy. As we shall see in Sect. \ref{seckine}, it is the region with the bluest velocities in \ha\, and \oiii$\lambda$5007.

Finally, at coordinates RA(J2000) = 11:52:37.8 and DEC(J2000)=-02:28:03.0, there is a local maximum in the \sii/\ha\, and \oi/\ha\, maps with line ratios barely distinguishable from their surroundings in the  \oiii/\hb\, line map, and completely undistinguishable in the  \nii/\ha\, map. This location is unresolved at our seeing, and it does not seem clearly associated with any knot of SF, according to the our continuum images, or the broad band images provided by \citet{Micheva13}, at higher spatial resolution.
We discuss the nature of this source in Sect.  \ref{secsnr}.

When looking at the position of each tile in the BPT diagram, the data points mostly fall below the theoretical borders proposed by \citet{Kewley01a}, even if in the diagram involving the \oi/\ha, data points are just at the border. This would suggest that this plasma is primarily photoionised by massive stars. However, this maximum theoretical borders were defined based on models covering up to much larger metallicites than the one of \object{UM\,462}. Theoretical borders based only on models at the metallicity, $Z=0.004$ (dotted-dashed lines in Fig. \ref{BPT}), closer to that of \object{UM\,462} would be more suitable. These are marked in the diagrams with indigo lines. %
They were derived using models take into account the impact of the binary star population, that  produces more Wolf-Rayet and hot stars at older ages \citep{Xiao18}.
While data points are clearly consistent with photoionisation by massive stars in the \nii/\ha\, and \sii/\ha\, diagrams, it seems that the   \oi/\ha\, diagram requires an additional mechanism (e.g. shocks) or specific configuration of the relative distribution of gas and stars is needed (e.g. a patchy distribution of the ISM  as  in the picked-fence scenario \citep[][]{Heckman01,Bergvall06,Ramambason20}. Either way, this is  more important in DIG-1 than in the main sites of SF, specially in the part between the horns which, as we shall see in Sect. \ref{seckine}, presents high velocity dispersion and different velocities depending on the ionic species.

\subsubsection{Emission line ratios mapping the ionisation structure \label{secionstruc}}

Beyond the line ratios involved in the BPT diagrams, some additional diagnostics have been proposed in the literature as proves of the ionisation structure. For example, \citet{Pellegrini12} used maps of the \oiii$\lambda$5007/\sii$\lambda\lambda$6716,6731 line ratio to assess the optical depth of the ionising radiation in individual \hii\, regions. In a way, this is a compact manner of jointly looking at the two axes of the \oiii/\hb\, versus \sii/\ha\, diagram. Besides, to avoid (or at least minimise) dependencies on elemental abundances, one could use ratios involving lines coming from different ionisation stages of a given element \citep[e.g.][]{Izotov16,Weilbacher18,Micheva19}. Three of those ratios are available within the MUSE spectral range,  namely: \siii$\lambda$9069/\sii$\lambda\lambda$6716,6731, \oiii$\lambda$5007/\oi$\lambda$6300, and \oiii$\lambda$5007/\oii$\lambda\lambda$7320,7331. The first two have been already used to explore the possibility of ionising  photons escaping from the galaxy \citep[e.g.][]{Ramambason20}. Regarding a ratio involving \oiii\, and \oii\, emission lines, the most commonly used is the one based on the much brighter  \oii$\lambda\lambda$3726,3728 doublet (unavailable here). Nonetheless, we were able to derive a line ratio map with the \oii$\lambda\lambda$7320,7331 in a reasonable portion of the galaxy.
The extinction corrected maps for all the four observed ratios are presented in Fig. \ref{mapU}. All the four maps basically replicate the structure described for the \oiii/\hb.

The upper row in Fig. \ref{diagRamambason20} presents the relations between those ratios involving ions of the same element. 
The \oii$\lambda\lambda$7320,7331 doublet is a relatively faint feature, detectable here because of the superb MUSE sensitivity.  As mentioned above, these lines are not typically used in the literature. Thus, to make the comparison with other works easier, 
we did not use the direct  \oiii$\lambda$5007/\oii$\lambda\lambda$7320,7331 ratio. Instead, we estimated  the flux in the  \oii$\lambda\lambda$3726,3728 doublet by scaling  the \oii$\lambda\lambda$7320,7331 with a factor of $\times48$. This factor was determined by calculating the ionic abundance for $O^+$ from \oii$\lambda\lambda$3726,3728 for varying fluxes that were compared iteratively  with abundances from \oii$\lambda\lambda$7320,7331 until convergence. 
Because this ratio contains a derived quantity relying on two faint lines and an assumption in the relation with the \oii$\lambda\lambda$3726,3728 lines, we also included in the lower row of Fig. \ref{diagRamambason20} equivalent diagrams, this time involving the \oiii$\lambda$5007/\sii$\lambda\lambda$6716,6731. This way,  we use a ratio that can be directly derived from strong measured lines, in this case at price of mixing species.

The trend in all the four diagrams is similar. Data points are organised along a (more or less) tight sequence and ordered according to the ionisation degree, with \hii-1, being the highest end, and the ear being the lowest end.
The \oiii$\lambda$5007/\oii$\lambda\lambda$3726,3728 ratio reaches values of up to $\sim$4.0. These large ratios are comparable to those measured in GP galaxies, Lyman break galaxies at $z\sim2-3$ and Lyman continuum leakers and $\sim1-2$ orders of magnitude larger than typical star-forming galaxies \citep{Nakajima14}. 
A galaxy with such large  \oiii$\lambda$5007/\oii$\lambda\lambda$3726,3728 is a good candidate to have a high escape fraction, $f_{esc}$, even if not all the galaxies with these ratios are necessarily LyC leakers.
The ratio for the integrated spectrum of DIG-1 is comparable to those measured in the individual tiles for the horns, and still relatively high (\oiii$\lambda$5007/\oii$\lambda\lambda$3726,3728$\sim$1), with the area between the horns and the ear having \oiii$\lambda$5007/\oii$\lambda\lambda$3726,3728 of $\sim$0.50 and $\sim$0.15, respectively.

   \begin{figure*}[th]
   \centering
\includegraphics[angle=0,trim=0 0 0 0, width=0.90\textwidth, clip=,]{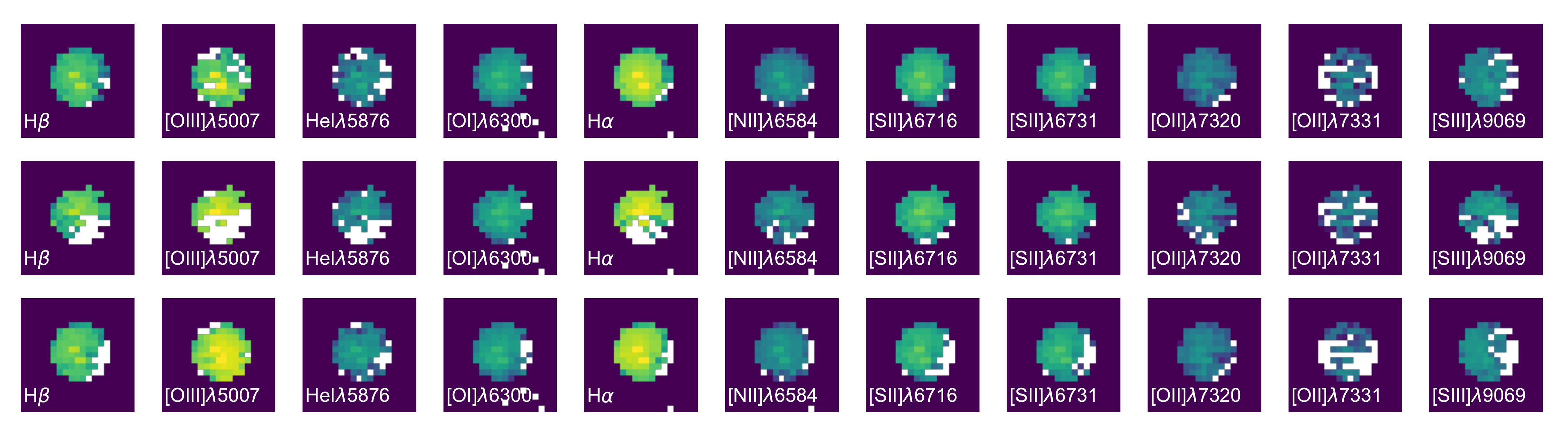}
   \caption{Stamps with the line emission mapping recovered for the point source, by means of the three interpolation methods (from top to bottom: linear, nearest, cubic).
The intensity stretch is in logarithmic scale, covers four orders of magnitude and it is the same in every stamp.
   Spaxels in white have negative fluxes and are artefacts created by the interpolation. 
  }
   \label{figsnrmaps}
    \end{figure*}

   \begin{figure*}[th]
   \centering
\includegraphics[angle=0,trim=05 0 0 0, width=0.19\textwidth, clip=,]{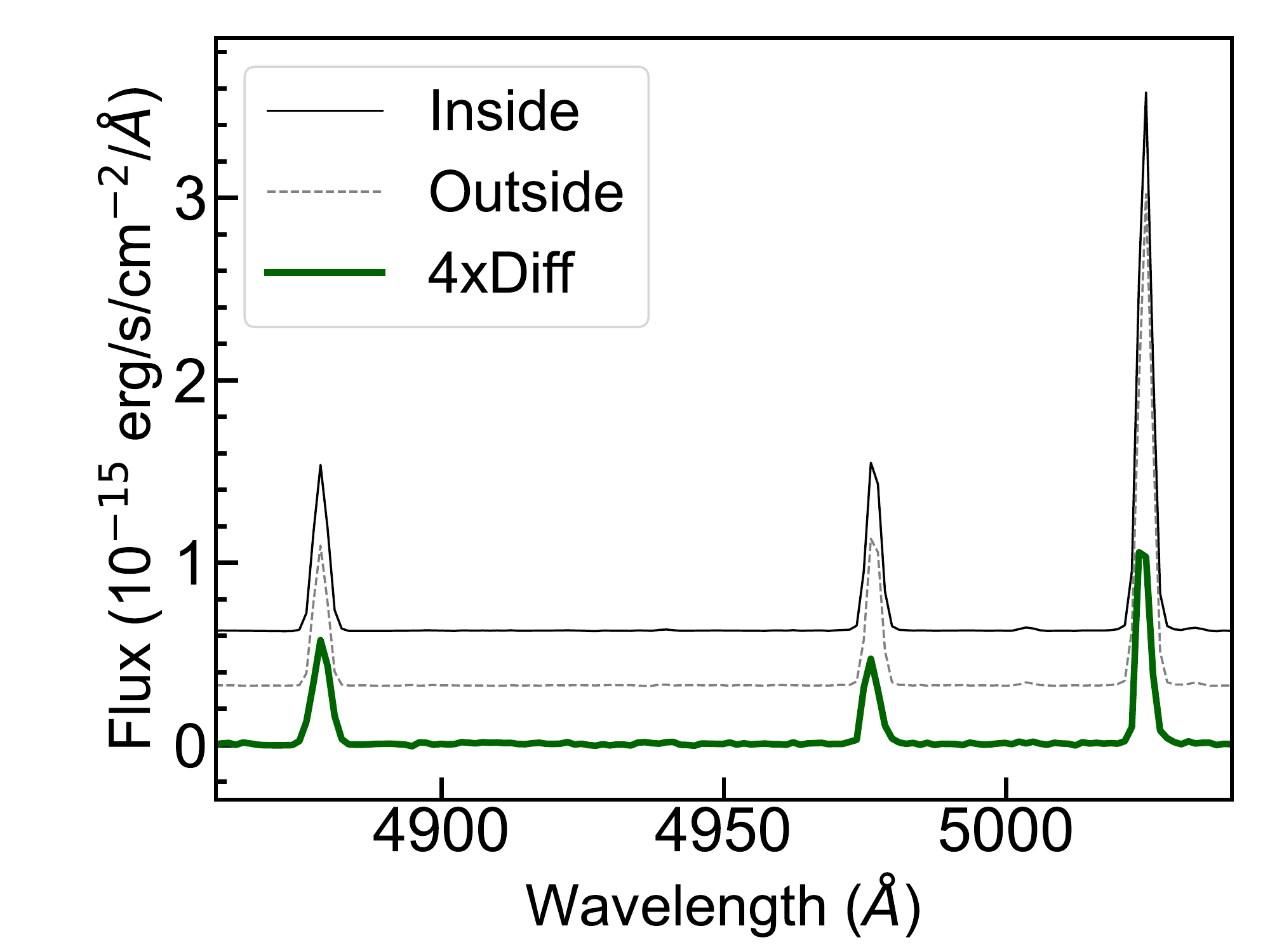}
\includegraphics[angle=0,trim=05 0 0 0, width=0.19\textwidth, clip=,]{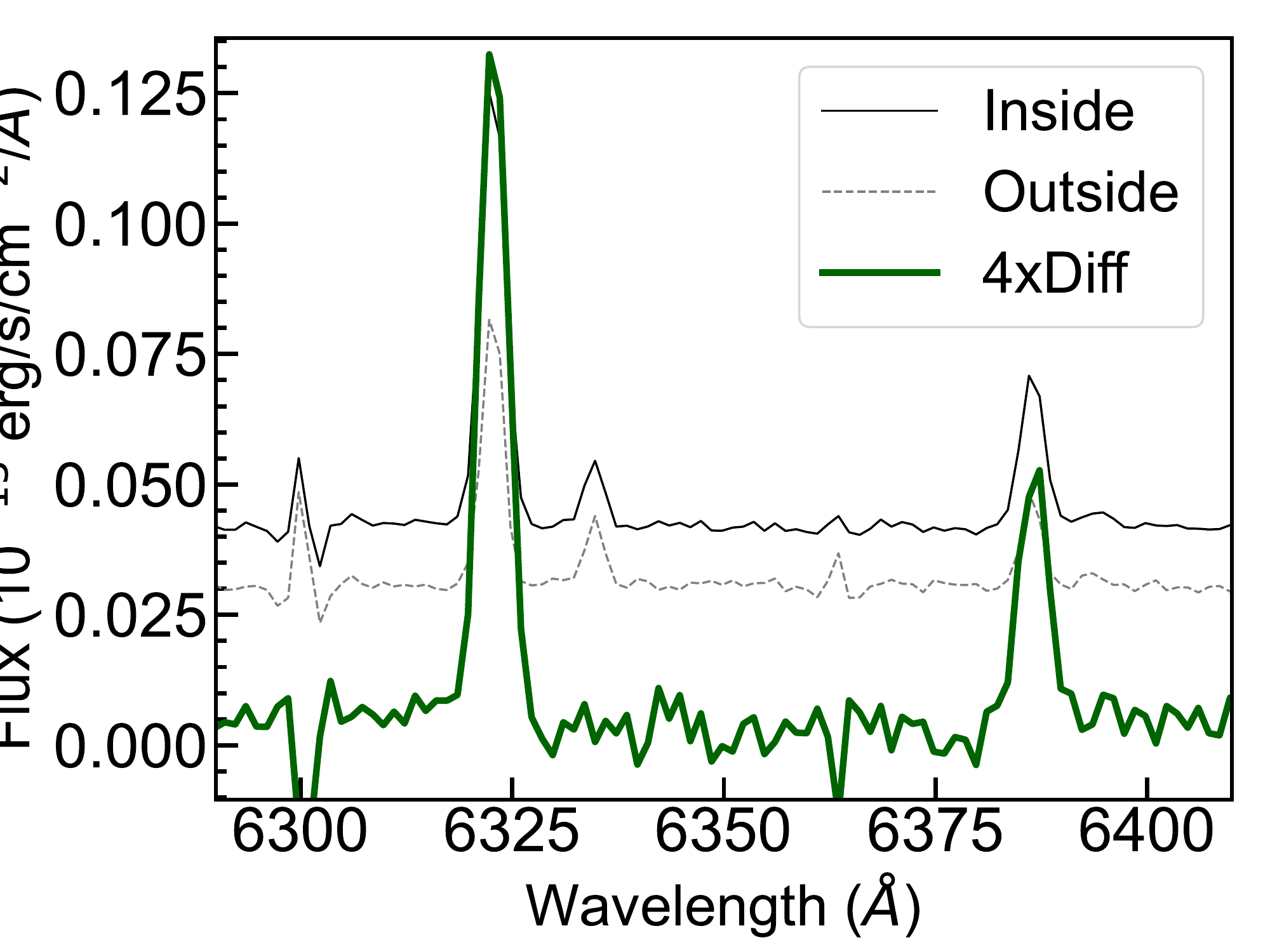}
\includegraphics[angle=0,trim=05 0 0 0, width=0.19\textwidth, clip=,]{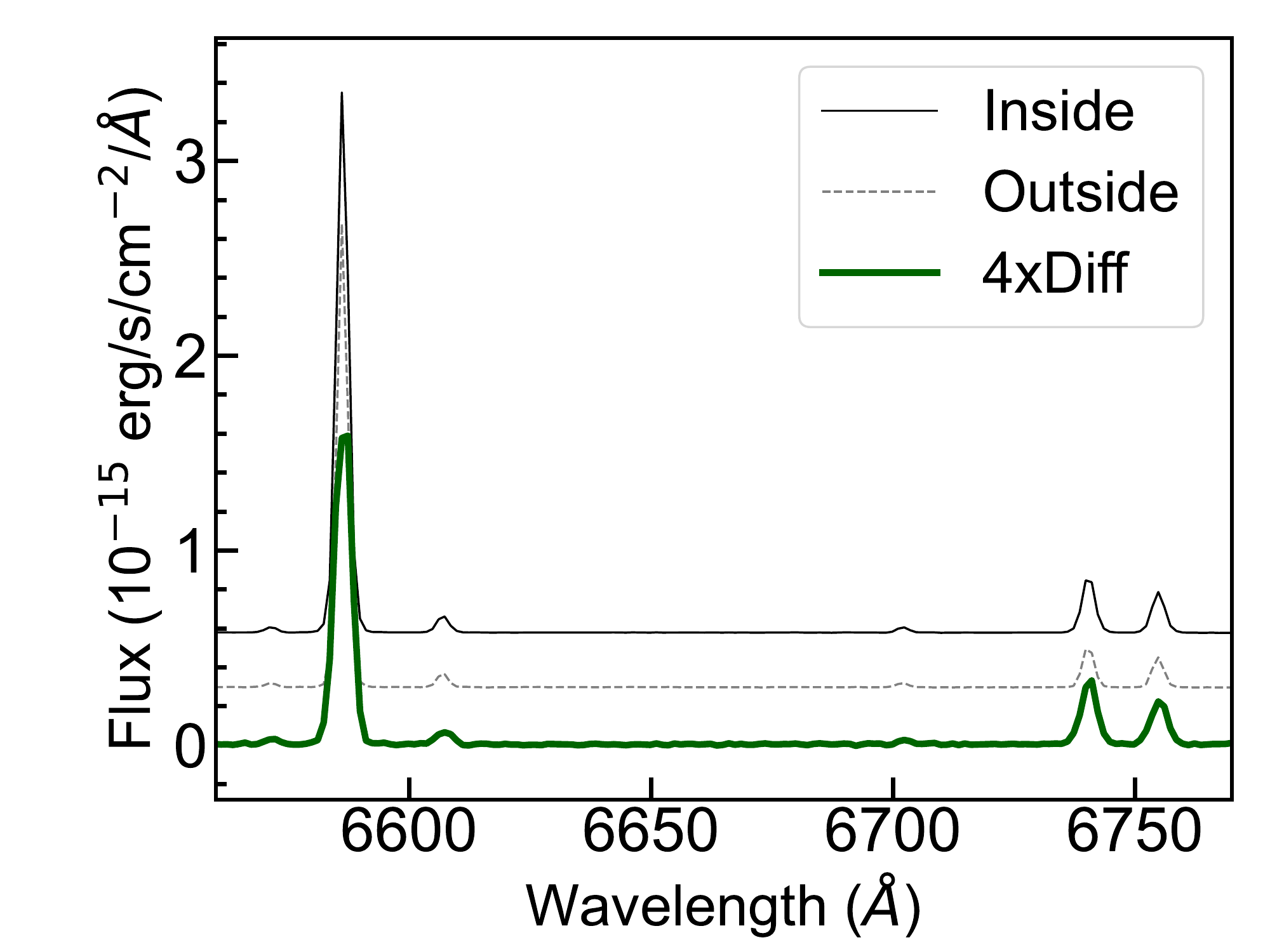}
\includegraphics[angle=0,trim=05 0 0 0, width=0.19\textwidth, clip=,]{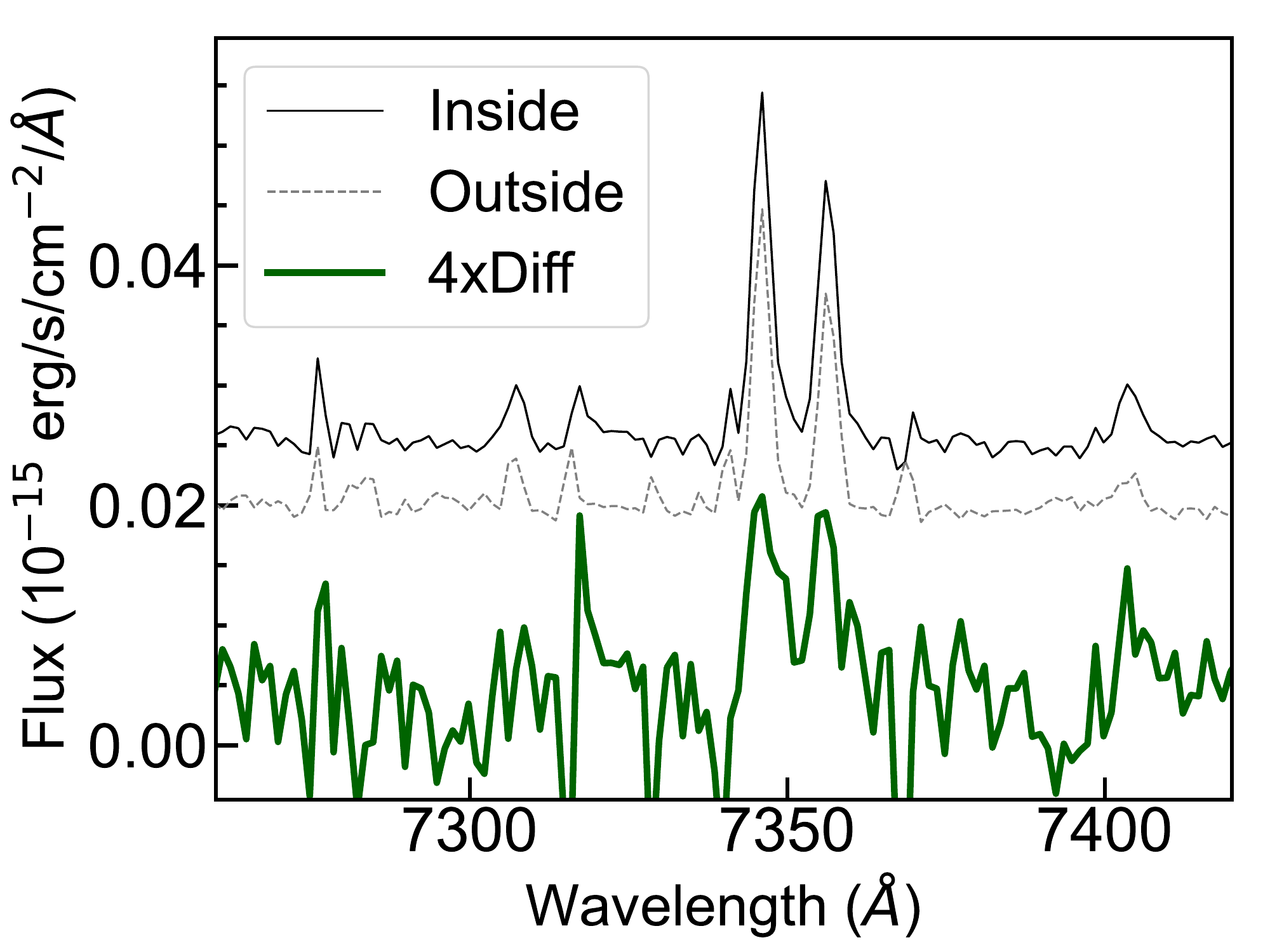}
\includegraphics[angle=0,trim=05 0 0 0, width=0.19\textwidth, clip=,]{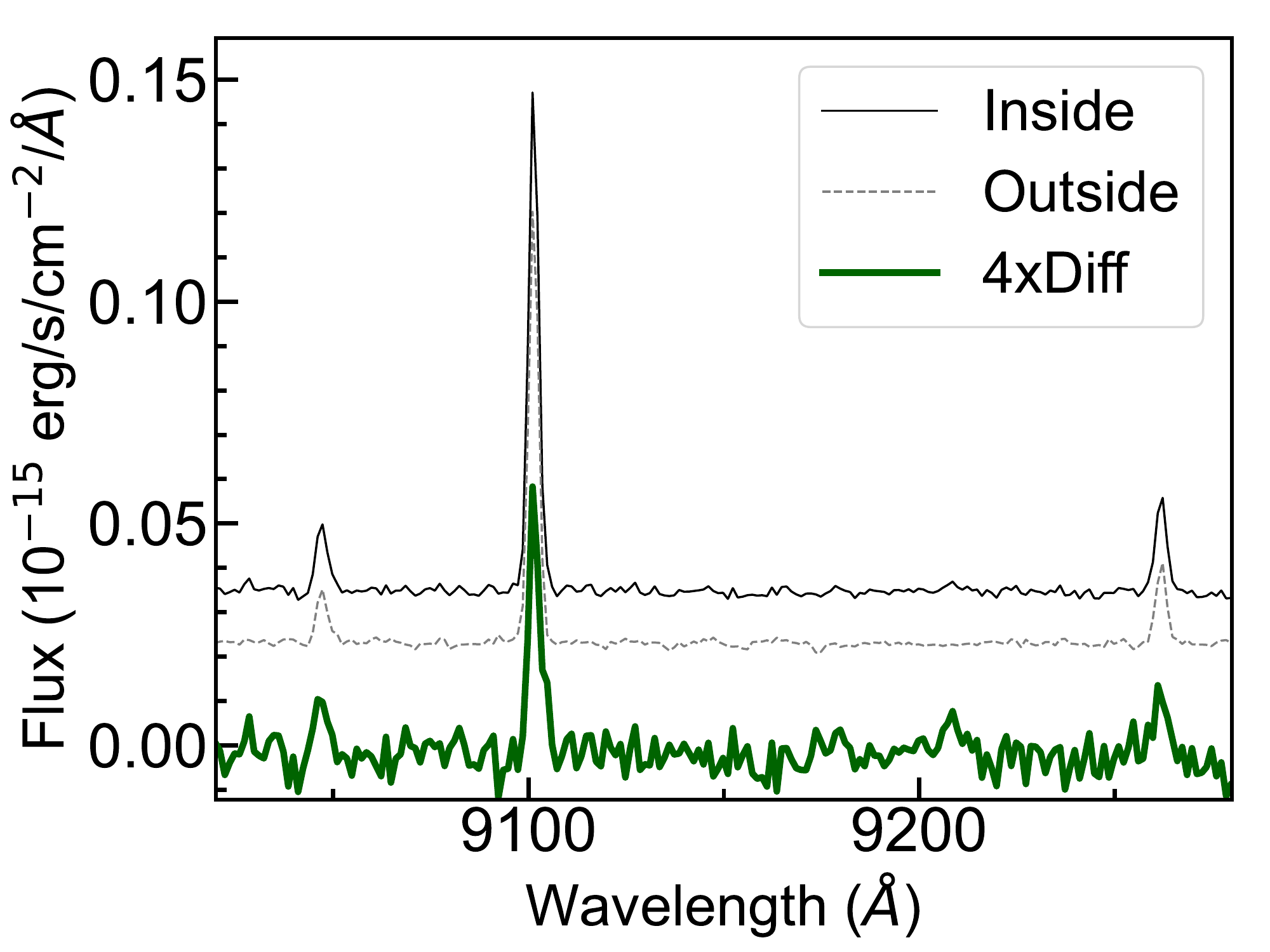}
     \caption{Selected windows for the spectrum of the SNR candidate. The recovered spectrum is displayed with a dark green thick continuum line.  Besides, the total spectrum at the location point source under study ('Inside', \emph{black continuous thin line}) and the spectrum used to subtract the background emission ('Outside', \emph{grey dashed thin line}) are also displayed. To better see all the three spectra, the recovered spectrum was scaled by a factor $\times$4, and the 'Inside' and 'Outside' spectra are  offset by  $\times$0.1 and $\times$0.2 the peak value of 'Inside' in the displayed spectral window. 
  }
   \label{figsnr}
    \end{figure*}

   \begin{figure}[th]
   \centering
   \includegraphics[angle=0,trim=80 12 60 35, width=0.35\textwidth, clip=,]{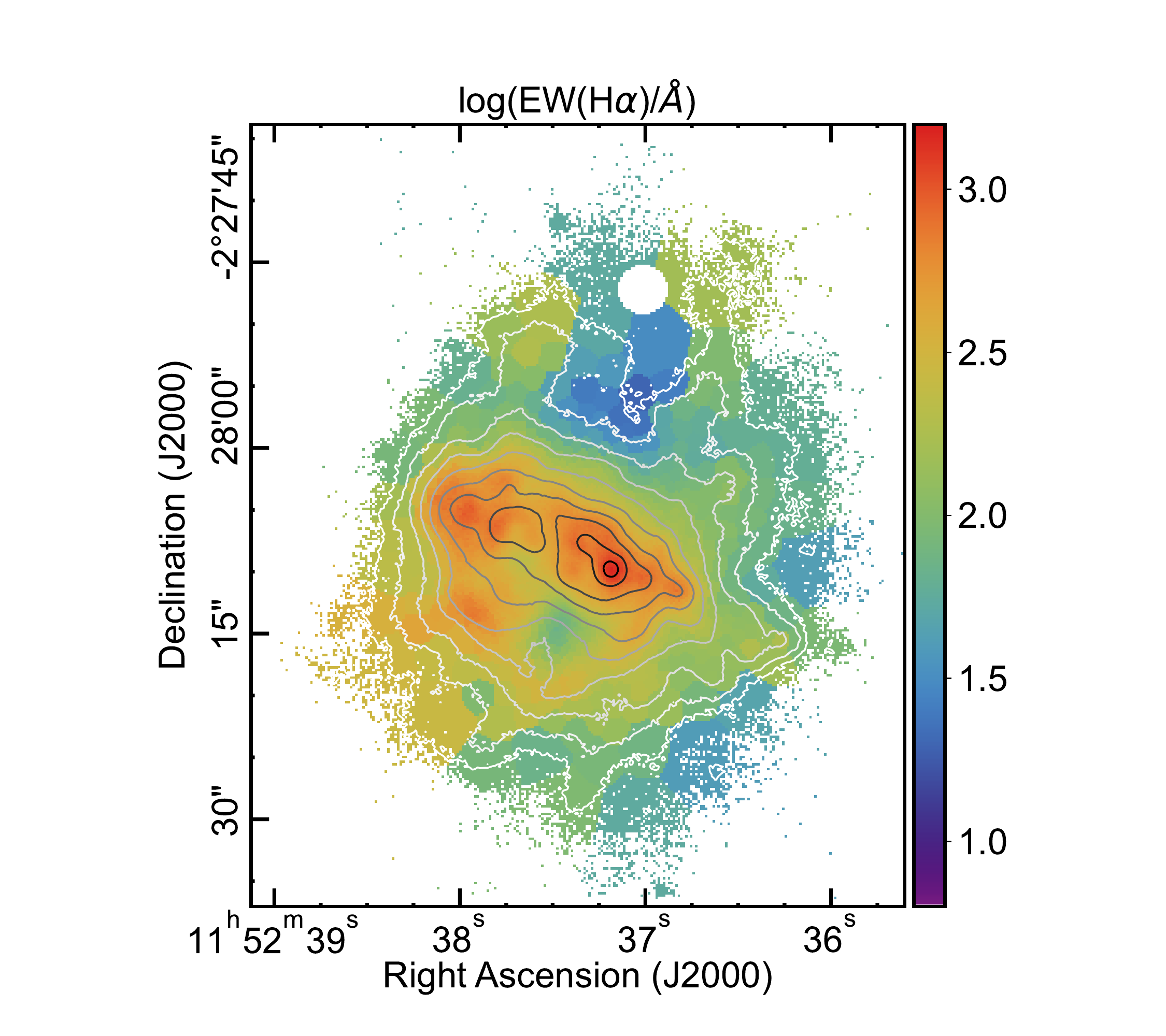}

   \includegraphics[angle=0,trim=80 12 60 35, width=0.35\textwidth, clip=,]{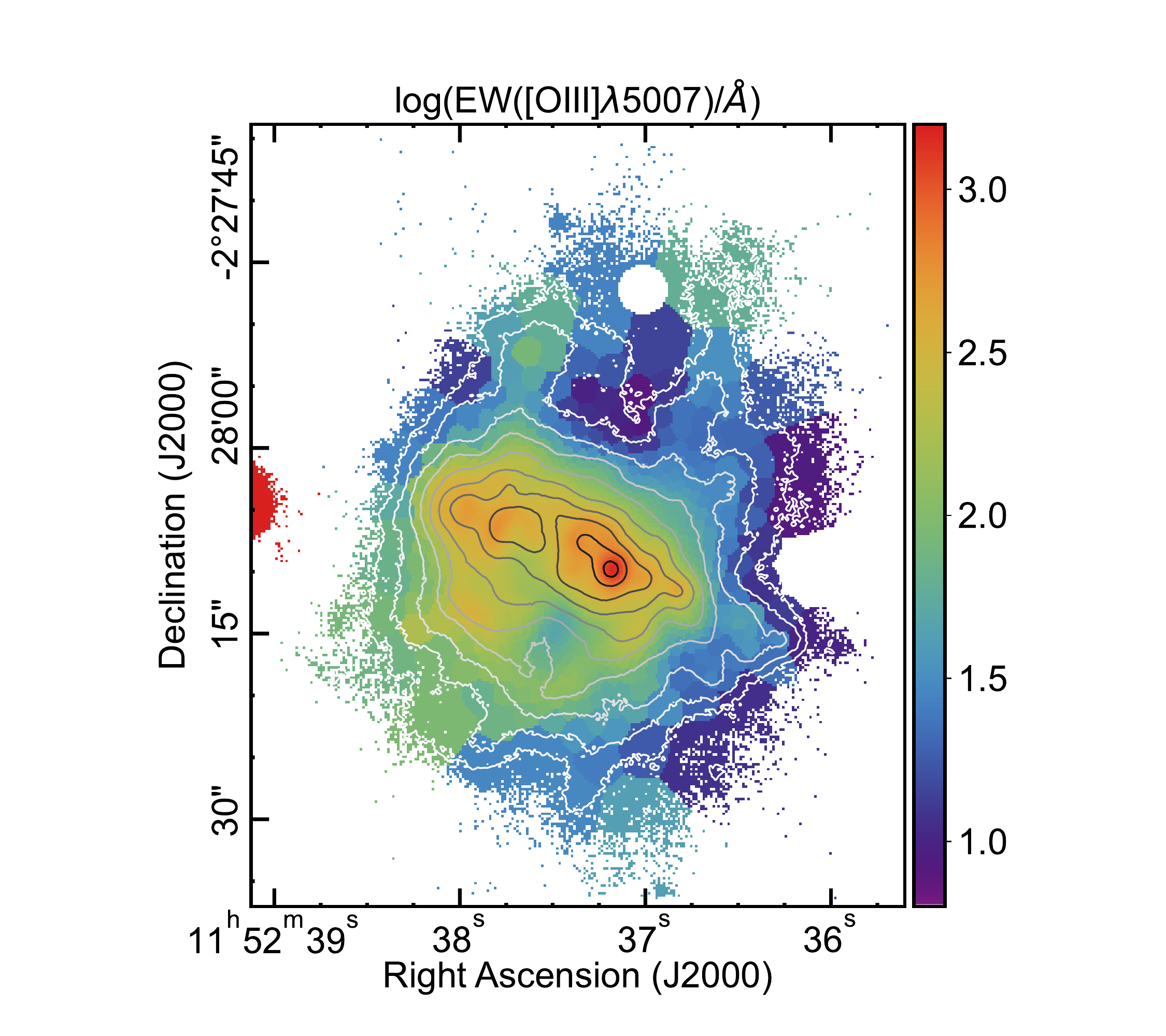}
    \caption{Maps with equivalent width for the two strongest lines used here, \ha\,  (\emph{top}) and \oiii$\lambda$5007 (\emph{bottom}).
For reference,  the map in \ha\, flux made by line fitting on a spaxel-by-spaxel basis is overplotted with ten evenly spaced contours (in logarithmic scale) ranging from 1.26$\times$10$^{-18}$~erg~cm$^{-2}$~s$^{-1}$~spaxel$^{-1}$ to 1.26$\times$10$^{-15}$~erg~cm$^{-2}$~s$^{-1}$~spaxel$^{-1}$. 
 }
   \label{mapew}
    \end{figure}

   \begin{figure*}[th]
   \centering
   \includegraphics[angle=0,trim=80 12 60 35, width=0.35\textwidth, clip=,]{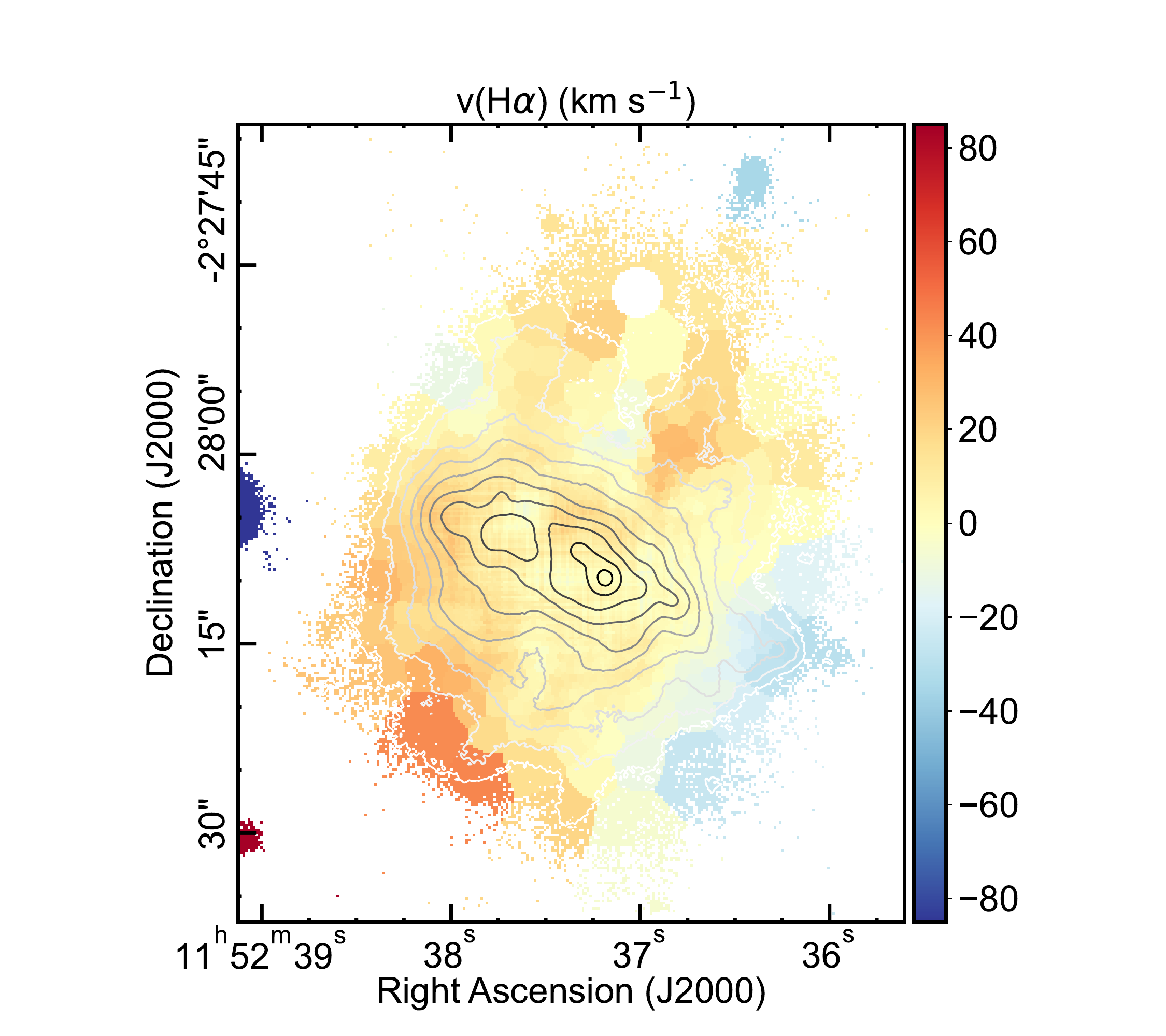}
   \includegraphics[angle=0,trim=80 12 60 35, width=0.35\textwidth, clip=,]{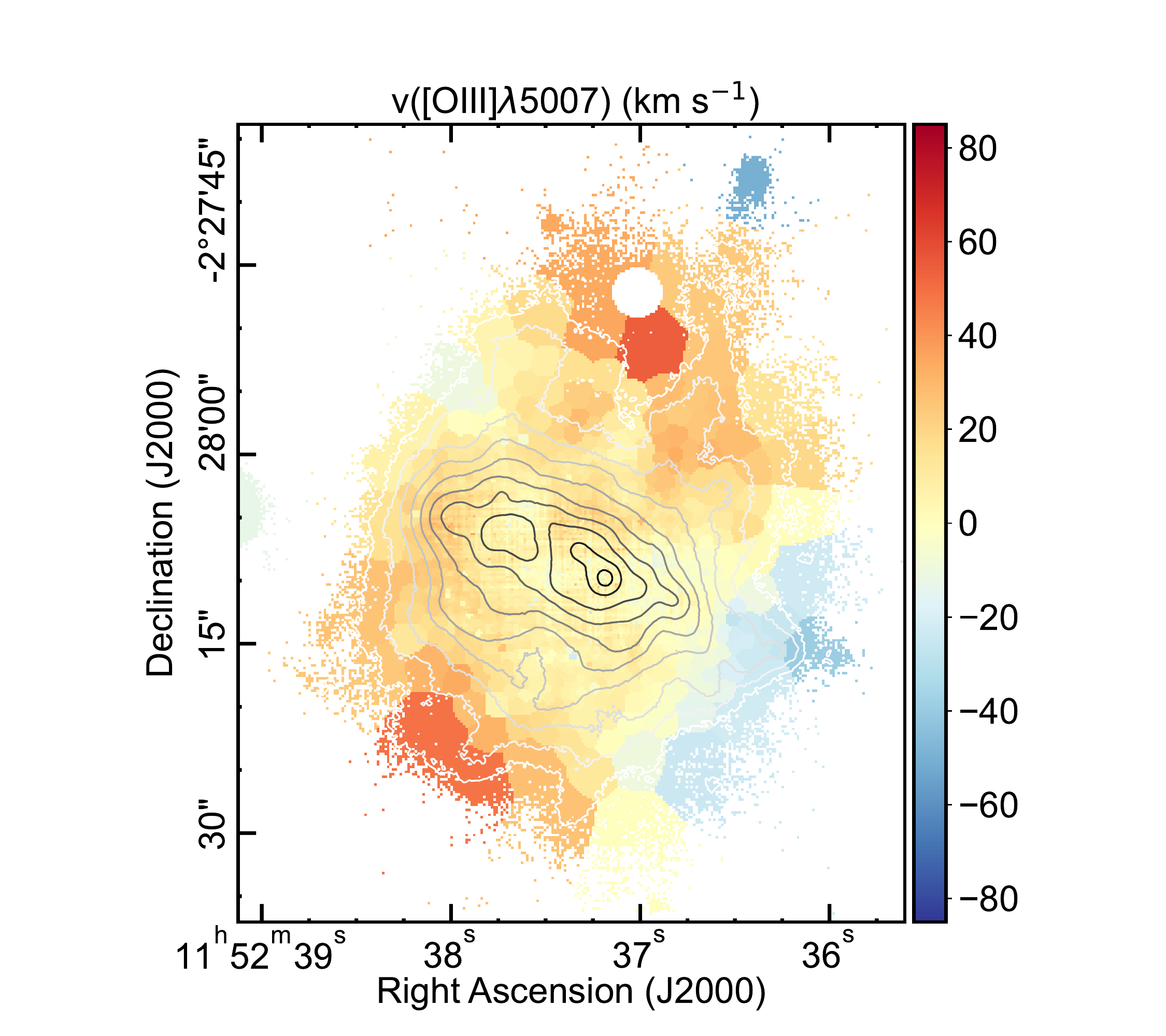}
   \includegraphics[angle=0,trim=80 12 60 35, width=0.35\textwidth, clip=,]{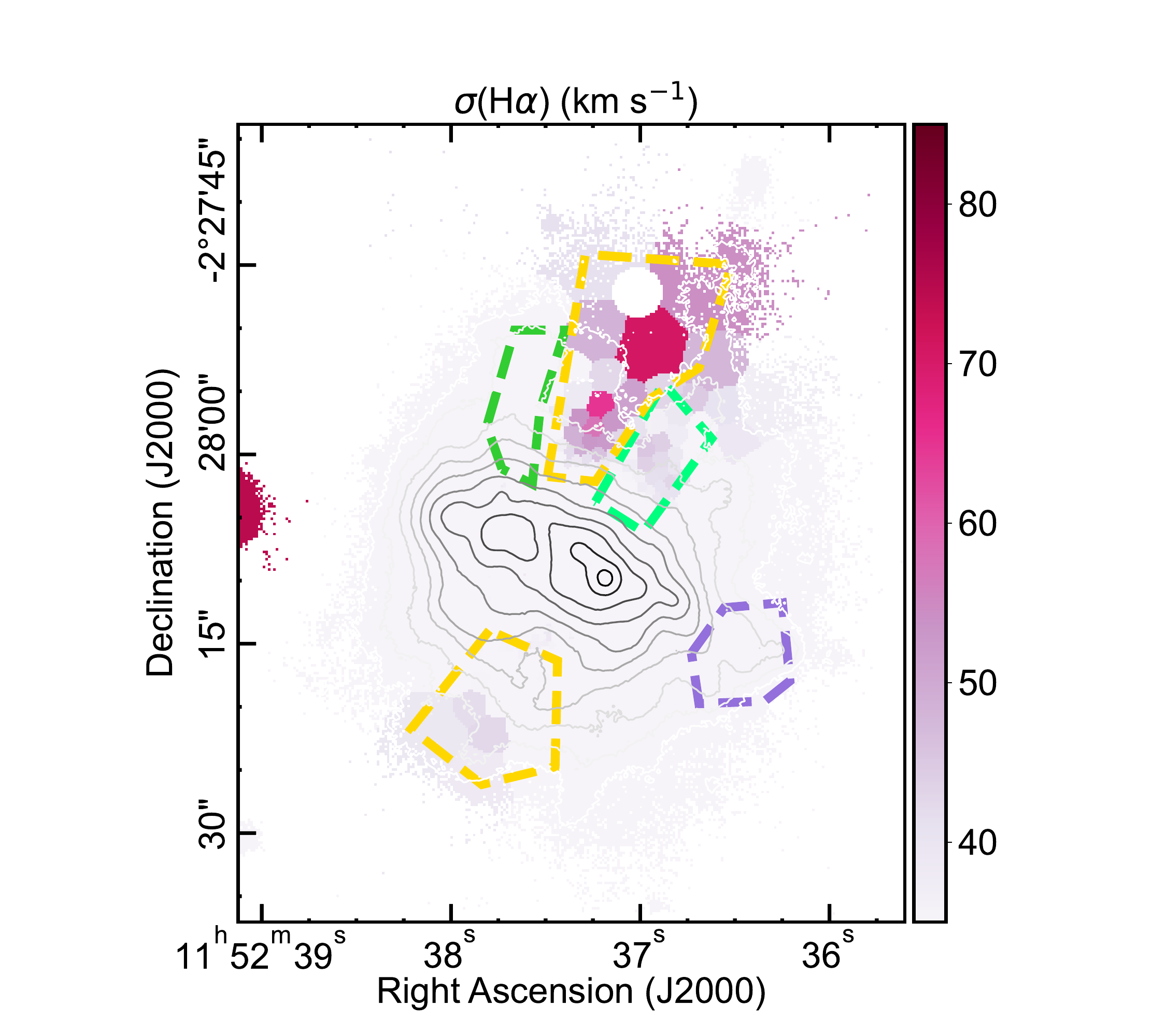} 
   \includegraphics[angle=0,trim=80 12 60 35, width=0.35\textwidth, clip=,]{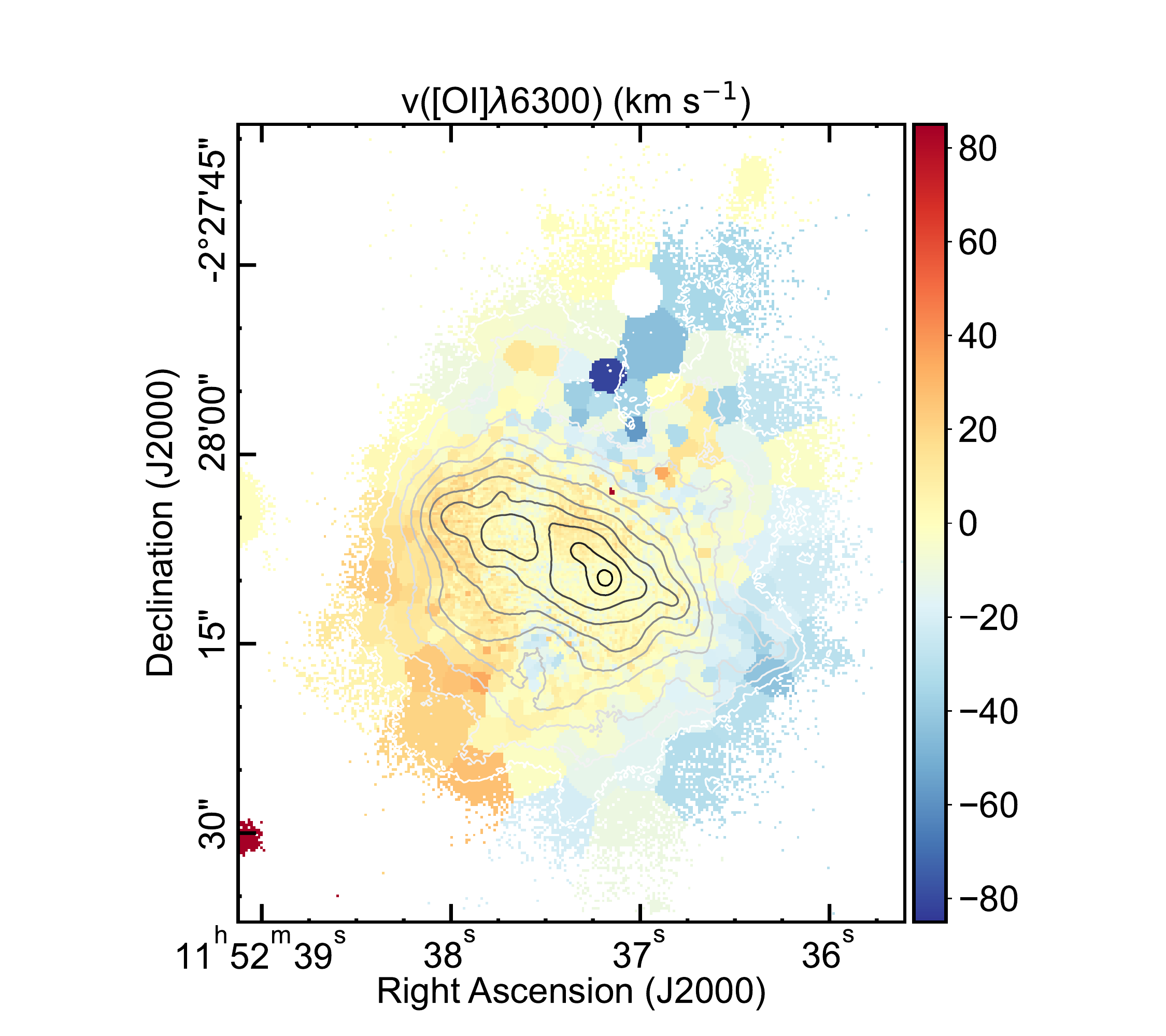}
    \caption{Representative maps related with the kinematics of the galaxy. From top to bottom and from left to right, they are:  %
    \ha\, velocity field, a  velocity field for a line tracing the high-ionisation zone (\oiii$\lambda$5007), \ha\, velocity dispersion map corrected for instrumental resolution, velocity field for a line tracing the low-ionisation zone (\oi$\lambda$6300).
Values for the velocities are presented as relative to the systemic velocity $v_{sys} = 1056$~km~s$^{-1}$ as derived from the redshift reported in Tab. \ref{basicdata}.
For reference,   the map in \ha\, flux made by line fitting on a spaxel-by-spaxel basis is overplotted with ten evenly spaced contours (in logarithmic scale) ranging from 1.26$\times$10$^{-18}$~erg~cm$^{-2}$~s$^{-1}$~spaxel$^{-1}$ to 1.26$\times$10$^{-15}$~erg~cm$^{-2}$~s$^{-1}$~spaxel$^{-1}$. 
   North is up and east towards the left. }
   \label{mapvel}
    \end{figure*}

\subsubsection{Line ratios at local maximum in \sii/\ha\, and \oi/\ha\, \label{secpointsource}}

In Sect.  \ref{secbpt}, we mentioned a location in the galaxy with a local maximum in the \sii/\ha\, and \oi/\ha\, line ratios but barely distinguishable from their surroundings in the  \oiii/\hb\, line map, and completely undistinguishable in the  \nii/\ha\, map (see Fig. \ref{mapBPT}).
We discuss its origin in Sect. \ref{secsnr}. Before that, the emission from this source, which is blended with the overall DIG emission in the galaxy needs to be isolated. We approached this challenge from two fronts, imaging and spectroscopy, making good use of the dual character of IFS data.

For the first approach, we used the reddening corrected flux images to create stamps around the location of interest, and masked that location with a circular mask. Then we created 'DIG stamps', where the emission in the masked area was estimated by 2D interpolation, and we recovered the emission of the source of interest by subtracting the interpolated stamp to the original one. We tried three interpolation methods: linear, nearest, and cubic. The outcome covers a range in satisfaction, depending on the emission line and the interpolation method. In general, it worked better for lines created by ions with low ionisation potential. Regarding the used method, cubic interpolation was the most prone to artefacts. The size of the stamp (3\farcs8$\times$3\farcs8) and the mask radius ($r=1\farcs0$ ) were chosen by trial and error looking for a combination minimising the number of spaxel with negative fluxes in the recovered stamps. A particularly challenging line was \oiii$\lambda$5007, very bright, but created by an ion with high ionisation potential. 
The maps for the most relevant lines are presented in Fig. \ref{figsnrmaps} and they confirm the flux excess in an unresolved source in all the lines, especially when looking at low-ionisation ones. We used these images to derive its 'DIG-cleaned' line ratios. The mean values of all the three interpolation methods are displayed in Fig. \ref{BPT} and  Fig. \ref{diagRamambason20} with a green solid circle in each graphic. The error bars represent the standard deviation. The large error bar in the \oiii/\hb\, (Fig. \ref{BPT}) and \oiii/\sii\, and \oi/\oiii\, line ratios (Fig. \ref{diagRamambason20}) reflect the difficulty of recovering a clean  artefact-free map for the \oiii$\lambda$5007 line. 
Values clearly move away from the \hii-like zone, and, when looking at the x-axes of the BPT diagrams, the lower the ionisation potential associated with the numerator is, the larger this displacement becomes. Likewise, Fig. \ref{diagRamambason20} shows that low ionisation lines are enhanced over high ionisation lines.

Besides, we recovered the whole spectra associated with the unresolved source. This is even more challenging, since one needs to deblend the information of the unresolved source and the overall DIG emission per wavelength bin. However, it can provide an independent way of estimating the line ratios, and potentially could reveal particularities in the line profiles that help to understand the nature of the unresolved source. 
As in the previous approach, we chose a stamp around the source of interest and extracted the spectrum in the same circular aperture defined above. Then, we estimated the spectrum of the underlying  DIG emission  as the median of the spectra in the remaining spaxels, scaled to the size of the circular aperture. Again, the specific size of the working stamp (4\farcs2$\times$4\farcs2) was chosen by trial and error. Selected portions of the recovered spectrum (and those for the circular aperture and background DIG) are presented in Fig. \ref{figsnr}. All the strong emission lines are well recovered, as well as some of the faint lines like the \oii\, lines at $\sim$7325~\AA\, (rest frame). Again, low ionisation lines (\sii\, lines, and specially the two \oi\, lines) are much stronger than in the underlying  DIG emission. Line ratios, as measured from this spectrum, agree with those plotted on the diagnostic diagrams in Figs. \ref{BPT} and \ref{diagRamambason20} within $\sim$0.03~dex. 
Besides, while velocity dispersion in the surrounding DIG is below the MUSE resolving power, we detect marginally broader lines in the recovered spectrum, at least in the reddest lines, where MUSE resolution is larger. We measured a $\sigma_{\rm{H}\alpha}$= 1.70~\AA\, that, once corrected from instrumental width ($\sigma_{ins}$=1.06~\AA\, in \ha), implies a linewidth of $\sigma\sim$50~km~s$^{-1}$.

\subsubsection{Equivalent width maps \label{secEW}}

An additional piece of the puzzle is the equivalent width ($EW$) for the strongest emission lines. In particular, hydrogen recombination lines can be used to get a rough estimate of the age of the youngest burst of SF. This will be discussed in Sect. \ref{secSB}. In our case, both, \ha\, and \hb\, could be used to this end. Since nebular \ha\, is $\gtrsim3$ times stronger than nebular \hb\, and stellar \ha\, can be $\sim2.0-2.5$ times fainter than nebular \hb,
the precise value depending on the metallicity and age of the stellar population \citep[see e.g.][]{GonzalezDelgado05}, we opted here for using $EW$(\ha) for that purpose. 
Additionally, the $EW$ can be used to pin-point those locations where there is an excess of ionised gas emission with respect to that of the underlying stellar population. In this manner, one can better delineate ionised gas structures like bubbles, cones, filaments, superwinds and others.

We present the map for the $EW$(\ha) in the top panel of Fig. \ref{mapew}. Not unsurprisingly, the highest values are found in the \hii-1 to  \hii-4  regions, in particular in \hii-1, with values of up to $\sim$1500~\AA, and \hii-4, with values of $\sim$800~\AA. These values, in particular that for region \hii-4, will be discussed in Sect. \ref{secSB}, together with the ionisation structure as traced by the ratios introduced in Sect. \ref{secionstruc}. Here we put the focus on the external parts (i.e. the substructure delineated in the DIG-1 region).

The $EW$(\ha) in DIG-1 is still high, ranging from $\sim10-30$~\AA\, in the region between the horns, to values of $\sim$300-600~\AA\, in two structures relatively symmetric to the horns with respect to the main body of the galaxy, and that bend towards each other and join at the most south-eastern edge of the galaxy. Hereafter, we  refer to this set of two structures as 'the beard'. Finally, the so-called horns and ear, present somewhat intermediate values of $\sim$100-200~\AA.

For completeness, we have also included $EW$(\oiii) map (bottom panel of Fig. \ref{mapew}). The galaxy presents a structure pretty similar to the one for $EW$(\ha), ranging from $\sim10$~\AA\,  between the horns to $\sim$1400~\AA, in \hii-1, just modulated by the ionisation structure. De facto, the joint use of these two maps would be almost  equivalent to the \oiii/\hb\, map presented in Sect. \ref{secionstruc}.

When comparing the ionisation structure and the $EW$ maps in here, a noteworthy finding is that both the horns and the beard present similar line ratios, yet while the horns define an open structure and have relatively low $EW$(\ha) values, the beard is a closed structure with relatively much higher $EW$(\ha). We discuss these differences in Sect. \ref{sechornsandbeard}, once the kinematic of the system has been introduced, and together with the rest of the observational evidence in these areas.

\subsubsection{Ionised gas kinematics \label{seckine}}

We recovered  kinematic information for the overall ionised gas (by means of e.g. \ha\, and \hb). Moreover, since we fit individually most of the strong emission lines, we were able to recover independent kinematic information for ions corresponding to high ionisation regions (by means of the \oiii\, lines) and low ionisation regions  (by means of the \oi, \nii, or \sii\, lines). The velocity fields for a representative subset of lines is presented in Fig. \ref{mapvel}. Besides, the figure also contains the velocity dispersion map  once corrected from instrumental resolution,  as derived from \ha.
The overall stellar velocity field (not shown) was provided by  \texttt{FADO}, and we used as a reference to which we can compare the velocities of ionised gas. It was completely unstructured, not showing any sign of rotation, any gradient, or any other coherent pattern. To reject the suspicion that this might have to do with the code used to model the stellar population, we derived the stellar velocity field  also using two additional codes with similar purpose, \texttt{ppxf} \citep{Cappellari17} and \texttt{PLATEFIT}  \citep{Brinchmann04}, with comparable outcome. Thus, we concluded that the stellar velocity field was simply a consequence of seeing the galaxy (almost) face-on.

The ionised gas presents a different situation. All the velocity maps in Fig. \ref{mapvel} present some structure, with higher (redder) velocities in the east side of the galaxy and lower (bluer) velocities in the west, with a $\Delta v\sim$40~km~s$^{-1}$.  In that sense these maps are reminiscent of the velocity field for \hi\, \citep{vanZee98} that, at a spatial resolution of $\sim$5$^{\prime\prime}$, was interpreted as compatible with solid-body rotation with receding velocities in the east and approaching velocities in the west/south-west.
However, the structure in all these maps is much more complex than simple rotation. In the \ha\, map, the highest (reddest) velocities are not in the east-west direction, as one would expect if the ionised gas were rotating in a similar manner as the neutral gas, but towards the south/south-east. Besides, the northern half of the galaxy presents a clear velocity stratification: the two horns have similar velocities in all the maps, while the space between the horns present redder velocities in the high ionisation lines (here \oiii$\lambda$5007, $v_{\oiii} - v_{\rm H \alpha} \sim 20-50$~km~s$^{-1}$) and bluer velocities in the low ionisation lines (here \sii$\lambda$6716 with    $v_{\rm H \alpha} - v_{\sii}  \lesssim 25$~km~s$^{-1}$ - not shown -, and \oi$\lambda$6300 with $v_{\rm H \alpha} - v_{\oi}  \lesssim 45$~km~s$^{-1}$). This area with velocity stratification presents velocity dispersions well above the MUSE instrumental width. Moreover, the level of stratification (i.e. difference between the velocities of several species) seems correlated with the velocity dispersion.
A similar situation but far less dramatic is also visible in the southern half of the galaxy. There, the region marked with a dashed yellow polygon in the velocity dispersion map has  $v_{\oiii} - v_{\rm H \alpha} \sim 10$~km~s$^{-1}$,  $v_{\rm H \alpha} - v_{\sii}  \lesssim 10$~km~s$^{-1}$  - not shown -, and \oi$\lambda$6300 with $v_{\rm H \alpha} - v_{\oi}  \lesssim 20$~km~s$^{-1}$, and velocity dispersions marginally above what one can resolve at the MUSE spectral resolution.
Both regions in the northern and southern half marked with the yellow polygons have  large \oi/\ha\, line ratios, well beyond the theoretical border proposed by \citet{Xiao18}.

Examples of velocity stratification in the literature are scarce. Velocities differences between the same lines under consideration have been identified in Orion \citep{GarciaDiaz08,Weilbacher15}. However, this samples areas within an \hii\, region itself and at the sub-pc level, much smaller scales than what we see in \object{UM 462}. \citet{MonrealIbero10a} also detected a velocity stratification in \object{NGC 5253}, this time at scales of $\sim$50-100~pc, more comparable to  (but still smaller than) the scales sampled here. In that case, the velocity stratification could be explained by a symmetric outflow centred at the main and youngest super star clusters.
This does not seem a plausible explanation here since the apex of the northern area with velocity stratification and high velocity dispersion seems located, not at  \hii-1, the peak of  \ha\, emission and youngest star cluster (see Sect. \ref{secstars}) but in between, \hii-3 and \hii-2, where the youngest stars look slightly older. Moreover, the observational evidence is different in the northern and the southern parts of the galaxy.
However, the high \oi/\ha\, line ratios together with the high velocity dispersion suggest that shocks are having an impact on the ISM in this area.
In Sect. \ref{sechornsandbeard}, we discuss a possible scenario able to explain the observed kinematics, together with the rest of observational evidence.

   \begin{figure}[th]
   \centering
   \includegraphics[angle=0,trim=80 12 60 35, width=0.35\textwidth, clip=,]{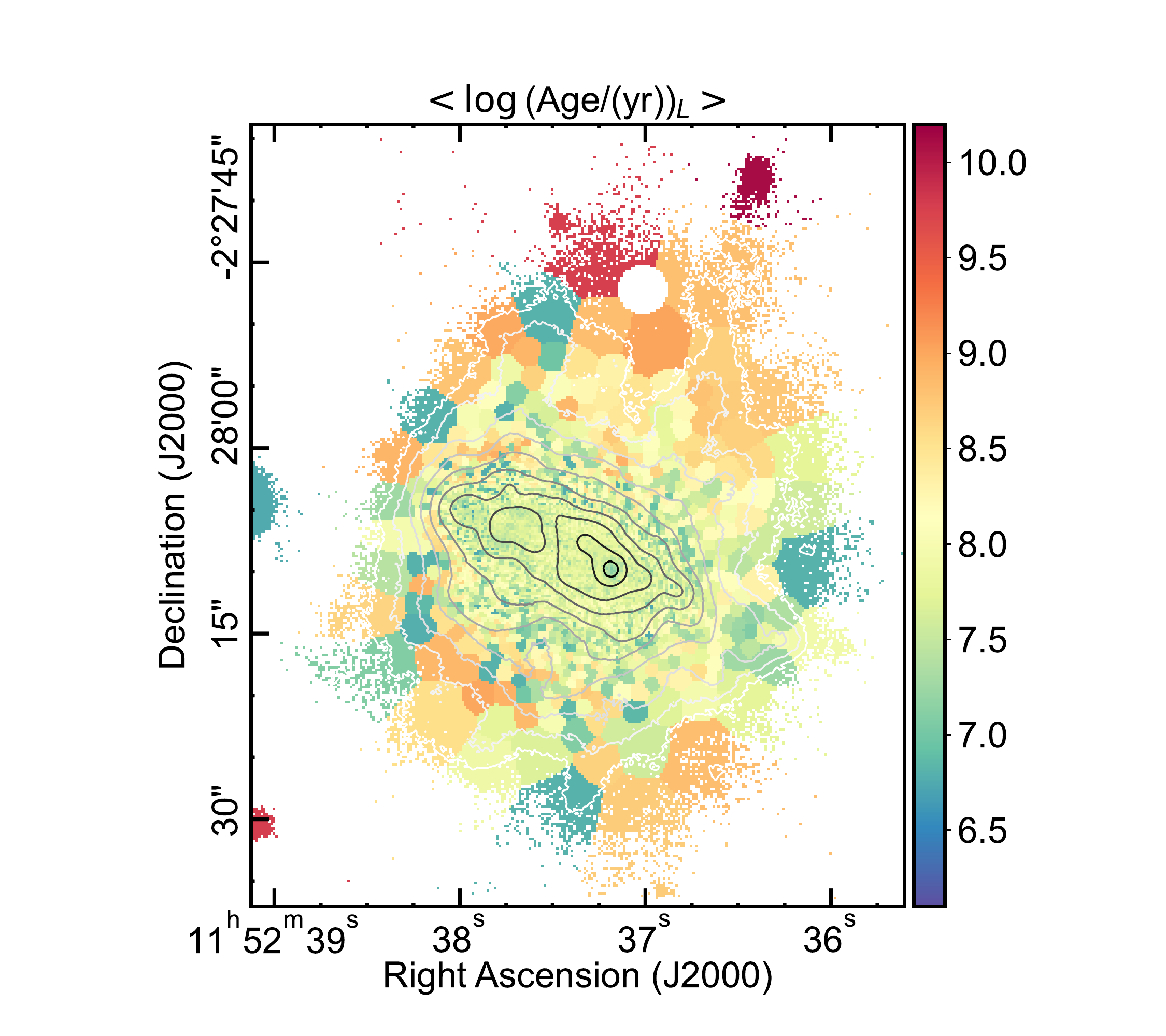}

   \includegraphics[angle=0,trim=80 12 60 35, width=0.35\textwidth, clip=,]{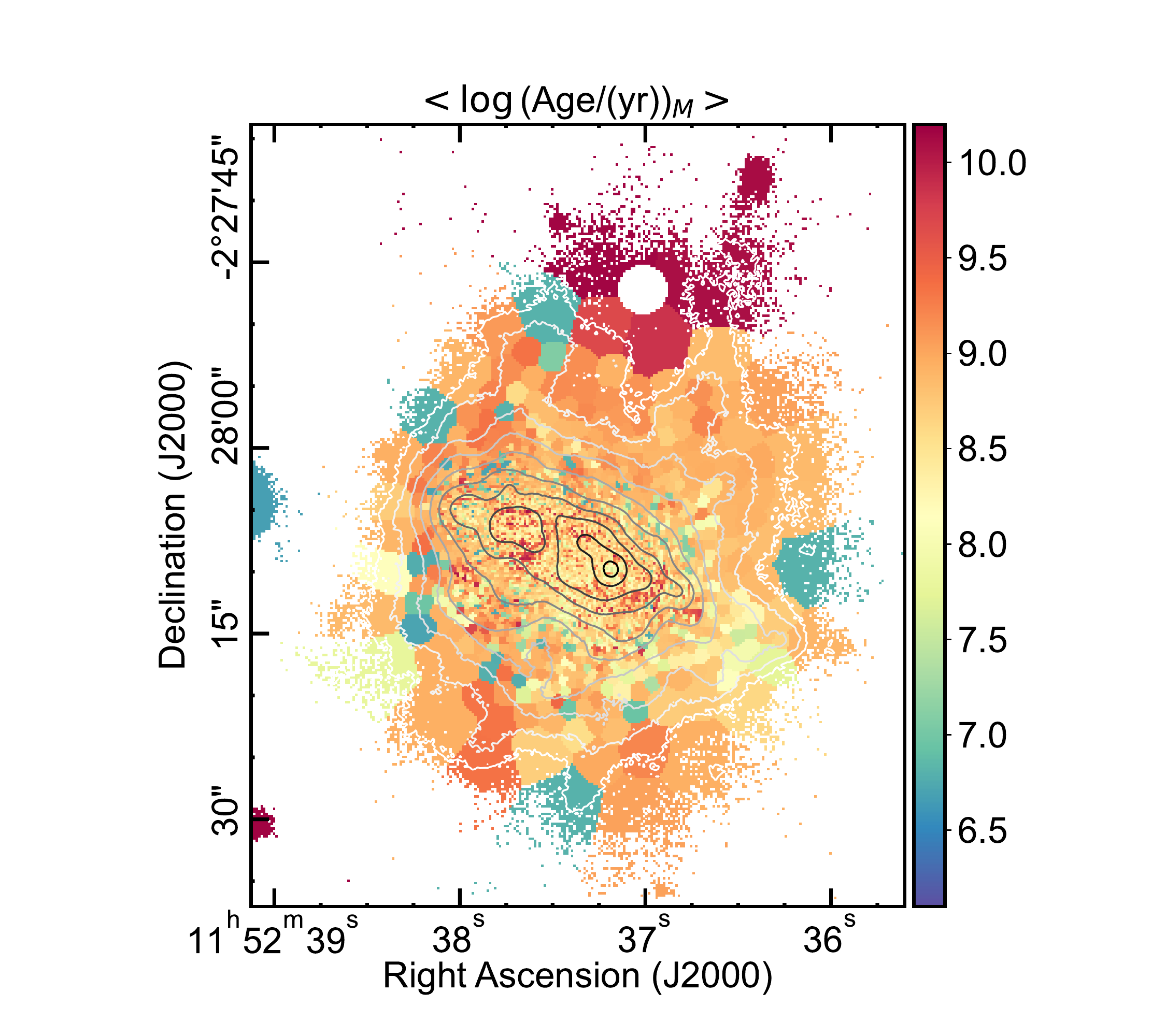}
    \caption{Maps with the mean age of the stellar population weighted by light (\emph{top}) and mass (\emph{bottom}).
For reference,   the map in \ha\, flux made by line fitting on a spaxel-by-spaxel basis is overplotted with ten evenly spaced contours (in logarithmic scale) ranging from 1.26$\times$10$^{-18}$~erg~cm$^{-2}$~s$^{-1}$~spaxel$^{-1}$ to 1.26$\times$10$^{-15}$~erg~cm$^{-2}$~s$^{-1}$~spaxel$^{-1}$. 
   North is up and east towards the left. }
   \label{mapage}
    \end{figure}

 \begin{figure}[ht]
 \centering
\includegraphics[angle=0,  trim=10 0 0 0, width=0.44\textwidth, clip=,]{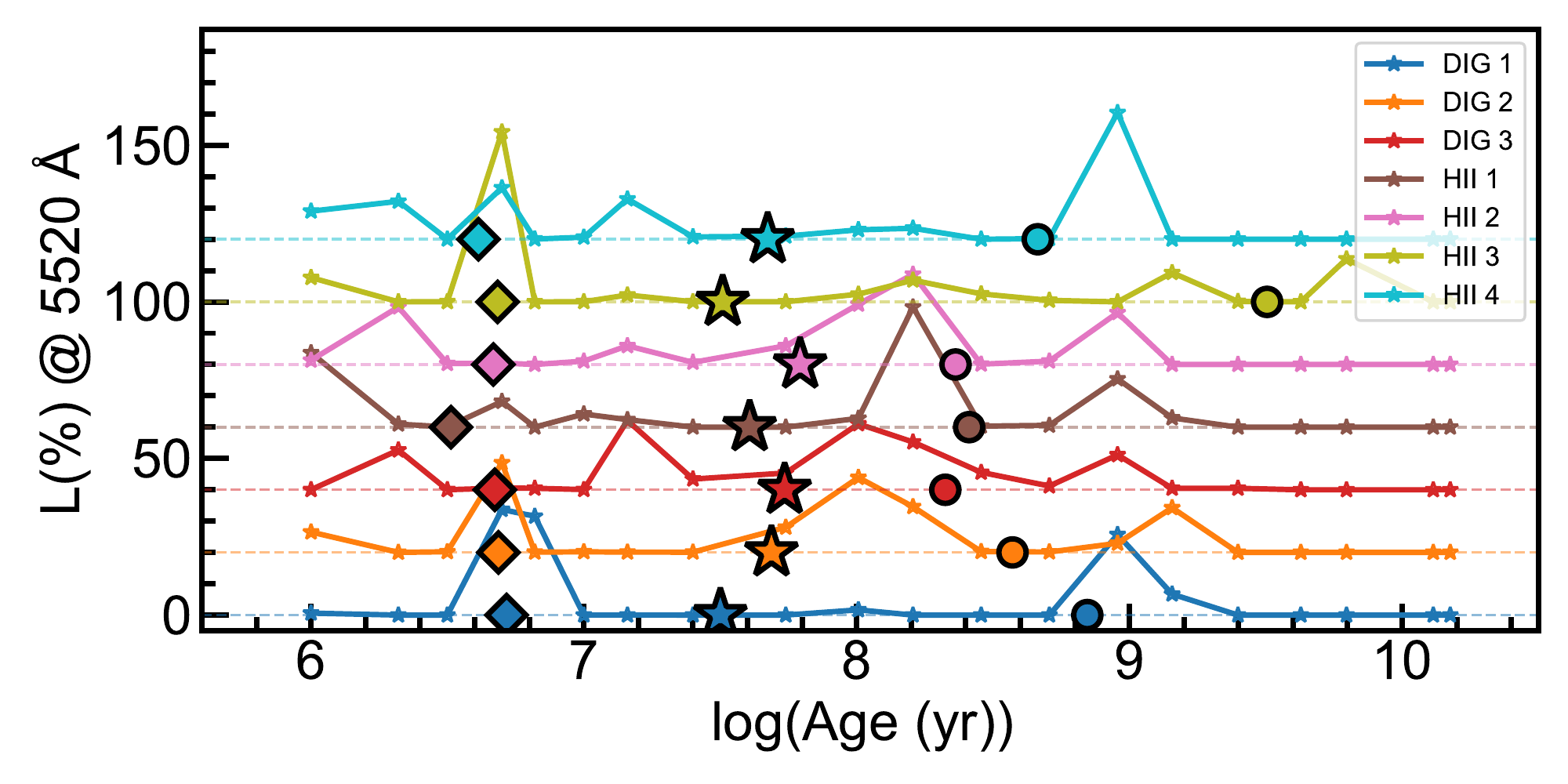}

\includegraphics[angle=0,  trim=10 0 0 0, width=0.44\textwidth, clip=,]{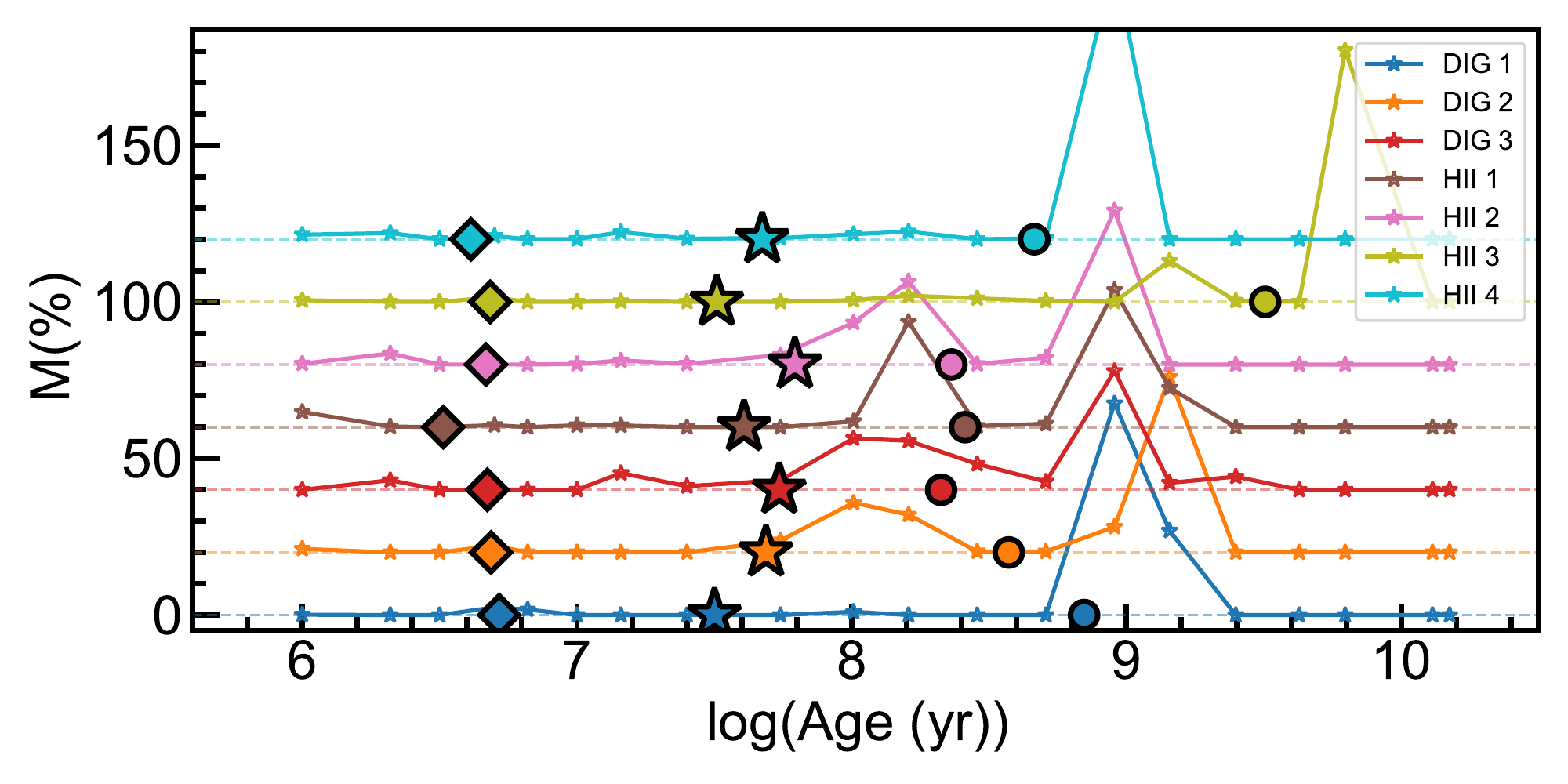}
\caption{
 Light (\emph{top}) and mass (\emph{bottom}) fractions per stellar population as provided by the \texttt{FADO} modelling for the seven areas considered here (\emph{small stars}). For a given age, the contributions of the two metallicities ($Z$=0.004 and $Z$=0.008) have been added.
 Likewise, to better visualise the light (or mass) fractions for all the seven areas, data have been offset by  +10$\,N$ with $N$ the spectrum number, as indicated by the  alphabetically ordering of the labels. Three characteristic values are shown on top of each function:  Mean stellar age logarithmic weighted by mass (\emph{circles}) and by light (\emph{stars}) together with the estimated age for an instantaneous burst of SF  (\emph{diamonds}), derived from the equivalent width in \ha, and the original STARBURST99 models \citep{Leitherer99}.
}
\label{SummaryPops}
\end{figure}

\subsection{The stellar populations \label{secstars}}

\subsubsection{General overview \label{secgenoverview}}

Along with the information for the ionised gas, the MUSE spectra also contain valuable information about the stars. 
Getting a finely tuned composition of the underlying stellar population, meaning a reliable and detailed 2D star formation history, is a complex task and beyond the goals of the work presented here. Instead, we present here a general overview of
the modelling by \texttt{FADO} to just get an overall idea of the stellar populations in the galaxy.
They are based on the outcome of the modelling described in Sect. \ref{secmodconto} but were stable against a selection of a slightly different (i.e. more reduced or extended) set of base spectra.
The most relevant information is summarised in Fig. \ref{mapage}, that contains maps of the average age weighted by light and mass,  and Fig.~\ref{SummaryPops}, that contains the light and mass fractions per stellar population  for each of the seven areas defined by \texttt{astrodendro}. Several pieces of information can be inferred from these figures.


\begin{table*}
\caption{Basic properties of the recent SF in the regions defined by \texttt{astrodendro}. \label{basicpropsSF}} 
\centering  
\small
\begin{tabular}{lccccccc}
\hline
Property     &                    DIG 1 &                    DIG 2  &                    DIG 3 &                    \hii\, 1 &                    \hii\, 2 &                    \hii\, 3 &                    \hii\, 4 \\
\hline
L(\ha)$^{(a)}$       &  112.49$\pm$0.56 &  22.48$\pm$0.10 &  58.71$\pm$0.28 &  84.41$\pm$0.44 &   4.57$\pm$0.03 &  38.30$\pm$0.16 &   3.51$\pm$0.01 \\
SFR$^{(b)}$      &   60.41$\pm$0.30 &  12.07$\pm$0.05 &  31.53$\pm$0.15 &  45.33$\pm$0.23 &   2.45$\pm$0.01 &  20.57$\pm$0.08 &   1.88$\pm$0.01 \\
$\Sigma_{SFR} ^{(c)}$  &    1.67$\pm$0.01 &  15.71$\pm$0.07 &  26.90$\pm$0.13 &  66.08$\pm$0.34 &  43.14$\pm$0.26 &  28.99$\pm$0.12 &  18.84$\pm$0.07 \\
Age$_{\rm SB 99}$  & 5.21  & 4.86 & 4.71 & 3.26 & 4.66 & 4.83 & 4.11 \\
\hline
\end{tabular}

$^{(a)}$In units of  10$^{-38} $erg s$^{-1}$; $^{(b)}$In units of  10$^{-3}$ M$_\odot$ yr$^{-1}$;  $^{(c)}$In units of  10$^{-2}$ M$_\odot$ yr$^{-1}$ kpc$^{-2}$; $^{(d)}$In Myr.
\end{table*}

Firstly, there is no need for a stellar population older than $\sim$1.2~Gyr to model the observed spectra anywhere but \hii-3, where  \texttt{FADO} sees a significant contribution in mass of a $\sim$4~Gyr stellar population. \citet{Micheva13} reported broad band colours consistent with an age of the stars in the underlying host galaxy of $\sim3-4$~Gyr and very low metallicity ($Z\sim0.001$).
This component extended well beyond the area covered by MUSE and it is likely included in the sky emission and thus, subtracted out in most of the apertures. The fact that this is still detected in \hii-3 suggests that this old population was more significant there but, given the degeneracies intrinsic to the modelling of the emission of the underlying population, this is a result that should be taken with caution.
Nevertheless, relatively old  (i.e. $\sim$1~Gyr) stars are found in any other aperture.
This $\gtrsim$1 Gyr stellar population  account for most (i.e. $\sim$60-70~\%) of the mass over the whole galaxy.

Secondly,  Fig. ~\ref{SummaryPops}  suggests a significant  ($\sim$10-30~\%)  contribution to the stellar mass of an intermediate age ($\sim$40-400~Myr) stellar population in apertures \hii-1, \hii-2, and their envelopes (DIG-3, but also DIG-2). This intermediate-to-young stellar population extends well in time in \hii-2.

Thirdly, \texttt{FADO} also recovers the very young stellar population responsible for the ionising photons. The luminosity of the galaxy in all the apertures is dominated by $\sim$6~Myr old stars, even if only contributes with $\lesssim$10\% of the \textsc{stellar} mass. 
Besides, most of the apertures but DIG-1 contain even younger stars up to the youngest population included in the modelling by \texttt{FADO}  (i.e. 1~Myr, here assimilated to on-going SF).
The contribution of this on-going SF is particularly important in \hii-1, and to certain extent in its neighbouring apertures (DIG-3 mostly, but also \hii-2), and is clearly seen in the  luminosity-weighted  age map.
Finally, this suggests that the most recent SF propagates from the outer to the inner parts of the galaxy (i.e. from DIG-1 to inner apertures), then from east to west, following a path roughly like \hii-3 $\rightarrow$ \hii-2 $\rightarrow$ \hii-1.
Region \hii-4 would be out of this sequence. It also formed stars that are not $\gtrsim$1 Gyr old. However, it seems to follow an independent story,  having formed stars at practically every age, including now, in a relatively continuous manner. 
\subsubsection{Starburst ages and star formation rates \label{secSB}}

The emission lines, and in particular \ha, can also be used to characterise the most recent SF.   We derived the star formation rates (SFR) for all the seven apertures using the \ha\, luminosity and the expression:
\begin{equation}
{\rm SFR} ({\rm M_\odot\, yr}^{-1}) = 10^{-41.27} \times L(H\alpha) \,  {\rm\, (erg\, s^{-1})}
\end{equation}
provided by \citet{Kennicutt12}. They are presented in Table  \ref{basicpropsSF}. 
As suggested by the results in Sect.  \ref{secgenoverview}, the SFR is the highest in  \hii-1. Then, it decreases from west to east and from the inner to the outer parts of the galaxy. We note that DIG-1, the aperture associated to the emission in the outermost parts of the galaxy, contributes with a far from negligible fraction ($\sim$35\%) to the total luminosity in \ha.

An alternative path to estimate the youth of the most recent burst of SF, beyond \texttt{FADO}, is using the \ha\, equivalent widths jointly with the STARBURST99 models\footnote{\texttt{https://www.stsci.edu/science/starburst99/docs/\\default.htm}} \citep{Leitherer99}. Here, we used the original models for an  instantaneous burst of SF,  an initial mass function with $\alpha$=2.35 and upper limit for the star mass of $M_{up} = 100$~M$_\odot$ as a baseline. We looked for the estimated age for metallicities $Z$=0.004 and $Z$=0.008. The average of both values is presented with diamonds in Fig. \ref{SummaryPops} and listed in Table~\ref{basicpropsSF}. Again, the extreme youth of the burst in \hii-1 stands out, and even if the burst is young in every aperture, the direction outlined above (increasing age from west to east, from inside to outside) is again outlined.

A comparison of the ages of the youngest bursts of SF at each location and the line ratios tracing the ionisation structure (Sect. \ref{secionstruc}) shows how ratios compatible with lower ionisation parameter are found in those locations with older stars, as expected. Not for the first time in this contribution,  \hii-4  departs from this sequence. It is the location with the second youngest burst and yet, it presents line ratios compatible with lower ionisation parameter than  \hii-2 and  \hii-3.
We posit here stochasticity as a possible cause of this observational evidence \citep[e.g.][]{Paalvast17,Krumholz15}. Determining an accurate stellar mass for the young burst at each of the \texttt{astrodendro} locations is beyond the scope of this paper. For Sect. \ref{secgenoverview}, we simply embraced those provided by \texttt{FADO}, which were enough to get an overview of the global characteristics of the stellar populations. Nonetheless, we can use an approximate estimate of the masses of the youngest stellar populations using the SFR and age provided in Table \ref{basicpropsSF} to explain \hii-4's rogue behaviour, assuming that the reported SFRs were constant over the age of the young bursts.
Such estimates are largely uncertain but still reliable enough for a discussion in terms of orders of magnitude. In that sense, while the stellar mass in \hii-1 and \hii-3 is of the order $\sim10^6$ M$_\sun$, that in  \hii-2 is of $\sim10^5$ M$_\sun$ and the one in \hii-4 is even less  ($\sim5\times10^4$ M$_\sun$ ). At that range of masses the initial mass function is not adequately sampled in the high-mass end. Thus, it may well happen that the star massive enough to provide the more energetic photons at that location simply did not manage to be born.

Adding up the contribution of  all the apertures, the total SFR is 0.17 $M_\odot$ yr$^{-1}$, larger but comparable with the value reported by \citet{Paudel18}. GP galaxies typically have SFRs $\sim3-30$~M$_\odot$~yr$^{-1}$ \citep{Cardamone09}, which is 1-2 orders of magnitude larger than the SFR here.
In every other aspect considered so far (e.g. gas metallicity, electron density and temperature and emission line ratios), \object{UM\,462} is similar to the GP galaxies. This suggests that GP galaxies could be the brightest and most vigorous end of a galaxy population at $z\sim$0.1-0.2 where objects like \object{UM\,462} may be common. In turn, \object{UM\,462} would be representative of the faint end of the GP galaxies distribution and thus allows us to study the processes taking place in faint GP galaxies.
Up to now, at $z\sim0.1-0.2$, studies similar to those existing for the brighter relatives were not possible but they may be common with the advent of the James Webb Space Telescope (JWST). Such studies will surely benefit from the results found for this and similar galaxies in the Local Universe.

 \begin{figure}[ht]
 \centering
 \includegraphics[angle=0,  trim=80 12 60 35, width=0.24\textwidth, clip=,]{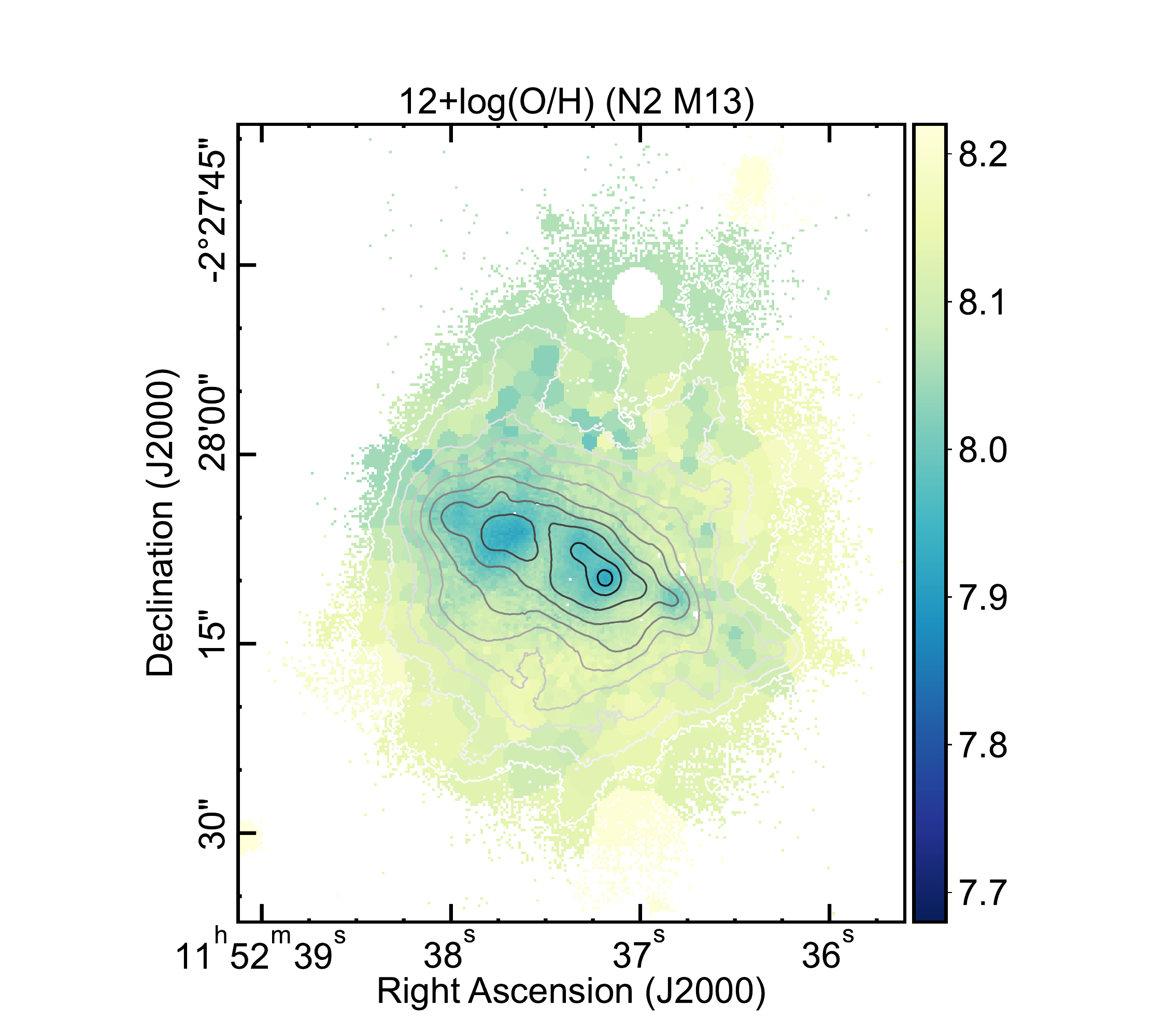}
\includegraphics[angle=0,  trim=80 12 60 35, width=0.24\textwidth, clip=,]{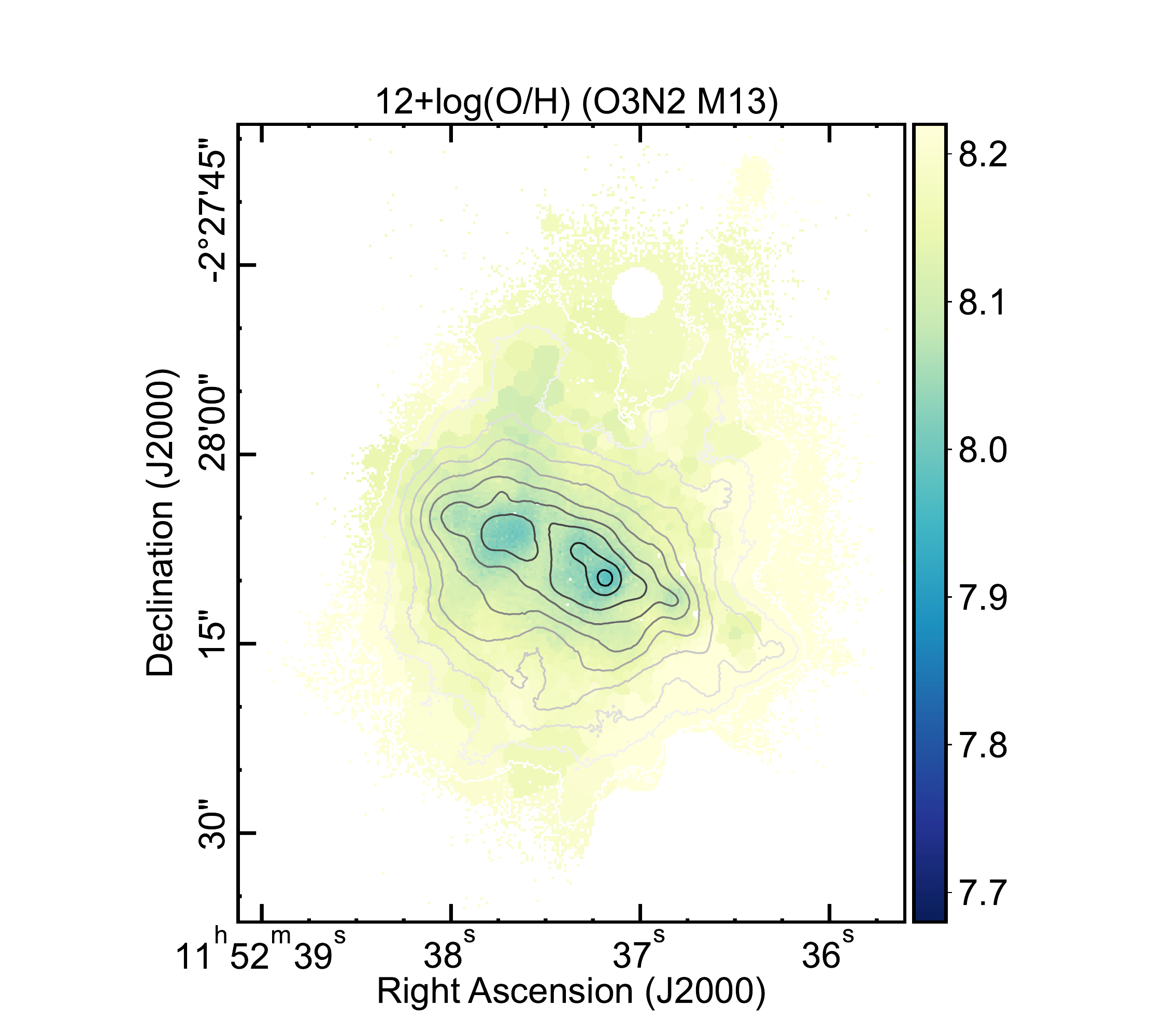}

\includegraphics[angle=0,  trim=80 12 60 35, width=0.24\textwidth, clip=,]{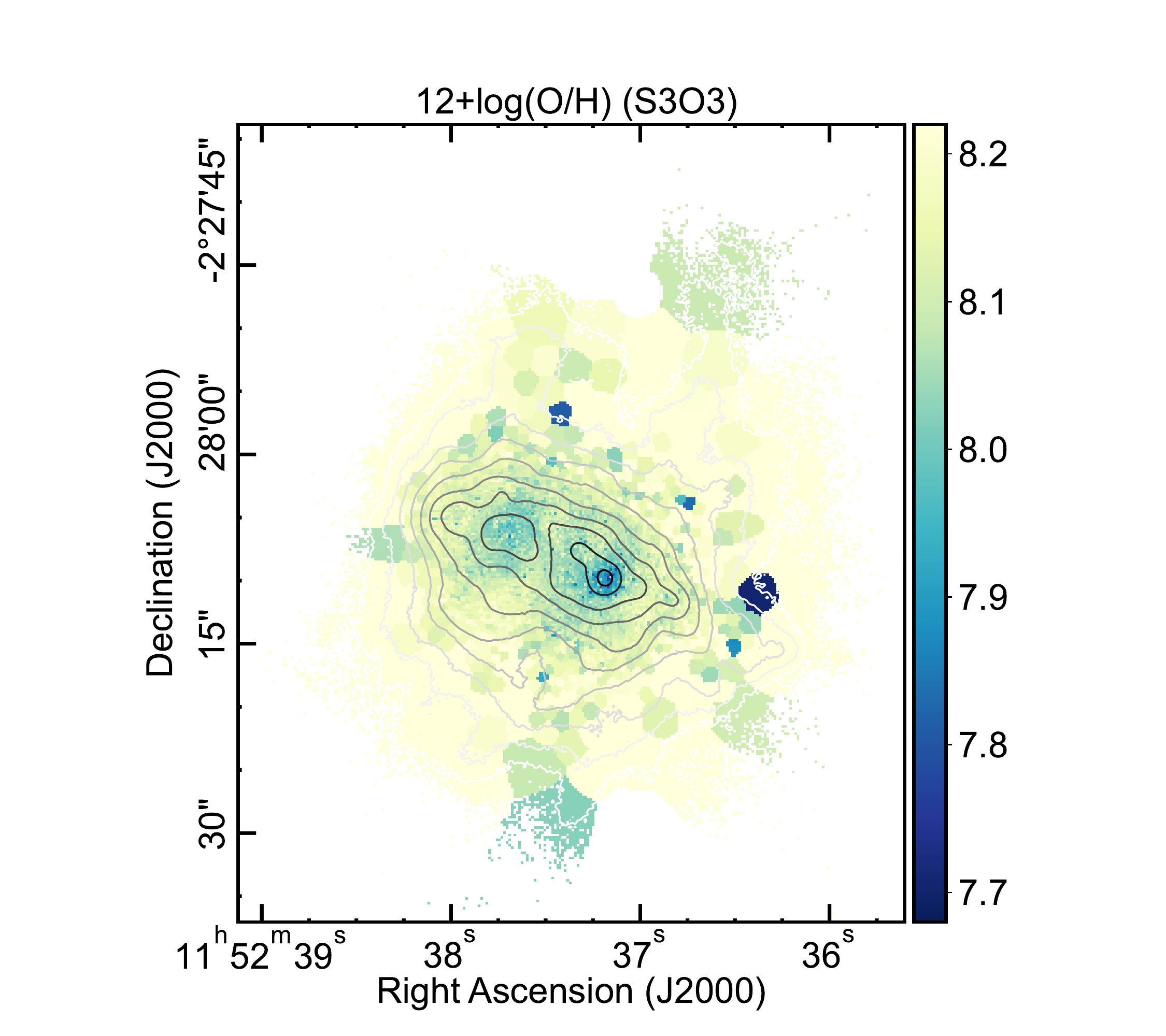}
\includegraphics[angle=0,  trim=80 12 60 35, width=0.24\textwidth, clip=,]{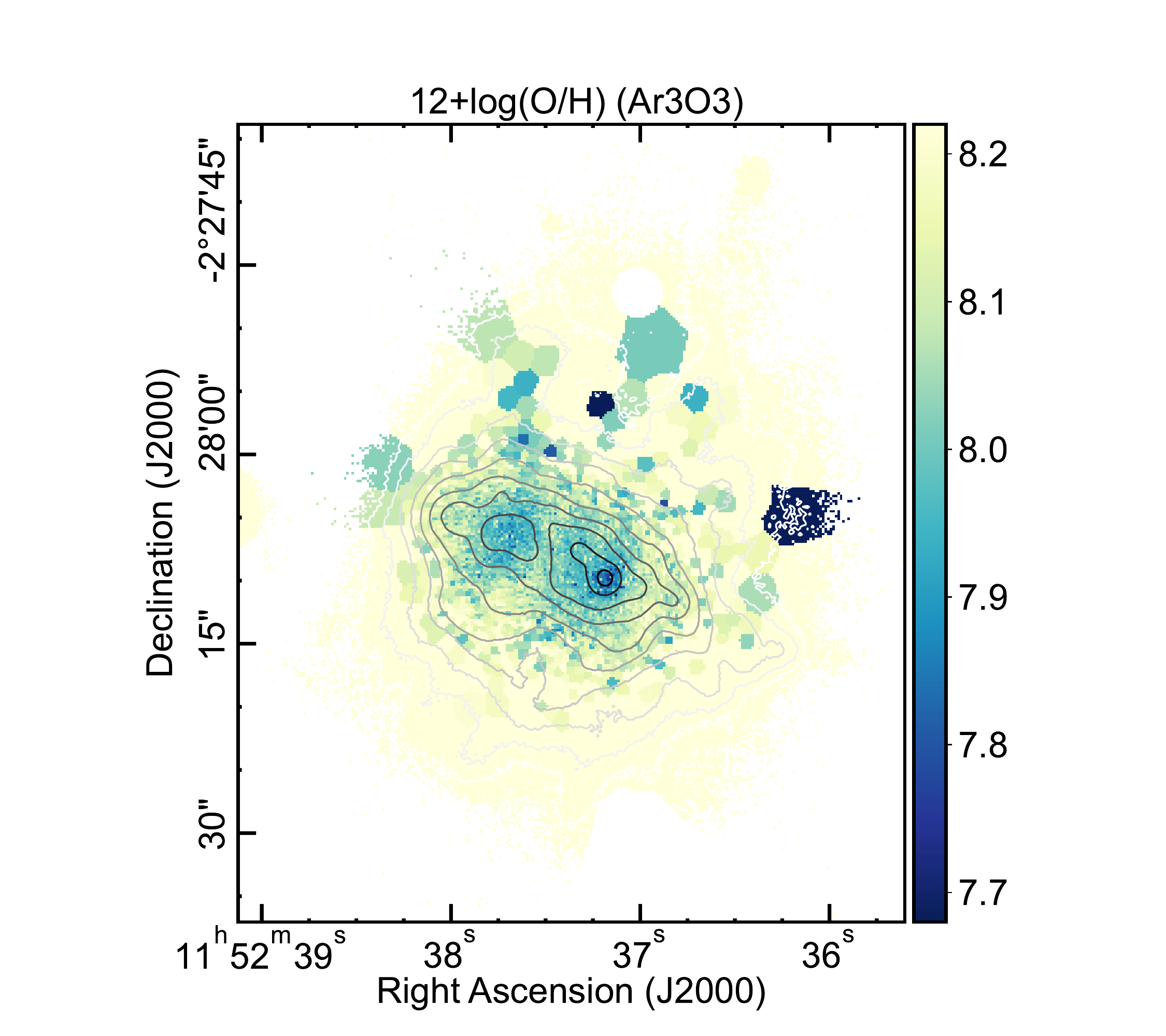}

\includegraphics[angle=0,  trim=80 12 60 35, width=0.24\textwidth, clip=,]{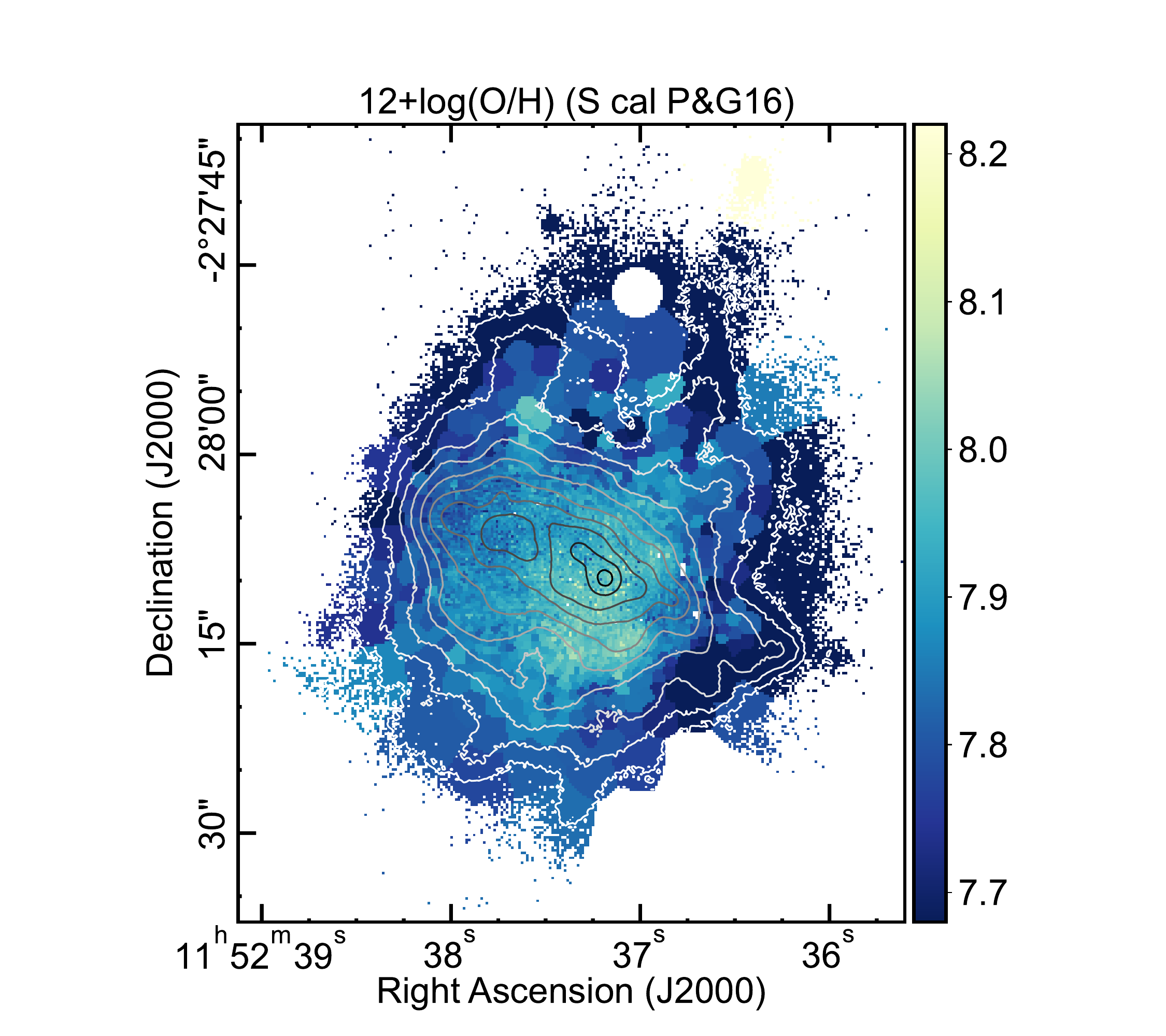}
\includegraphics[angle=0,  trim=80 12 60 35, width=0.24\textwidth, clip=,]{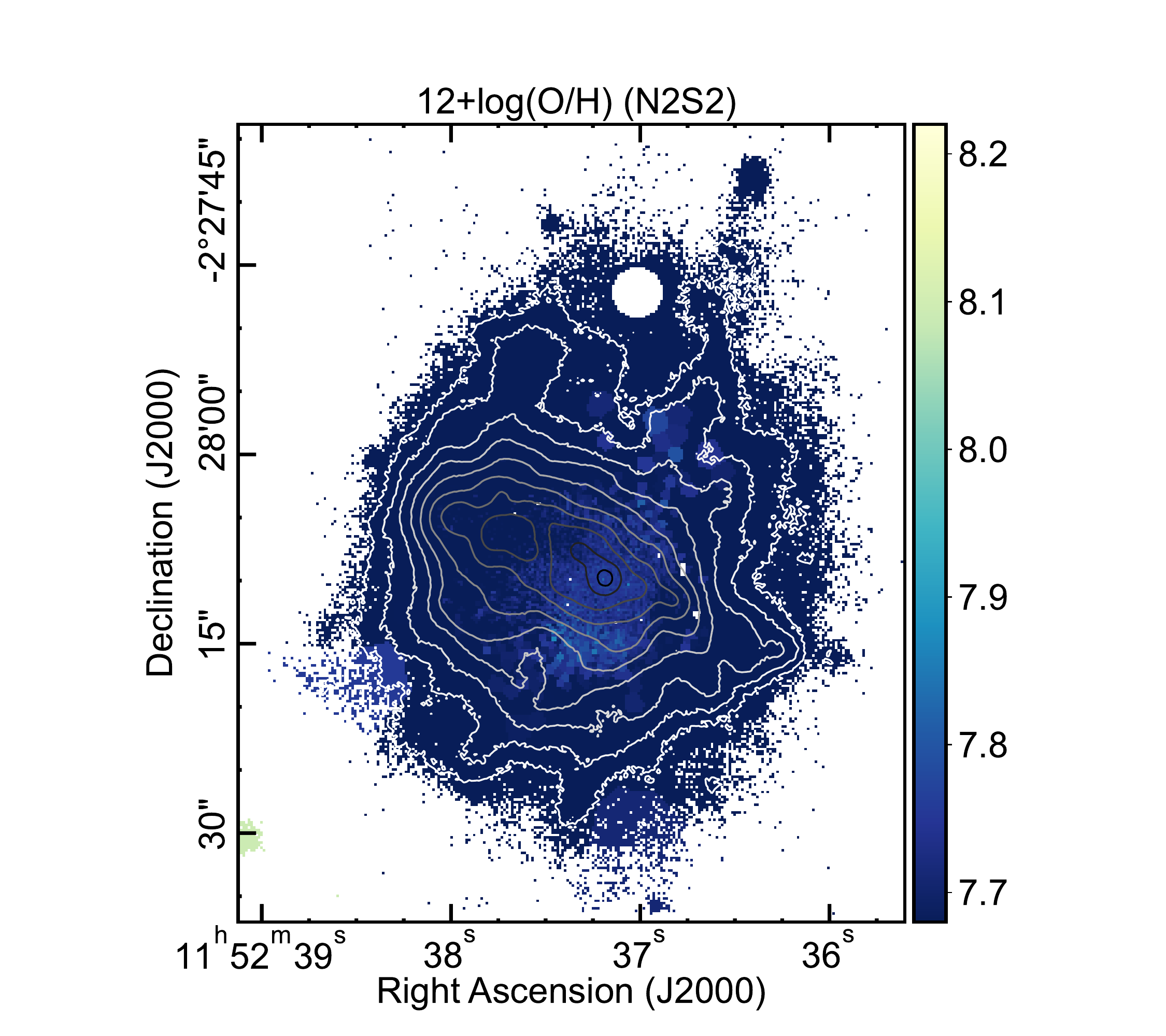}
\caption{\label{mapOtot}
Maps of total oxygen abundance derived by means of different strong line methods.
\emph{Upper row:} Calibrations by \cite{Marino13} for the N2 (\emph{left}) and the O3N2 (\emph{right}) methods. 
\emph{Middle row:} Calibrations by  \cite{Stasinska06} for the  S3O3  (\emph{left})  and Ar3O3  (\emph{right})  methods.
\emph{Lower row:} S calibration by \citet{Pilyugin16}   (\emph{left})  and N2S2 calibration by  \cite{Dopita16}  (\emph{right}).
}
\end{figure}

 \begin{figure}[ht]
 \centering
 \includegraphics[angle=0,  trim=0 0 0 0 0, width=0.24\textwidth, clip=,]{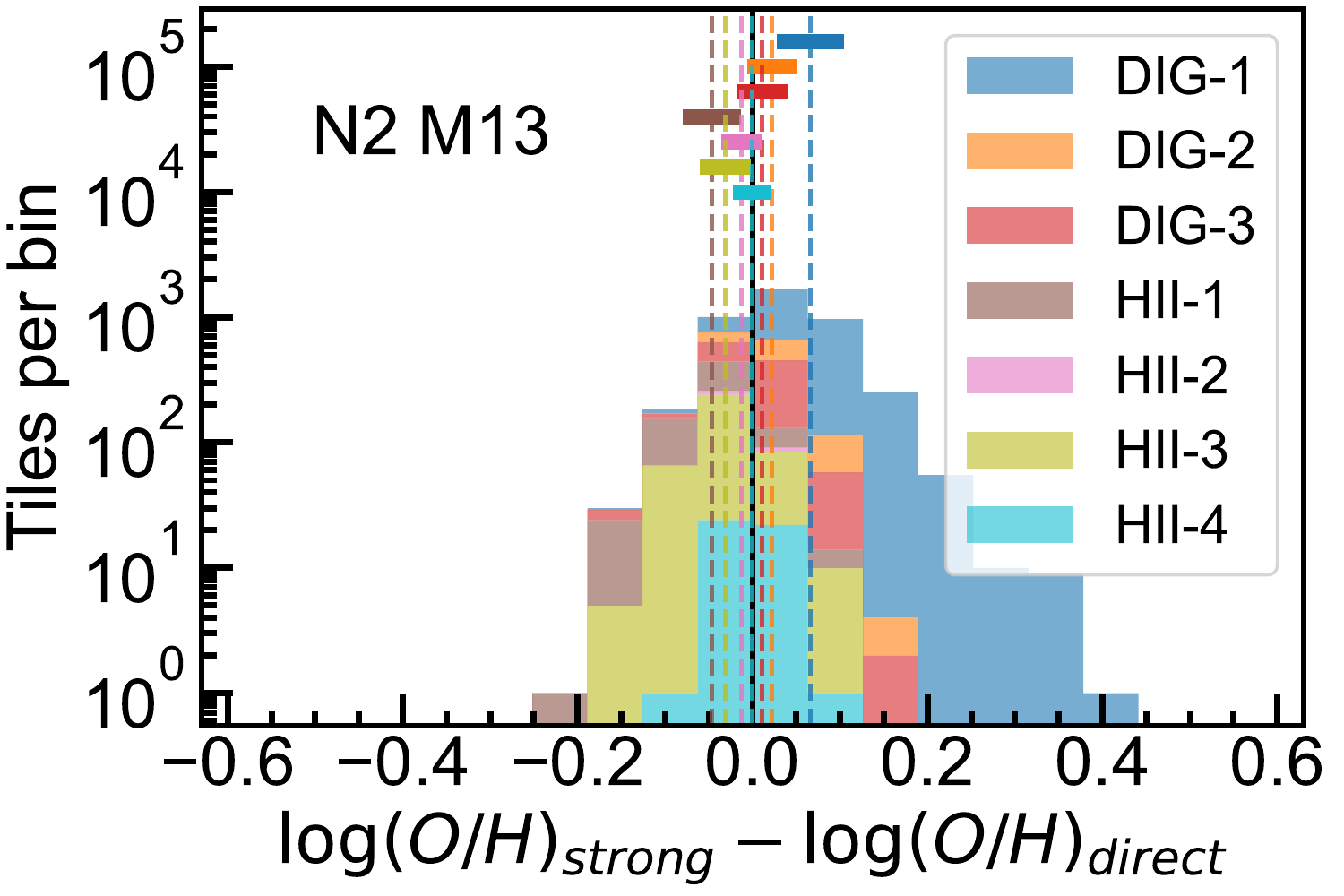}
 \includegraphics[angle=0,  trim=0 0 0 0 0, width=0.24\textwidth, clip=,]{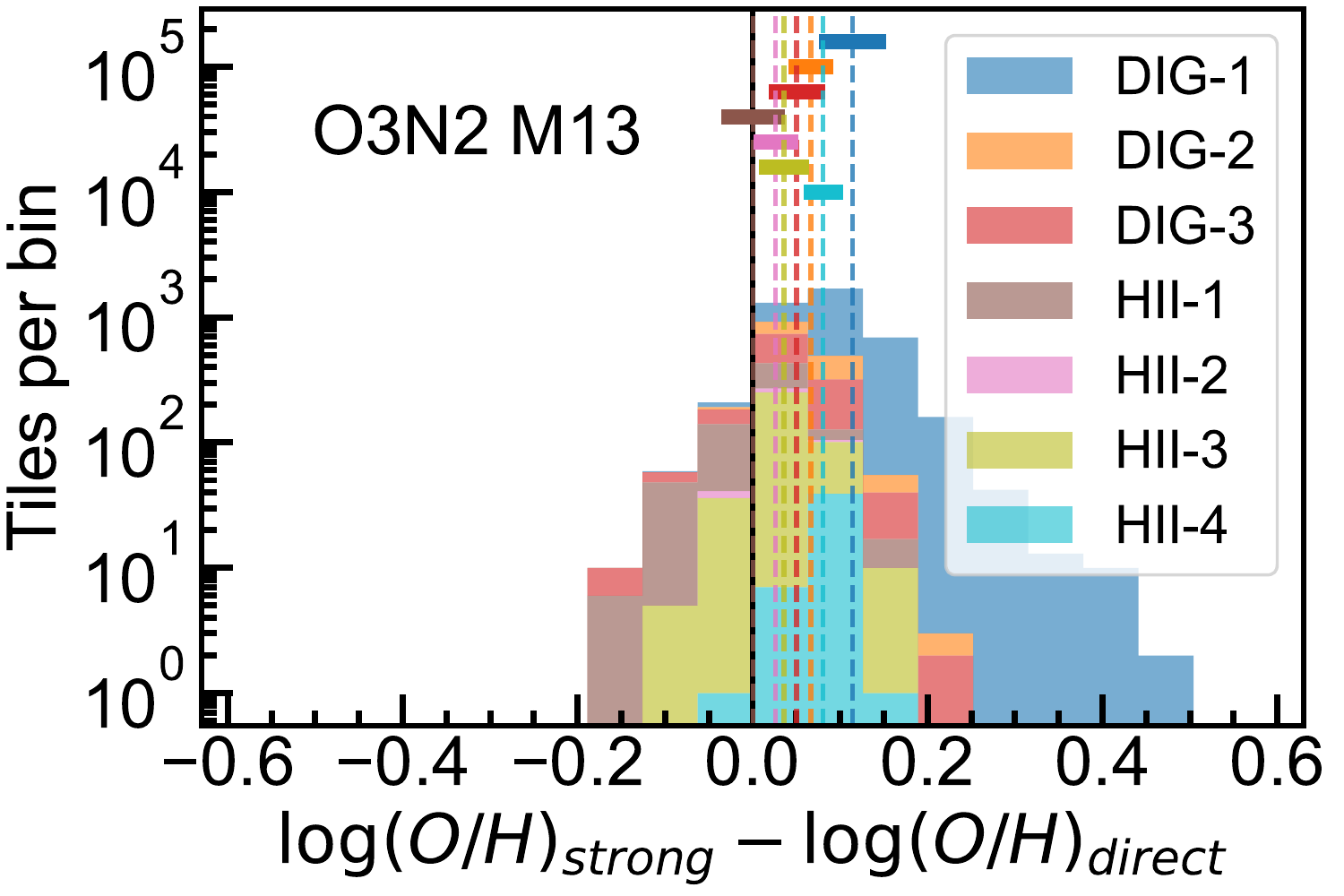}

 \includegraphics[angle=0,  trim=0 0 0 0 0, width=0.24\textwidth, clip=,]{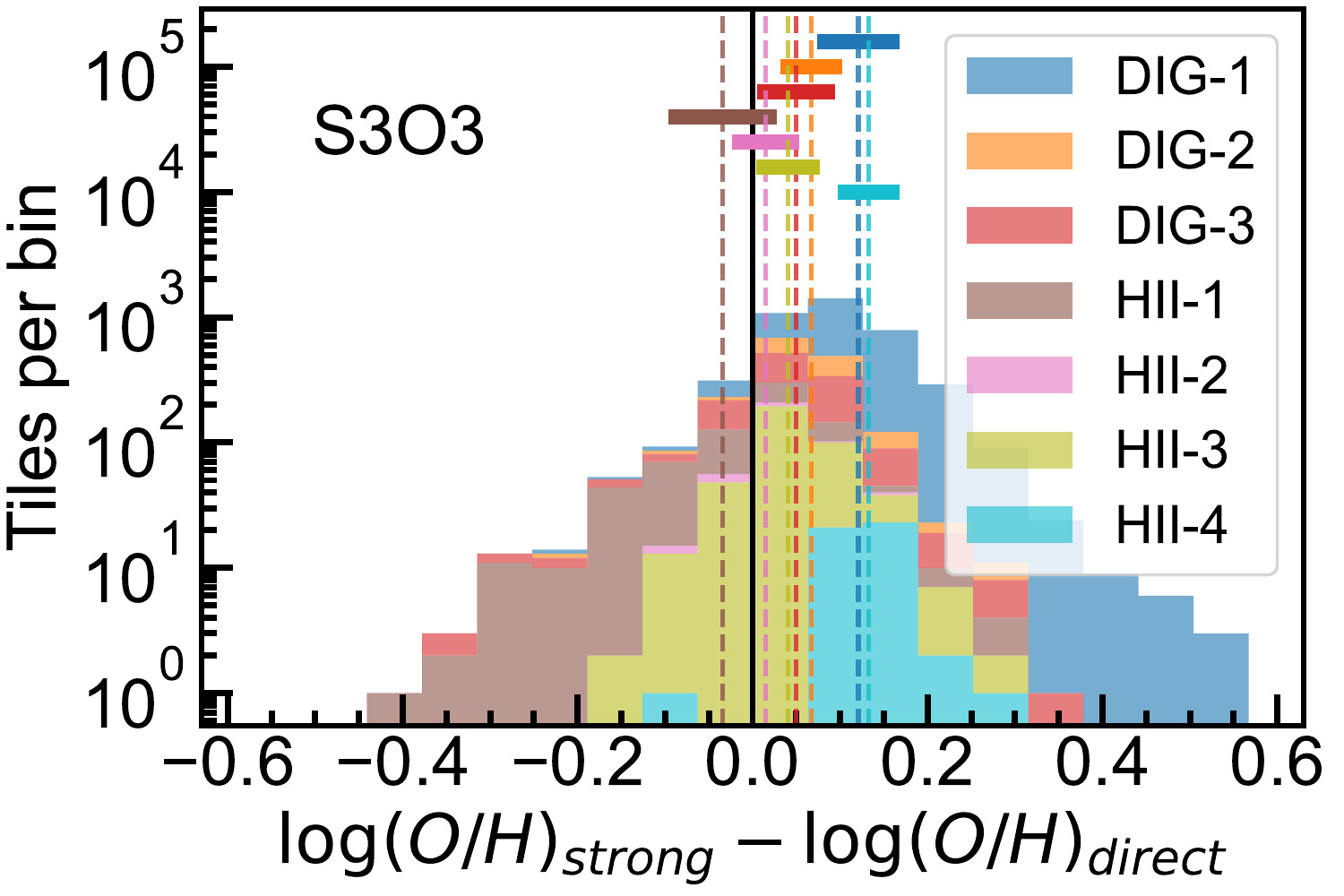}
 \includegraphics[angle=0,  trim=0 0 0 0 0, width=0.24\textwidth, clip=,]{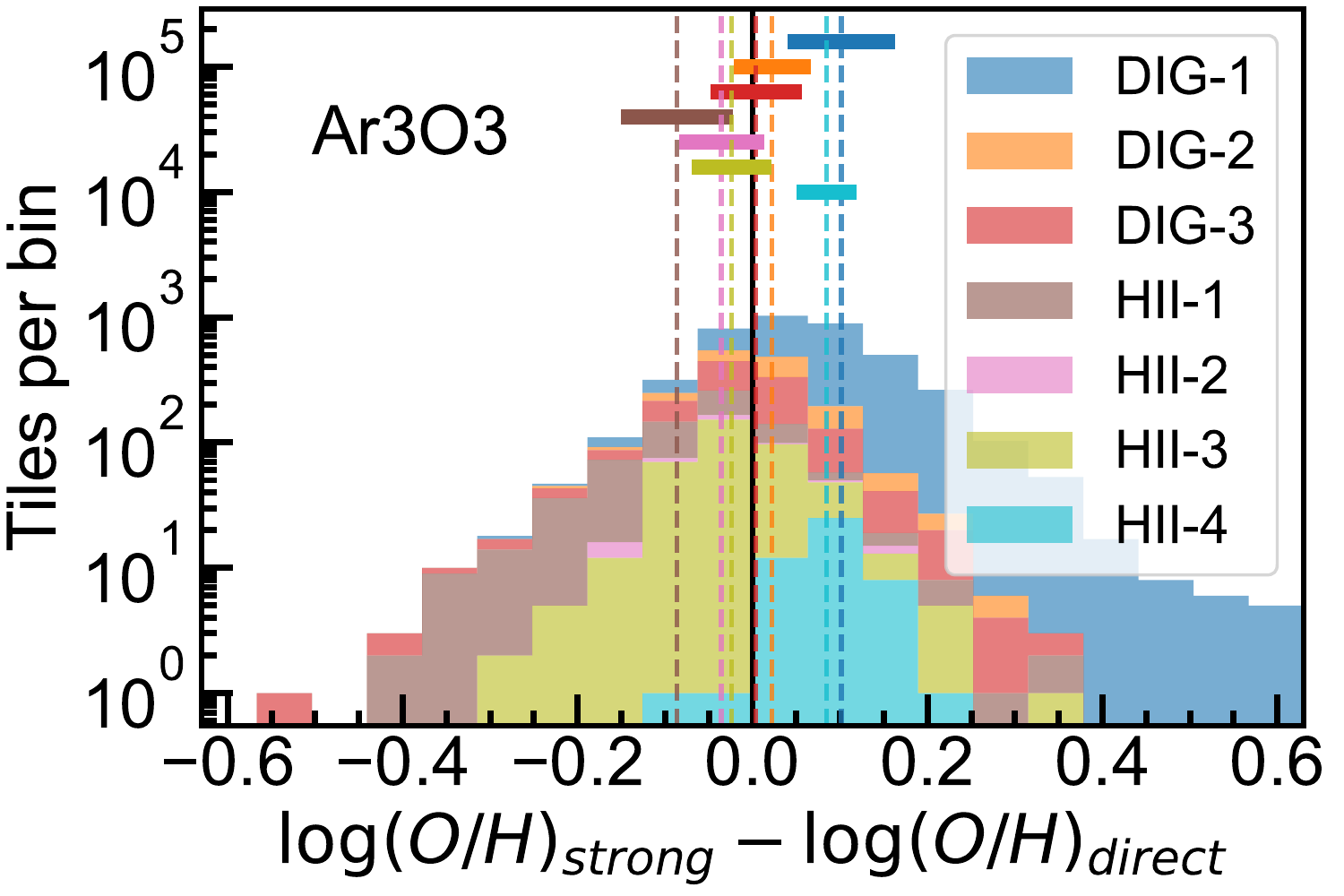}

 \includegraphics[angle=0,  trim=0 0 0 0 0, width=0.24\textwidth, clip=,]{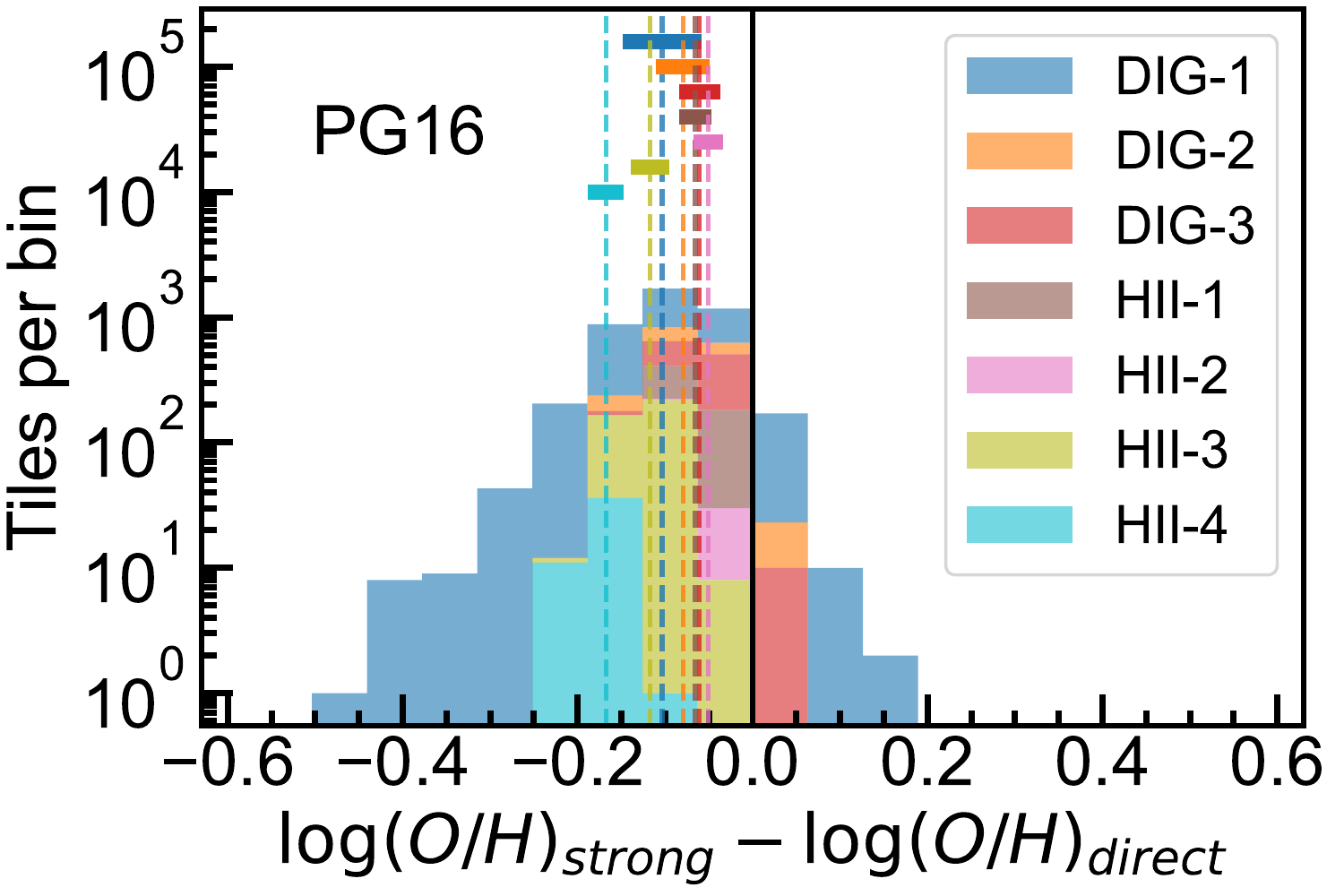}
 \includegraphics[angle=0,  trim=0 0 0 0 0, width=0.24\textwidth, clip=,]{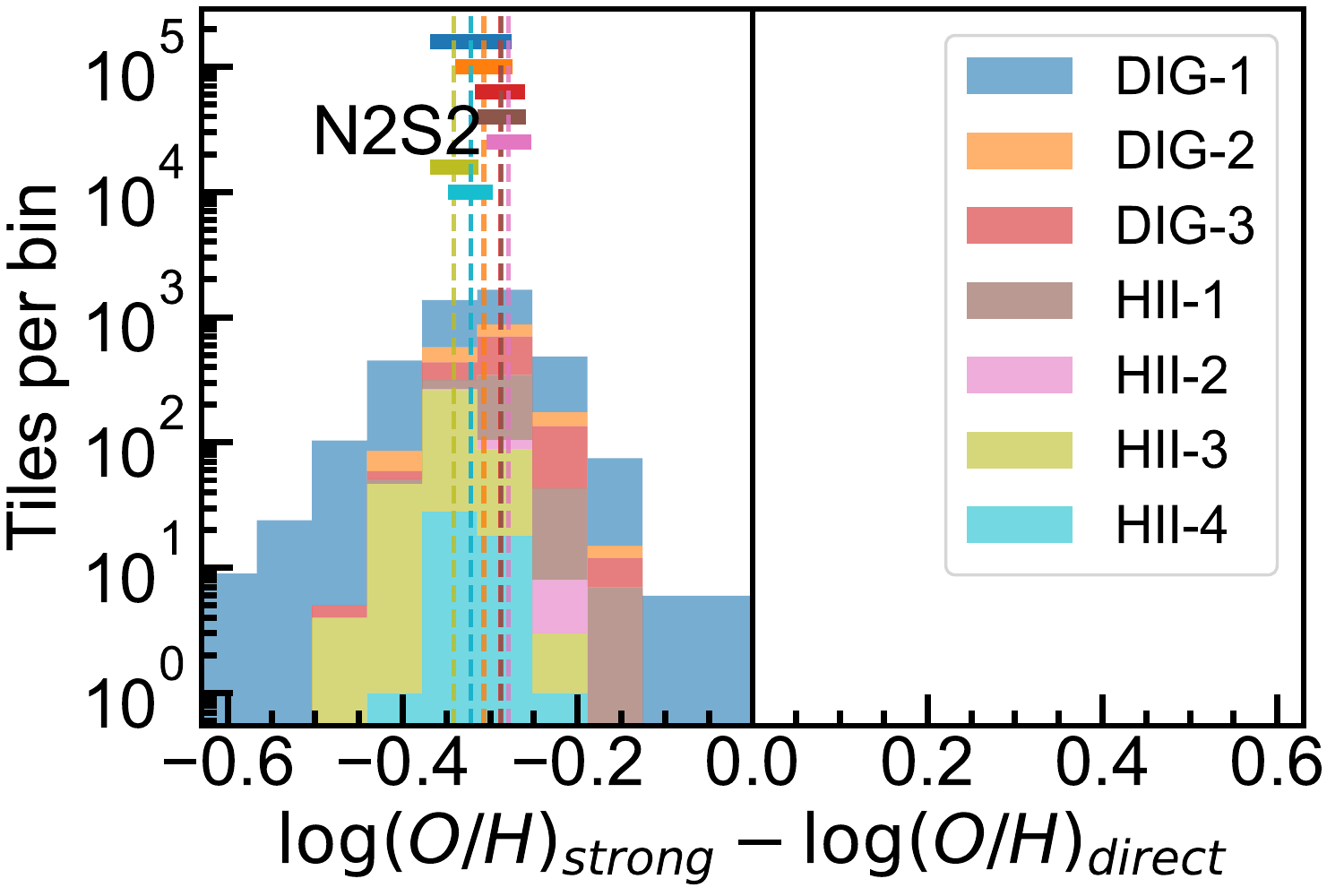}

\caption{\label{diffOhist}
Histograms with the difference between the oxygen abundance, 12+$\log$(O/H), as derived from different strong line methods and the direct method, in the different regions defined with \texttt{astrodendro}. Vertical dashed lines denote the mean of a given distribution while the length of thick horizontal bands is the standard deviation. The order of the rows and columns is the same as in Fig. \ref{mapOtot}.
}
\end{figure}

\section{Discussion \label{secdiscussion}}

\subsection{Strong line methods to derive oxygen abundance \label{secstronlinemethods}}

Deriving abundances by means of the direct method in extragalactic sources is certainly challenging, since it needs a determination of $T_e$ which, in turns, depends on faint auroral lines (e.g. here \nii$\lambda$5755 or \siii$\lambda$6312), specially in low surface brightness diffuse component of the ionised gas. Instead, it is more common to use certain combinations of strong emission lines with more or less well established empirical and/or theoretical calibrations: the so-called strong line methods. Several of the lines involved on these methods are observed all over the galaxy. Thus, these data constitute a good opportunity to assess their performance.

One of the most used indexes is N2=$\log$(\nii$\lambda$6584/\ha), first proposed by  \citet{StorchiBergmann94}. It seems handy for studies at high redshift, since it involves only two lines that are very close in wavelength. However, it should be handled with care since it is sensitive to relative atomic (nitrogen) abundances, in particular at low metallicity \citep{MoralesLuis14}.
Another common index is the O3N2=$\log$((\oiii$\lambda$5007/\hb)/(\nii$\lambda$6584/\ha)), proposed by \citet{Alloin79}. At least in very nearby \hii\, regions, it seems less sensitive to variations in the ionisation parameter \citep[e.g.][]{MonrealIbero11} but it  involves a larger number of lines spread over a wider spectral range. 
The oxygen abundance maps for these indexes using the calibration proposed by \citet{Marino13} are presented in the upper row of Fig. \ref{mapOtot}. 
Similar maps are obtained with other calibrations \citep[e.g.][]{Bian18,PerezMontero21}.

Besides, \citet{Stasinska06} proposed two additional indexes that behaved well when determining the overall metallicity of the galaxy (Ar3O3=$\log$(\ariii$\lambda$7135/\oiii$\lambda$5007) and S3O3=$\log$(\siii$\lambda$9069/\oiii$\lambda$5007)),  even if their uncertainty is somewhat larger ($\sim$0.3~dex).
The oxygen abundance maps using these indexes are presented in the central row of Fig. \ref{mapOtot}.
All the four maps are presented with a scale covering the same range as the oxygen abundance map in Fig. \ref{mapOionic}, derived through the direct method, to make easier the comparison between these figures.
Given the strength of the involved lines, uncertainties for these calibrators are dominated by the calibration method at play, not by the line measurements and are within the range of $\sim0.05-0.30$~dex, depending on the index.

Overall, the metallicity distribution in all these four maps is radically different to the one presented in Fig. \ref{mapOionic}. While the galaxy is seen as presenting an homogenous abundance in oxygen when the direct method is used, it would have been interpreted as chemically inhomogeneous (to a different degree) if any of these four strong line methods had been used.

Assuming the metallicity derived from the direct method as the 'true metallicity', we can evaluate the accuracy of each of these methods  in the different parts of the galaxy by comparing the metallicity as derived from a strong line method against the one from the direct method. This is shown as histograms in Fig. \ref{diffOhist}. In general, all the four methods tested here predict reasonably well the metallicities in the \hii\, regions. This is not surprising since they were calibrated using \hii\, regions (or at the very least, whole galaxies dominated by the emission of their \hii\, regions). However, there are nuances. The dispersion is clearly lower for the N2 and O3N2 indexes than for the S3O3 and Ar3O3 ones. In particular, these two last indexes overestimate the metallicity in \hii-4 by $\gtrsim$0.1~dex.
This discrepancy is even stronger when looking at the tiles corresponding to the DIG apertures, in particular DIG-1, that traces the outermost part of the galaxy. On average, all the ratios predict abundances $\gtrsim$0.1~dex larger in DIG-1, but the further from the main site of SF one goes, the larger the difference becomes, reaching values of $\gtrsim0.2$~dex in the N2 and O3N2 methods, and  $\gtrsim0.3$~dex in the S3O3 and Ar3O3 in a significant area of the galaxy.

The underlying cause of these discrepancies can be explained by the same effect, even if the impact depends on the ratio one is looking at.
ISM in the DIG is expected to be dominated by low excitation lines, thus it is not surprising that those indexes involving only high excitation lines (i.e., the S3O3 and Ar3O3, with ionisation potentials of 23.34 for S$^{++}$ eV,  27.63 eV for Ar$^{++}$,  and 35.12 eV for O$^{++}$) do not adequately trace the metallicity there. That metallicity is preferentially overestimated can be understood by the fact that O$^{++}$ has larger ionisation potential than S$^{++}$ or Ar$^{++}$. Of course, when looking along the line of sight of a spaxel in the DIG, contribution of the different zones  (using Garnett's nomenclature) with the whole range of ionisation degree is expected. However, the relative contribution of the high-ionisation (and better traced by O$^{++}$ over S$^{++}$ or Ar$^{++}$) zone is expected to be less important in the DIG, implying higher S3O3 or Ar3O3 indexes that would translate in higher metallicities.

Something similar (but more subtle since also a low excitation line is involved, \nii$\lambda$6584) seems at play in the other two indexes. Here, the discrepancies between the direct and the strong line methods, when looking at the DIG are smaller ($\lesssim$0.1 dex). In the particular calibrations tested here the N2 index provides metallicities in slightly better agreement than the O3N2 index.

The last rows in Figs. \ref{mapOtot}  and \ref{diffOhist} contains the results for two additional indexes using different set of lines.
On the left column on that row, we present the results for the $S$ calibration as proposed by \citet{Pilyugin16}, that makes use of a combination of oxygen, sulphur and nitrogen lines. This index was conceived to take into account the ionisation parameter, and it is, de facto, the one that finds the most uniform metallicity distribution of the indexes tested here, with differences $<$0.2~dex between \hii-1 and \hii-4 and comparable metallicities in the \hii\, region and the surrounding DIG. However, it seems to globally underestimate the metallicity by $\sim$0.1~dex.
This may be attributed to the somewhat low relative nitrogen-to-oxygen abundance, that is is at the lower end of the $\log$(N/O) values used by  \citet{Pilyugin16} to calibrate this index.

Finally,  on the right column on that row, we present the results for an index proposed by  \citet{Dopita16}, and that makes use of  a combination of the \nii, \sii, and \ha\, emission lines (here referred as N2S2 index). This was, in principle, a handy choice of lines for high redshift studies since they fall within a short spectral range ($<300$~\AA), thus being independent of any extinction correction and relatively easy to observe with a single observational set-up. However, this is the index that clearly perform the worst of all those tested here, underestimating the metallicity by $\sim$0.4~dex. This index was calibrated using photoionisation models. Large discrepancies between the abundance values produced by empirical and theoretical calibrations of strong line methods are well acknowledged in the literature \citep[e.g.][]{Kewley08}, and in general, an empirical metallicity scale is preferable \citep{Pilyugin16}. The experiment presented here with this one galaxy supports this statement. A possible origin for this discrepancy can be the relative abundances assumed for the calibration of N2S2 index. At the oxygen abundance  for \object{UM\,462}, a $\log(N/O)\sim-1.4$ was assumed for the calibration, larger by $\sim$0.3~dex than the   $\log(N/O)$ measured here.

Having a reliable diagnostic to determine metallicities at any point of a galaxy is essential to identify chemical inhomogeneities that can be associated with in and outflows of material. The choice of the index, systematics of the specific calibration, the characteristics of the target under study (i.e. \hii\, regions or DIG, moderate star-forming galaxy or extreme starburst), and the level of contrast one wants to achieve when detecting chemical inhomogeneities, are all aspects that should be taken into consideration.
Ultimately, this has an impact on our understanding of the mechanisms that regulate the processes of galaxy enrichment and chemical evolution.
The exercise presented here, involving a single galaxy and a single metallicity, illustrates that there is still some way to go. A comparison between the direct method and the $S$ calibration in \object{Haro~11} points towards the same direction \citep{Menacho21}. In that sense, a similar exercise on a sample of galaxies covering a range of metallicities seems imperative. In parallel, new techniques that do not rely on the detection of auroral lines to obtain an estimate of the electronic temperature reveal themselves as  promising alternative \citep[e.g.][]{Kreckel22,Fernandez23}.

\subsection{A supernova remnant candidate \label{secsnr}}

In Sect.  \ref{secpointsource}, we measured the intrinsic line ratios and line width of a local maximun in the \sii/\ha\, and \oi/\ha\, maps. Here, we discuss the nature of the source causing this emission. There are several mechanisms that can produces line ratios departing from the typical values expected from \hii\, regions, and with enhanced forbidden line ratios.
Because of its point-like nature, we can reject those usually associated with extended DIG emission like post-AGB stars \citep{Binette94}.
Likewise, we can also discard structures larger than $\sim$150~pc like large bubbles and shells.
Besides, an active galactic nucleus can be rejected right away because of its non-nuclear nature.
Possible point-like causes for the observed line ratio are planetary nebulae (PNe) and Supernova remnants (SNRs), both of stellar origin or,  being more specific, the corpse left after the death of different kinds of stars. 
PNe have been detected with MUSE in galaxies at distances similar to that of \object{UM\,462}, and even beyond \citep{Kreckel17,Roth21}, and do have strong forbidden lines. However, this applies to both, high and low ionisation lines. Looking at  the original BPT diagrams, PNe typically have $\log$(\nii/\ha)$\sim$0.0-0.5~dex and $\log$(\oi/\ha)$\sim$-0.5-0.0~dex, much larger than the values we measured for our source, but  admittedly in the same direction.
However, the most prominent optical line in a PN is usually \oiii$\lambda$5007. In the original BPT diagrams, PNe typically occupied the locus with $\log$(\oiii/\hb)$\gtrsim$1, larger than any ratio we measure at any place in \object{UM\,462}. This is even more true in the emission attributed to the source under discussion.

A last possibility is that  this emission could be caused by a SNR. This, in principle, should be detectable with MUSE data at these distances \citep{CidFernandes21}.
The expected size of a SNR is $\sim$40 pc \citep{Roth18}, thus compatible with an unresolved source at the distance of \object{UM 462}. Historically, a $\log$(\sii$\lambda\lambda$6716,6731/\ha) = $-$0.4 have been considered the canonical  lower limit for a typical SNR spectrum \citep{Mathewson73,Fesen85}.  
We measured a $\log$(\sii$\lambda\lambda$6716,6731/\ha)$\sim-0.49$~dex. Thus, strictly speaking, our line ratios do not comply with this criterion, although they are quite close.  However, this criterion is based on only one ratio
and seems too strict. As discussed by \citet{Kopsacheili20}, it might miss SNRs with relatively low shock velocities, characteristic of e.g. older
SNRs.
These authors derived additional criteria involving sets of two or three line ratios by means of machine learning techniques. Our pair of  $\log$(\sii$\lambda\lambda$6716,6731/\ha) and  $\log$(\oi$\lambda$6300/\ha) are compatible with the criterion provided by these authors. Besides, assuming a relation between the \oii$\lambda\lambda$7320,7331 and \oii$\lambda\lambda$3726,3729 as in Sect.  \ref{secbpt}, our line ratios also comply with the criterion proposed by \citet{Fesen85} based on the \hb, \oii$\lambda\lambda$3726,3729, and \oi$\lambda\lambda$6300,6363 emission lines.
Our  \nii$\lambda\lambda$6548,6584/\ha\, line ratio is somewhat lower than the values reported in these works. However, metallicity may play a role here. 
\citet{Payne07,Payne08} present  \nii$\lambda\lambda$6548,6584/\ha,  \sii$\lambda\lambda$6716,6731/\ha,  \oi$\lambda\lambda$6300,6363/\ha\,   line ratios for a sample of SNR in the Magellanic Clouds, the closest star-forming galaxies with metallicities similar to that of UM\,462. Their ratios compare well with those measured here. 
Besides,  we measure a  ratio of the two sulphur lines of 1.35$\pm$0.02, comparable to the line ratios measured for SNR in the Magellanic Clouds \citep{Payne07,Payne08}. The corresponding density, as derived in Sect.  \ref{secphysprop}  is $n_e(\sii)=100\pm^{20}_{15}$~cm$^{-3}$, smaller but comparable to the median $n_e(\sii)$ of 530~cm$^{-3}$ reported by \citet{Payne08}.  We could also estimate an electron temperature from the sulphur lines of $T_e$=11\,000$_{+2900}^{-3500}$~K, somewhat lower than the one determined for the environmental ISM.
All in all,  the most plausible of all the ionisation mechanism considered to explain the emission at RA(J2000) = 11:52:37.8 and DEC(J2000)=-02:28:03.0  is a SNR. Interestingly enough, such a SNR, just at the outskirts but not in the regions \hii-1 - \ldots - \hii-4 supports the scenario suggested  in Sect. \ref{secgenoverview} where the most recent SF propagates from the outer to the inner parts of the galaxy (and then from east to west).

\subsection{The presence of Wolf-Rayet stars in UM~462}

In Sect. \ref{secstars}, we presented an overview of the stellar populations and saw how the \texttt{FADO} modelling suggested that the most recent SF propagates from east to west.
An additional tracer of the youngest stellar populations is the presence of spectral features characteristics of Wolf-Rayet (WR) stars. These are massive (M$\gtrsim$25~M$_\odot$) stars at certain short-duration stages after abandoning the main sequence \citep[see][for a review]{Crowther07}
and can be used as a precise clock to estimate the age of young stellar population.

   \begin{figure}[th]
   \centering
   \includegraphics[angle=0,trim=80 12 60 35, width=0.35\textwidth, clip=,]{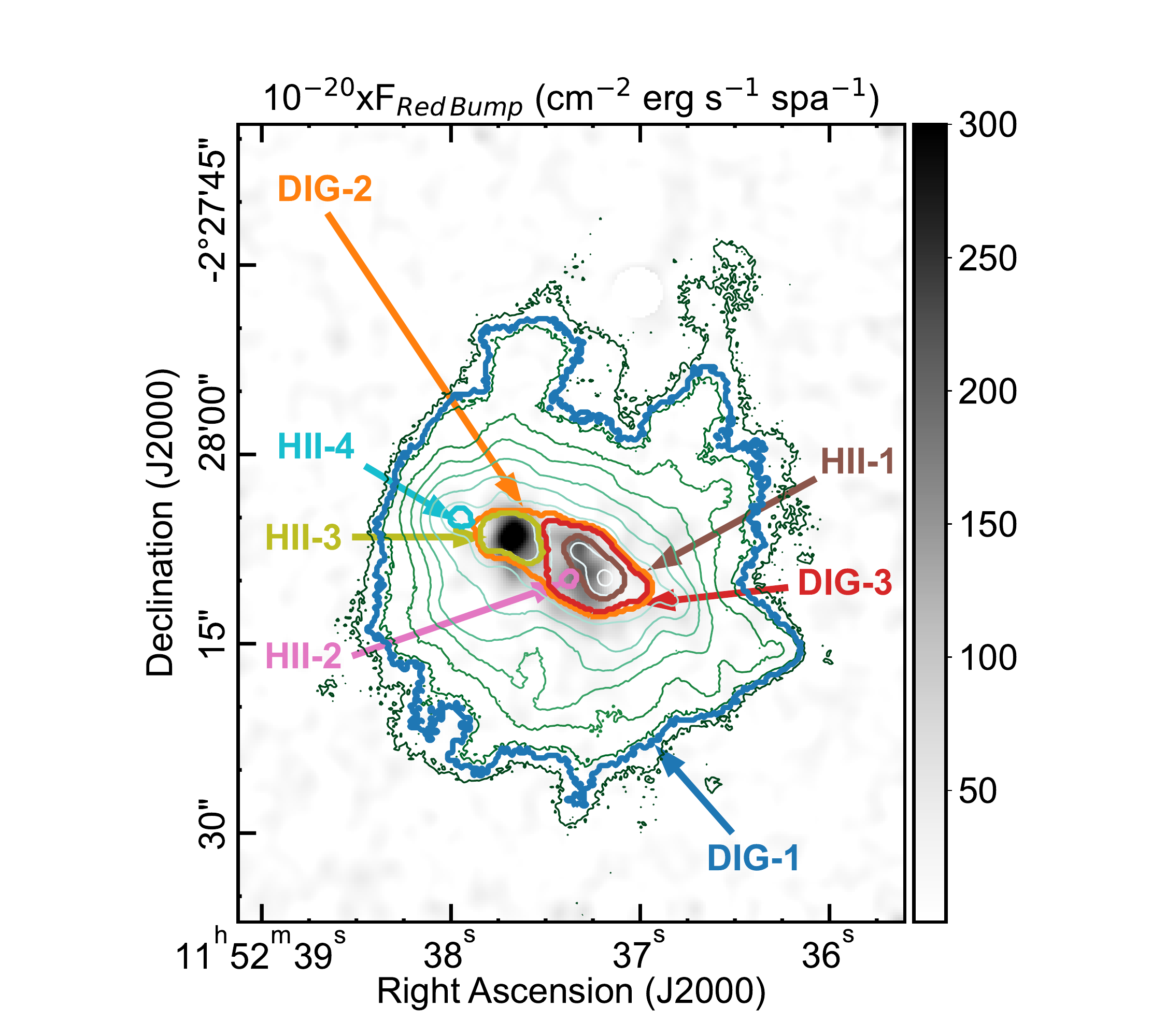}
      \caption{
   Map showing the location with excess of emission at the wavelengths corresponding to the red bump ($5760-5860$~\AA, rest-frame). For reference, 
   the map in \ha\, flux made by line fitting on a spaxel-by-spaxel basis is overplotted with ten evenly spaced contours (in logarithmic scale) ranging from 1.26$\times$10$^{-18}$~erg~cm$^{-2}$~s$^{-1}$~spaxel$^{-1}$ to 1.26$\times$10$^{-15}$~erg~cm$^{-2}$~s$^{-1}$~spaxel$^{-1}$. Likewise, the different regions and areas discussed along the work are marked with the same colour and line code as in Fig. \ref{astrodendrostruc}.
   North is up and east towards the left. 
 }
    \label{figwr}
   \end{figure}
   
\begin{figure}[h]  
\centering
   \includegraphics[angle=0,trim=0 0 0 0, width=0.32\textwidth, clip=,]{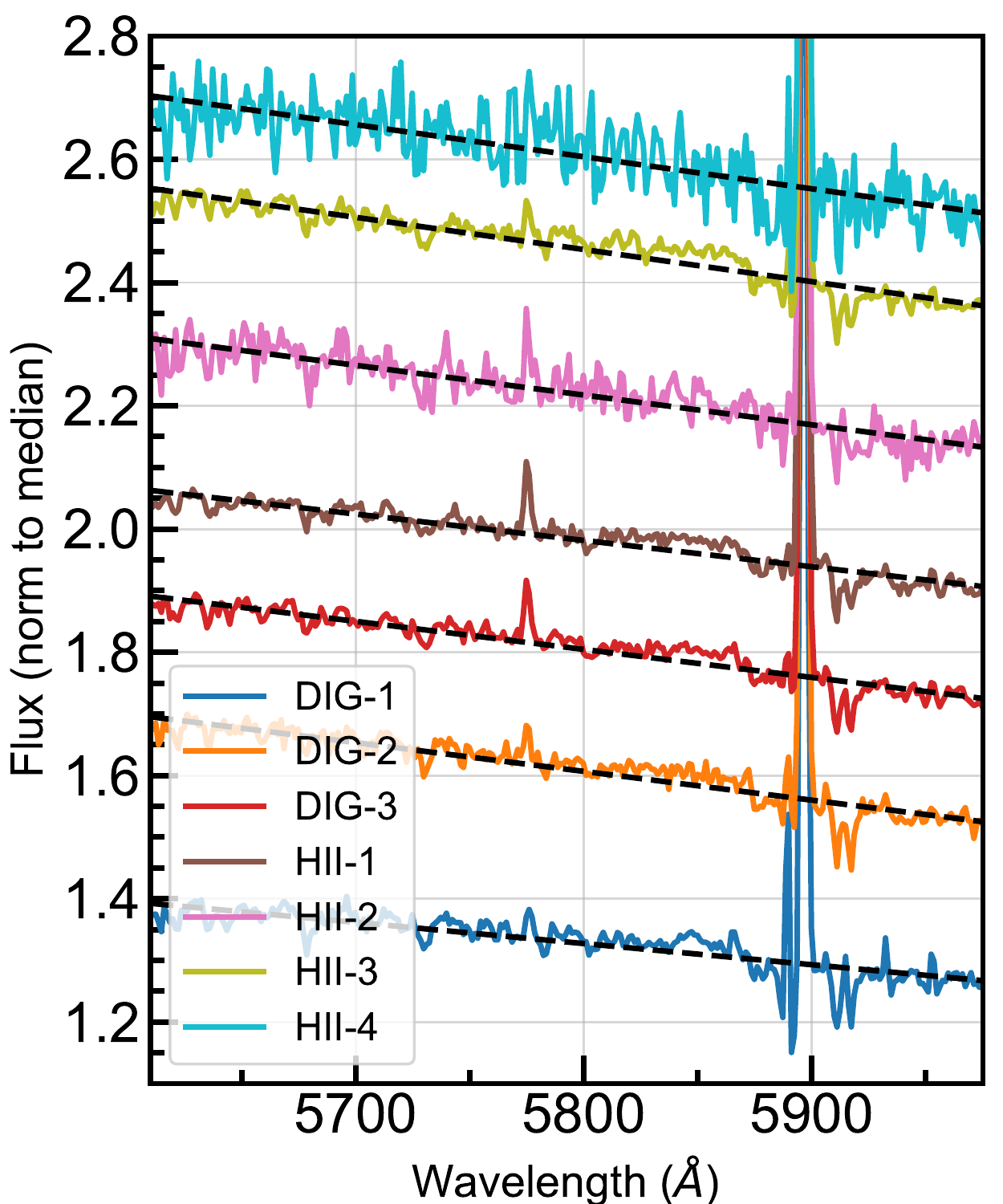}
         \caption{ Zoom of the same spectra displayed in Fig. \ref{spectraregions} in the spectral range corresponding to the red bump. To guide the eye, a linear fit to the local continuum is also displayed with a black dashed line for each spectrum.
 }
   \label{figwrspec}
    \end{figure}

The taxonomy of WR stars is rather complex and relies on the characteristics (presence, relative strength and shape) of certain emission features in the stellar spectra.
When studying unresolved stellar populations, the classification is simplified to the maximum, and usually only two main classes are considered: nitrogen WR stars (WN stars, with strong helium and nitrogen lines),  where the star shows the products of the CNO burning, and carbon (or oxygen) stars (WC and WO stars, with
strong helium, and carbon or oxygen lines), where the star shows the products of He burning.
During its life, a given massive star may or may not pass by some of the stages associated to the different types of WR stars, depending on a variety of aspects such as its initial mass, metallicity, amount of rotation or whether it is an isolated star or in a binary system \citep{Crowther07}.
Both, binarity and rotation, favour the appearance of WR features, lowering the lower mass limit for an star to go through the WR phases  and making them last longer \citep[e.g.][]{Meynet05,Eldridge08,Georgy12}.

Spectral features characteristics of WR stars are fainter at lower metallicities \citep{Hadfield07} but by no means inexistent. The most important WR features used in extragalactic astronomy are a bump around 4650~\AA, i.e. the blue bump, which is mainly but not always characteristic of WN stars, and a fainter bump around 5808~\AA, i.e. the red bump, characteristic of WC stars. There are plenty of examples for detection of these features in BCDs or galaxies with metallicities similar to that of \object{UM\,462} \citep[e.g.][]{Guseva00,LopezSanchez10}.

Given the youth of the most recent burst of SF, one would expect the presence of such features in  \object{UM\,462}. However, this has not firmly proven so far. On the one hand, based on the detection of a broad \heii$\lambda$4686 emission line, as well as the detection of the Si\,\textsc{iii}$\lambda$4565 and the \hei/N\,\textsc{ii}$\lambda$5047 spectral features,  \citet{Guseva00} suspect the presence of WR in the galaxy. On the other hand, \citet{James10} did not find evidence of those features in their data, even if they used several extraction apertures in an attempt to minimise the dilution of the WR features by the stellar continuum.

The strongest of the WR features, the blue bump, is not covered with the present data. Nonetheless, we can contribute to clarify this issue by looking for the red bump. To do so, we pinpointed the WR emission following a similar methodology as that described by \citet{MonrealIbero17b}. Basically, we created a map for the red bump emission by simulating the action of narrow tunable filters at the bands proposed by \citet{Brinchmann08}, corrected for the redshift of the galaxy, and applied those to the Galactic extinction corrected data cube.
We did not convolve the resulting image with a Gaussian filter since nothing was gained with that because of the poor seeing of the data.

The resulting map is presented in  Fig.~\ref{figwr}. There is a clear flux excess at the location of aperture \hii-3, as well as in some areas of \hii-1, and DIG-3. Additionally, there might be some marginal flux excess in the part of DIG-1 closest to the centre of the galaxy. A zoom of the spectra presented in Fig.~\ref{spectraregions} in the spectral range of the red bump is presented in Fig.~\ref{figwrspec}. To guide the reader, we added a 1-degree polynomial fit to the local continuum in each spectrum. The bump is clearly seen in apertures \hii-3, \hii-1, and DIG-3. Besides, there may be a marginal detection in DIG-2 and DIG-1. Thus, these data support the presence of WR stars in the galaxy. Because of the poor seeing and the inaccessibility  of the blue bump, we did not attempt to further isolate the WR emission and estimate the number of WR stars.

A point worth mentioning is that most of the red bump emission is not associated with the youngest burst of SF, \hii-1, but to the somewhat older \hii-3. Differences in the distribution of the blue and red bumps emission have been used in the past to trace the propagation of the most recent  SF within a galaxy with the youngest stellar populations at those places where only the blue bump is detected, then going towards somewhat older stellar populations in those places where both bumps (or only the red bump) are detected \citep{Westmoquette13,MonrealIbero17b,Gunawardhana20}. In that sense, the red bump detection together with the results presented in Sect. \ref{secgenoverview} and \ref{secSB} motivate us to hypothesise  the presence of WR stars, and in particular WN stars, in \hii-1, and claim loudly for 2D spectroscopic data at high spatial resolution to confirm (or reject) that hypothesis. That is: mapping the blue bump in the centre of \object{UM\,462}, and carrying out a proper census of the WR population in the galaxy would allow to strength (or reject) the scenario of east-west propagation of the SF.

\subsection{The horns, the beard and the story behind them\label{sechornsandbeard}}

In Sect. \ref{secbpt}, we introduced  two $\sim$1~kpc-long  structures towards the north of the galaxy, clearly delineated in the maps with ratios involved in the BPT diagram. We nicknamed them 'the horns'. Similarly, we identified a closed, almost loop-like structure in Sect. \ref{secEW}. We named it  'the beard'. Here, we discuss the origin of both structures based on the join observational evidence collected through this work.

\paragraph{The horns:}
With an angle between them of $\sim$25$^{\circ}$, these two structures frame the only area in the galaxy with velocity dispersions clearly high enough to be measured at the MUSE spectral resolution and displaying velocity stratification. They have relatively high $EW$(\ha), thus suggesting larger concentration of ionised gas, and line ratios tracing relatively high ionisation parameter (even if not as high as the main site of SF).

We posit here that this set of structures represents the effects of the stellar feedback in the galaxy at a relatively advanced stage, as  the 2D projection (in the plane of the sky) of a fragmented super-bubble, where the galaxy has already lost part of its ISM. The wall of such a fragmented super-bubble is seen edge-on at the location of the horns, while in between the horns, we see the integrated emission of the back and front side of the wall, probably dominated by the emission of the front side.

This would be consistent with higher velocity dispersions just in the middle point between the horns then decreasing when moving in the E-W direction towards the horns, exactly as observed. In this scheme, one would expect higher ionisation species being preferentially found in the inner side of the wall, while lower ionisation species would  preferentially be located in the outer side of the wall. Under the assumption that the emission in between the horns is dominated by the front side of this wall, the kinematics, with redder velocities for \oiii$\lambda$5007 and bluer velocities for \oi$\lambda$6300, supports this picture.

The scenario presented here is only sketched but already allows to make some predictions that can be used to plan spectroscopic observations at high spectral resolution, able to reinforce, refine, or refute it.
These would not need to map the whole area but should sample, at least, the horns and the area in between at one or two locations, and provide information for at least one hydrogen recombination line, one line tracing high excitation (e.g. \oiii) and one line tracing low excitation (e.g. \nii, \oi, or \sii). For the scenario proposed here to be valid, such observations should deliver line splitting, with larger split in the middle point between the horns, and decreasing when moving towards the horns, where there should be no splitting at all. Likewise, this splitting should be larger for the low ionisation lines than for the high ionisation lines. Finally, for a given line the blue component should be stronger than the red component.

Perhaps the closest example to the stratification found here is presented by \citet{Menacho21}, who also found kpc-scale areas in \object{Haro 11} with velocity difference between the \ha\, and  \nii\, emission lines.  These were also explained as  perhaps remnants of a kpc-scale superbubble whose break out might have created a filamentary structure.
The findings in this area of \object{UM 462}  suggest a similar mechanism at play here, where the system has managed to carve a passage through which, perhaps, ionising photons from the youngest generation of stars could escape from the galaxy.

\paragraph{The beard:} This structure presents some similarities with the horns (e.g. line ratios), but its interest lies precisely in the differences: i) it is a closed structure (as opposed to open); ii) it is somewhat smaller in size; iii) it has $EW$(\ha) values larger than the horns by a factor $\sim$2-3; iv) the area encircled by the structure have velocity dispersions just marginally above what it is possible to resolve with MUSE; v) it is in a location of the galaxy with much more atomic gas \citep{vanZee98}.

Here we propose that this could be a location where the effects of the stellar feedback are at a somewhat earlier stage than in the northern part of the galaxy. The ionised gas structure fruit of the stellar feedback is still (almost) unbroken and the (ionised) gas is still retained within the galaxy.
An observational experiment as the one designed above could also be set up for this structure. However, reinforcing, refining of refuting this proposal would be more difficult than the one for the northern part of the galaxy since velocity differences between the components are expected to be smaller, and thus a putative line splitting more difficult to detect.

When comparing the observational evidence presented here and found in GP galaxies, an outstanding difference is the lack of clear broad emission line wings in \object{UM\,462} while these seem common in the GP galaxies \citep{Amorin12}. There are three non-exclusive reasons that may explain this difference. On the one hand, if  \object{UM 462} actually represents the  faint end of the GP distribution, it might well be that the effects of the stellar feedback,  even if existing, are not as vigorous as in  GP galaxies. On the other hand, given the spatial resolution at which GP galaxies have been observed so far, the observed emission lines in these galaxies are integrated over considerable areas, and these broad components may just be the sum of the complex large scale gas motions within them.
High fidelity spatially resolved observations at high-spatial resolution of GP galaxies, now possible with the JWST, could shed light on the sort of (kinematic) substructures at place in GP galaxies.
The third possibility is that these broad components were actually present in \object{UM~462}, just challenging to be rescued. As we mentioned in Sect. \ref{seclinefitting}, both \oiii\, emission lines display a broad component at their base in the seven \texttt{astrodendro} spectra. However, this is at a level comparable to the wings of the MUSE LSF, and thus not attributable with certainty to the target. Delicate experiments as done with PMAS data for  \object{Mrk 71}/\object{NGC 2366}  \citep{Micheva17}, or by propagating the shape of the LSF (as measured from the arc lines) to the apertures under study, could shed light into this issue and constitute a nice avenue for future research.

\section{Conclusions \label{secconclusions}}

We present here a unique detailed spectroscopic study on a nearby BCD galaxy resembling the GP galaxies, \object{UM 462}. For that, we made use of MUSE data mapping most of the galaxy (i.e. an area of $\sim55^{\prime\prime}\times40^{\prime\prime}$ or $\sim$3.8~kpc$\times$2.8~kpc) with deep wavelength coverage from 4750 \AA\, to 9300 \AA. This allowed us to characterise the physical and chemical properties of the ionised gas, as well as get an overview on the properties of the stellar population in the galaxy.
Our main results are:

\begin{enumerate}

\item The \ha/\hb\, line ratio is consistent with no extinction in most of the galaxy.  The only exceptions are the main site of SF with c(\hb)$\sim$0.4 and some secondary sites of SF with c(\hb)$\sim$0.2. As expected, the overall stellar population suffers from less extinction than the ionised gas.

\item Electron densities as traced by the \sii\, emission line ratio are below the low density limit all over the galaxy but in the main site of SF, with $n_e$(\sii)$\sim$100~cm$^{-3}$. Although uncertainties are large, electron densities as traced by the \cliii\, line ratio seemed higher but again  below the low density limit for this diagnostic.

\item Electron temperatures by means of the \siii, \nii, and \hei\, lines at specific apertures were derived. Median $T_e$ decreases according to the sequence \siii$\rightarrow$\nii$\rightarrow$\hei. Furthermore, $T_e$(\siii) values are $\sim$13\,000~K, and uniform within the uncertainties over an area of $\sim20^{\prime\prime}\times8^{\prime\prime}$ ($\sim$1.4 kpc$\times$0.6kpc).  This value is comparable to the $T_e$'s  measured in GP galaxies. 

\item We derived several ionic and elemental abundances. Regarding helium, ionic abundance, He$^{+}$/H$^{+}$, ranges from $\sim7.7\times10^{-2}$ to $\sim8.2\times10^{-2}$, which is compatible with the values previously reported. Values become lower when going towards the outermost part of the galaxy, suggesting a larger contribution of neutral helium there.

\item Regarding light elements, we mapped oxygen ionic and elemental abundances by means of the direct method. Total oxygen abundances were 12+$\log$(O/H)$\sim$8.02, in agreement with the previous reported values. Within the uncertainties, there is no O/H inhomogeneity in \object{UM 462}. We also derived abundances for nitrogen, sulphur, argon, and chlorine in selected apertures, which were compatible with values available in the literature. Regarding nitrogen, we mapped its relative to oxygen abundance, N/O. We found a mild ($\lesssim$-0.2 gradient of $\log$(N/O) crossing the galaxy in the north-east to south-west direction (i.e. \hii-4 with  $\log$(N/O)$\sim-1.8\rightarrow$\hii-1 with  $\log$(N/O)$\sim-1.6$. This is of the order of the uncertainties, and thus still compatible with homogeneity in nitrogen.

\item Several strong line methods to derive metallicities were put to test by comparing their predictions with those by the direct method. In general, they performed well at the \hii\, regions and not that well at the DIG. The N2, O3N2, S3O3, and Ar3O3 indexes detected chemical inhomogeneities, to a different extent depending on the index, not present according to the direct method. The so-called S calibration provided a somewhat more homogeneous oxygen abundance, but it was lower by $\sim$0.1~dex  than the one derived by the  direct method.
The only tested index based on models, N2S2, largely underestimated the metallicity in the galaxy (i.e. by $\sim$0.4~dex).  A similar exercise on a sample of galaxies covering a range of metallicities seems imperative to identify the best metallicity diagnostic valid all over the galaxy, since this ultimately has implications on our ability to identify chemical inhomogeneities and to understand the galaxy chemical evolution.

\item We mapped several strong line ratios used for plasma diagnostic.  Those  involved in the BPT diagrams are mostly compatible with gas photoionised by massive stars. However, a systematic excess in the \oi/\ha\, ratio is found. This  suggests an additional mechanism or a specific configuration of the relative distribution of gas and stars.We note that
\nii/\ha, \sii/\ha, and \oi/\ha\, line ratios were the lowest in the main sites of SF, and they increased  when going towards the external parts of the galaxy. Furthermore, \oiii/\hb\, presented the reversed distribution. The galaxy presented two structures, referred to as 'the horns' here, towards the north with the highest \oiii/\hb\, ratios, with the exception of the main SF sites.

\item We also mapped other strong line ratios routinely used in the literature to map the ionisation structure, namely \oiii/\sii, \oiii/\oi, \oiii/\oii,\, and \siii/\sii. They closely follow the \oiii/\hb\, map, with larger ratios at the main site of SF and decreasing outwards. We note that  \oiii/\oii\, in particular can reach values of up to $\sim$4.0, which is well in the range of the Lyman break galaxies at $z\sim2-3$ and the Lyman continuum leakers. 

\item We present maps for $EW$(\ha) and  $EW$(\oiii). Unsurprisingly, both present very high values that can reach up to $\sim1500$~\AA at the \ha\,  peak. The maps allowed us to delineate the ionised gas structure in the external part of the galaxy. In particular, we identified two structures roughly symmetric to the horns, with respect to the main body of the galaxy, that bend towards each other and form a close structure, called here 'the beard'.

\item The velocity fields of the strongest lines present an overall  structure similar to the velocity field for atomic hydrogen with receding velocities in the east and approaching velocities in the west and south-west and a east-west velocity difference of $\Delta v \sim$40~km~s$^{-1}$. However, it is much more complex than simple rotation. The area between the horns presents a velocity stratification with redder velocities in the high ionisation lines and bluer velocities in the low ionisation lines. In addition, it is the only area in the galaxy with velocity dispersions clearly above the MUSE instrumental width.

\item The modelling of the stellar continuum suggests that most of the mass of the galaxy ($\sim$60-70~\%) can be attributed to a relatively old ($\gtrsim$1~Gyr) stellar population.
There is also a significant ($\sim$10-30~\% in mass) contribution of an intermediate age ($\sim$40-400~Myr) stellar population in the inner part of the galaxy. However, the luminosity of the galaxy is dominated by a young (i.e. $\sim$6~Myr) stellar population, even if it contributes only  $\lesssim$10\% to the stellar mass. The ages of these young stars in the selected apertures suggest that the most recent SF propagates from the outer to the inner parts of the galaxy, and most importantly from east to west. The \ha\, luminosity implies a total current SFR of 0.17 M$_\odot$~yr$^{-1}$.

\item We identified a point source presenting a local maximum in the \sii/\ha\, and \oi/\ha\, maps. We isolated the emission at the point source from the emission of the surrounding DIG. Given the upper limit for the physical size, its line ratios, and line widths ($\sigma\sim$50~km~s$^{-1}$), the most plausible point-like source causing this emission is a SNR.

\item We confirm the presence of WR stars in the galaxy by means of the red bump. Notably, the red bump emission does not peak at the same location as the maximum in \ha\, and where the most recent SF takes place, but towards the east, in what we named here \hii-3. Given the suggested east-to-west SF propagation, we postulate the presence of WN stars at the maximum in \ha, which may be detected by deep and high spatial resolution observations of the blue bump.

\item We discuss the horns and the beard structures and posit that they represent two snapshots of the stellar feedback in action, more advance in the case of the horns, and at somewhat earlier stage for the beard. We suggest that the so-called horns represent the walls of a fragmented super-bubble that creates a passage where, perhaps, ionising photons from the youngest generation of stars could escape from the galaxy.

\end{enumerate}

This ensemble of results exemplifies the potential of 2D detailed spectroscopic studies of dwarf star-forming galaxies at high spatial resolution  to learn about the mechanisms at work in first galaxies, as well as to understand the biases inherent to similar studies on those actual galaxies, where such a level of detail, understood as the number of spectral features available as well as spatial resolution, is unattainable. 
JWST is already obtaining integrated spectra of primeval galaxies similar to the one studied here \citep{Schaerer22,ArellanoCordova22}. In the not so distant future, thanks to its IFU mode, it will be able to map the physical and chemical properties of those same galaxies. Likewise, the cohort of extremely large telescopes will be equipped with IFS-based instruments in the near-infrared (i.e. HARMONI at the ELT \citep{Thatte20}, IRIS at the TMT \citep{Larkin16}, and GMTIFS at the GMT \citep{Sharp16}. This implies that in the not so distant future, we will be able to spectroscopically map the primeval galaxies at exquisite spatial resolution for which the present and similar works will undoubtedly constitute a key referent.

\begin{acknowledgements}
The authors want to thank Jorge S\'anchez Almeida for discussions on different aspects of the work presented here.
AMI also thanks Bego\~na Garc\'{\i}a-Lorenzo  and acknowledges  the Spanish Ministry of Science, Innovation and Universities (MCIU), and the Agencia Estatal de Investigación (project PID2019-107010GB-100) for the support during the initial phases of this project.
%
PMW gratefully acknowledges support by the BMBF through the ErUM programme
(project VLT-BlueMUSE, grant 05A20BAB).
Based on observations collected at the European Organisation for Astronomical Research in the Southern Hemisphere under ESO programme 0101.A-0282(A).
We are also grateful to the communities who developed the many python packages used in this research, such MPDAF \citep{Piqueras17}, Astropy \citep{AstropyCollaboration13}, numpy \citep{Walt11}, scipy \citep{Jones01} and matplotlib \citep{Hunter07}.

\end{acknowledgements}

%
%

\bibliographystyle{aa}
\bibliography{mybib_aa}


\end{document}